\documentclass[a4paper, 12pt]{scrartcl}   
\pdfoutput=1
\usepackage[margin=2.7cm, bmargin=4cm]{geometry}
\usepackage[T1]{fontenc} 
\usepackage[ansinew]{inputenc} 
\usepackage{amsthm}
\usepackage{amssymb}
\usepackage{graphicx} 
\usepackage{subfigure} 
\usepackage{amsmath}  
\usepackage{braket} 
\usepackage{lmodern}
\usepackage{mathrsfs}
\usepackage{bbm}
\usepackage[bookmarks,colorlinks=green,linkcolor=blue]{hyperref}
\usepackage{pdfsync}
\usepackage{fancyhdr}
\usepackage{authblk}
\usepackage{cleveref}
\usepackage{bbding}
\usepackage{bbold}
\usepackage{niceframe}
\usepackage{xcolor}
\usepackage{overpic}
\usepackage[framemethod=tikz]{mdframed}
\usepackage[sort&compress]{natbib}
\setcitestyle{square,numbers}

\hypersetup{
	colorlinks,
	linkcolor={red!50!black},
	citecolor={blue!50!black},
	urlcolor={blue!50!black}
}
\usepackage[titletoc]{appendix}

\numberwithin{equation}{section}

\title{Can Gravitational Instantons Really Constrain Axion Inflation?}
\author{Arthur Hebecker$^{1}$, Patrick Mangat$^{1}$, Stefan Theisen$^{2}$, \\ Lukas T. Witkowski$^{1}$}
\affil{\small $^{1}$ \emph{Institut f\"ur Theoretische Physik, Universit\"at Heidelberg}, \\ \emph{Philosophenweg 19, D-69120 Heidelberg, Germany} \\ 
	$^{2}$ \emph{Max-Planck-Institut f\"ur Gravitationsphysik}, \\ \emph{Albert-Einstein-Institut, 14476 Golm, Germany}}
\date{22 July 2016}

\begin{document}

\maketitle
\normalsize
\thispagestyle{empty}

\begin{abstract}
	
Axions play a central role in inflationary model building and other cosmological applications. This is mainly due to their flat potential, which is protected by a global shift symmetry. However, quantum gravity is known to break global symmetries, the crucial effect in the present context being gravitational instantons or Giddings-Strominger wormholes. We attempt to quantify, as model-independently as possible, how large a scalar potential is induced by this general quantum gravity effect. We pay particular attention to the crucial issue which solutions can or cannot be trusted in the presence of a moduli-stabilisation and a Kaluza-Klein scale. An important conclusion is that, due to specific numerical prefactors, the effect is surprisingly small even in UV-completions with the highest possible scale offered by string theory. 

As we go along, we discuss in detail Euclidean wormholes, cored and extremal instantons, and how the latter arise from 5d Reissner-Nordstr\"om black holes. We attempt to dispel possible doubts that wormholes contribute to the scalar potential by an explicit calculation. We analyse the role of stabilised dilaton-like moduli. Finally, we argue that Euclidean wormholes may be the objects satisfying the Weak Gravity Conjecture extended to instantons.

\end{abstract}

\newpage
\setcounter{page}{1} 
\tableofcontents

\newpage

\section{Introduction and Summary of Results}
Slow-roll inflation relies on flat scalar potentials, making axion-like fields ideal inflaton candidates. This is especially true in the context of large-field inflation. The latter is of particular interest since, on the one hand, it is arguably the most natural form of inflation and, on the other hand, it will be discovered or experimentally ruled out in the foreseeable future.

The flatness of axion potentials (we denote the axion henceforth by $\theta$)  is protected by a shift symmetry which is only broken non-perturbatively, i.e.~by instantons. However, possible problems with consistently embedding axionic models of inflation in quantum gravity are an issue of continuing concern \cite{9502069, 0303252, 0601001,12035476,14022287,14095793, 14123457,150300795,150303886,150304783,150307853,150307912,150400659, 150403566,150603447,150800009,150906374,150907049,151002005,151105119, 151200025,151203768,160206517,160505311,160608437,160608438,160706105}. In particular, the focus has recently been on the Weak Gravity Conjecture \cite{0601001}. In the context of axions, it states that with growing axion decay constant $f_{\text{ax}}$ the action $S$ of the `lightest' instanton decreases, such that the flatness of the potential is spoiled by corrections $\sim \exp(-S)$.

However, the Weak Gravity Conjecture has not been firmly established. In particular, its validity remains unclear outside the domain of UV completions of quantum gravity provided by the presently understood string compactifications. This is even more true for the extension to axions. Moreover, the prefactors of the $\exp(-S)$ corrections mentioned above may be parametrically small, especially if SUSY or the opening up of extra dimensions come to rescue just above the inflationary Hubble scale. 

Thus, it is useful to pursue the related but complementary approach of constraining axionic potentials on the basis of gravitational instantons. Indeed, the very fundamental statement that quantum gravity forbids global symmetries is, in the context of shift symmetries, explicitly realised by instantonic saddle points of the path integral of Euclidean quantum gravity. These are also known as Giddings-Strominger wormholes \cite{Giddings:1987cg}. If, as proposed in \cite{150303886}, gravitational instantons yield significant contributions to the axion potential, some models of natural inflation would be under pressure (at least those with one or only few axions like alignment scenarios), while axion-monodromy inflation models seem to be unaffected.\footnote{Natural inflation \cite{Freese:1990rb} with one axion requires a transplanckian axion field space. Ideas for realising natural inflation in a subplanckian field space of multiple axions were proposed in \cite{0409138, 0507205, 14047496, 150301015}. For models implementing these ideas see e.g.~\cite{07103883, 9804177, 14012579, 14037507, 14044268, 14045235, 14046209, 14046988, 14047127, 14047773, 14047852, 14060341, 14072562, 14091221,14104660, 14114768, 150301777, 150302965, 150307183, 150307912, 160506299}. Axion monodromy inflation was introduced in \cite{08033085, 08080706} (for a field theory implementation see \cite{08111989,11010026}).  A realisation of this idea with enhanced theoretical control is $F$-term axion monodromy \cite{14043040,14043542,14043711}. For further work in this context see \cite{09121341, 14053652, 14057044, 14097075,14112032,14115380, 150301607, 150402103, 150507871, 151001522, 151008768, 151108820, 151204463, 160701680}.} 
It is our goal to study the effect of Euclidean wormholes and that of related instantonic solutions in detail. In particular, in the spirit of what was said above, we want to be as model-independent and general as possible, ideally relying only on Einstein gravity and the additional axion. The goal is to constrain large classes of string models or even any model with a consistent UV completion.  As we go along, we will however be forced to consider certain model-dependent features and take inspiration from the known part of the string theory landscape. 

The aim of this paper is thus to determine the strongest constraints on axion inflation due to gravitational instantons. One important aspect of our analysis is that -- to be as model-independent as possible  --  calculations are performed in an effective 4-dimensional Einstein-axion(-dilaton) theory. However, this theory is only valid up to an energy-scale $\Lambda$ and, for consistency, we have to make sure that our analysis only includes gravitational instanton solutions within the range of validity of our effective theory. 

This leads to the following challenge pointed out in \cite{150307912} (see also \cite{9502069}) and which we will repeat here. Given an energy cutoff $\Lambda$, gravitational instantons within the range of theoretical control contribute at most as $\delta V \sim e^{-S} \sim e^{-M_p^2 / \Lambda^2}$ to the axion potential. Then, gravitational instantons are dangerous for inflation if their contribution to the potential is comparable to the energy density in the inflationary sector, i.e.~$\delta V \sim H^2$. If the cutoff $\Lambda$ is not much above $H$ gravitational instantons are clearly harmless. However, if $\Lambda$ is close to $M_p$ gravitational instantons can easily disrupt inflation. As a result, the importance of gravitational instantons for inflation hinges on a good understanding of the scale $\Lambda$ where the 4-dimensional Einstein-axion(-dilaton) theory breaks down. 

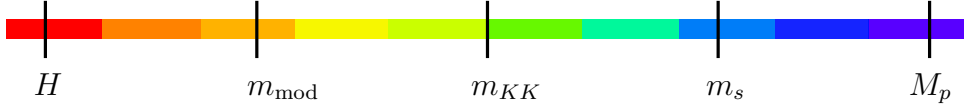
\begin{figure} 
	\centering
	\begin{overpic}[width=0.9\textwidth]{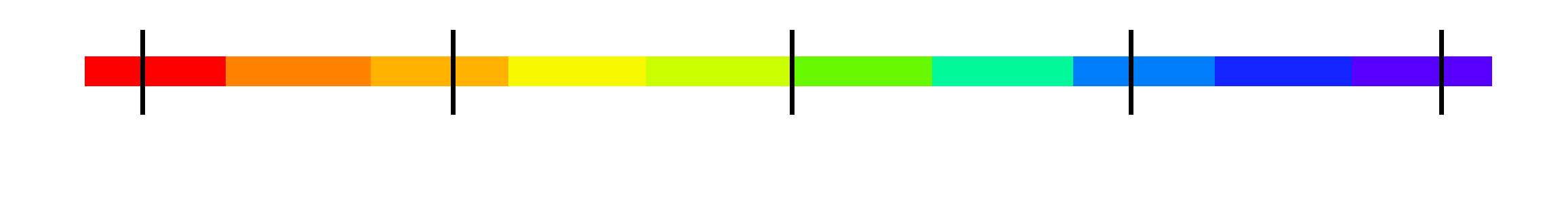}
		\put (8,2) {$H$} \put (28,2) {$m_{\textrm{mod}}$} \put (49,2) {$m_{KK}$} \put (71,2) {$m_s$}  \put (90,2) {$M_{p}$} 
	\end{overpic}
	\caption{Hierarchy of scales in a string model of inflation.}	
	\label{Fig:scales}
\end{figure}

To arrive at a quantitative expression for $\Lambda$ requires some knowledge about the UV completion of our theory. Here, we take string theory as our model of a theory of quantum gravity, i.e.~we assume that the effective Einstein-axion(-dilaton) theory is derived from string theory upon compactification. String compactifications give rise to a hierarchy of scales as shown in \autoref{Fig:scales}. Inflation is assumed to take place below the moduli scale $m_{\textrm{mod}}$ where only gravity and one or more axions are dynamical. Above $m_{\textrm{mod}}$ further scalars in the form of moduli become dynamical. As a result, if we want to work with a Einstein-axion theory the cutoff $\Lambda$ is the moduli scale. 

Here, we want to do better. An analysis using 4-dimensional gravitational instantons can in principle be valid up to the Kaluza-Klein (KK) scale $m_{KK}$, at which a description in terms of a 4-dimensional theory breaks down. However, to be able to go beyond $m_{\textrm{mod}}$ we have to allow for dynamical moduli. Hence, for this purpose Einstein-axion theories are insufficient and we have to study gravitational instantons in Einstein-axion-moduli theories instead.

These considerations give rise to the following structure of our paper. We start by recalling the Giddings-Strominger or Euclidean wormhole solution \cite{Giddings:1987cg} in \autoref{Section:Gravitational Instanton Solutions }. This is a classical solution of the axion-gravity system which gives space-time a `handle' with cross-section $S^3$. In fact, this solution can be interpreted as a real saddle point of the path integral only in the dual 2-form theory. We take some care to describe the relevant subtleties of the dualisation procedure in \autoref{Subsection: Dualisation}. Subsequently, we generalise to the case with an additional dilatonic scalar in \autoref{GravInstScalar}. Now extremal as well cored instanton solutions \cite{0406038,150906374} also exist. The situation with a dilaton is important for us as a model of the realistic string-phenomenology case with light moduli. \autoref{Subsection: Integration constant} focusses on the way in which cored and extremal gravitational instantons may arise from a Euclidean black 0-brane in an underlying 5d theory. In this way we obtain a UV-completion of cored and extremal gravitational instantons, which can then be understood by parameters of the 5d theory.  

\autoref{Section:Instanton Potentials from Wormholes} is devoted to the crucial issue whether a scalar potential is induced by Euclidean wormholes.  We will provide an explicit computation of the contributions to the axion potential from Euclidean wormholes. Thereby, we describe how to circumvent a recent counter-argument given in \cite{150906374}, suggesting that Euclidean wormholes could not break the axionic shift-symmetry. Thus, we stress that Euclidean wormholes are by no means less important than cored or extremal gravitational instantons.  

In \autoref{Section:Cored Grav Inst} we calculate the instanton actions for Euclidean wormholes as well as for cored and extremal gravitational instantons. We also give a quantitative answer to the question which gravitational instantons can be trusted within our effective theory with cutoff $\Lambda$. The result is as follows. As in the case of gauge instantons one can associate gravitational instantons with an instanton number $n$. Given an energy cutoff $\Lambda$ one can then only trust gravitational instantons with a sufficiently high instanton number $n \gg f_{\textrm{ax}} M_p / \Lambda^2$, where $f_{\textrm{ax}}$ is the axion decay constant.\footnote{Note that this implies that we neglect potentially more severe, but incalculable contributions due to instantons with low instanton numbers.}

In \autoref{Section:Moduli stabilisation} we take first steps towards studying gravitational instantons in the presence of dynamical moduli. We argue that the case with one light modulus coupled to the Einstein-axion theory can be modelled by an Einstein-axion-dilaton theory with massless dilaton. For one, in \autoref{sec:addmass} we show that for our purposes the modulus potential can be neglected if there is a sufficient hierarchy between the modulus mass and the cutoff $\Lambda$. The reason is that deep inside the `throat' of a gravitational instanton the modulus mass only gives a subleading contribution to the stress-energy tensor, while curvature and gradient terms dominate. As this region is also the source of the dominant part of the instanton action, we conclude that the action obtained for a massless modulus will remain a good approximation even in the massive case. We then motivate our restriction to moduli with dilatonic couplings. This implies that the modulus $\varphi$ is coupled to the axion $\theta$ through the kinetic term for the axion as $e^{\alpha \varphi} (\partial \theta)^2$. In \autoref{Subsection: Dilaton coupling from string theory} we review that dilatonic couplings arise frequently in string compactification. 

In \autoref{Section:Inflation} we analyse possible constraints for inflation due to gravitational instantons. To this end we identify the instantons with the largest contributions to the axion potential in \autoref{sec:mostdanger}. We arrive at the strongest constraint if the cutoff $\Lambda$ is as high as possible. In \autoref{sec:selfdual} we identify the highest possible cutoff $\Lambda_{\textrm{max}}$ for an effective 4-dimensional theory arising from a string compactification. This is given by the KK scale of a compactification with smallest possible compactification volume, which we take as the self-dual volume under T-duality. Unfortunately, there is an ambiguity in this definition of $\Lambda_{\textrm{max}}$ up to factors of $\pi$, which can be crucial. We then determine the maximal contribution $\delta V$ to the axion potential due to gravitational instantons and compare this to the scale of inflation in models of large-field axion inflation. Our main result is as follows. We find that gravitational instantons do not give rise to strong \emph{model-independent} constraints on axion inflation. Extremal gravitational instantons may be important for inflation, but this is model-dependent, as the size of their contribution depends on the value of the dilaton coupling $\alpha$.  

Last, in \autoref{Section: WGC} we record some observations regarding the Weak Gravity Conjecture (WGC) \cite{0601001} in the context of gravitational instantons. We pick up the idea from \cite{150906374} that extremal instantons play the role extremal charged black holes for the WGC. We then find that the WGC appears to be satisfied due to the existence of Euclidean wormholes. This either hints at a realisation of the WGC in the context of gravitational instantons, or implies  a different definition of the WGC in the presence of wormholes.

We summarise our findings in \autoref{Section:Conclusions} and point out directions for future work. Various appendices contain detailed computations on which some of our results are based, or clarify subtleties which are not absolutely essential for the understanding of the main body of the paper.  

Overall, our analysis leaves us with the following: a semi-classical approach to quantum gravity via gravitational instantons does not give rise to strong constraints for large-field inflation. Thus, if quantum gravity has anything to say about large field inflation, the quantum part will have to speak.

\section{Gravitational Instanton Solutions} \label{Section:Gravitational Instanton Solutions } 

A model of axion inflation will necessarily involve an axionic field coupled to gravity. One feature of such a system is that it may allow for \textit{gravitational instantons}, i.e.~finite-action solutions to the equations of motion of the Euclidean axion-gravity theory. 

Our starting point is the Euclidean action for an axionic field $\theta$ coupled to gravity, which takes the form ($M_p=1$)
\begin{equation} \label{Action with axion}
	S = \int d^4x \sqrt{g} \left[-\frac{1}{2}R + \frac{1}{2}K g^{\mu\nu} \partial_{\mu} \theta \partial_{\nu} \theta \right] \ .
\end{equation}
The prefactor $K$ can in principle depend on further fields. In this section we ignore the Gibbons-Hawking-York boundary terms, because we will be focussing on the dynamics of the system. Instead of working with the axionic field $\theta$, one can write the action in terms of the dual 2-form $B$ and its field strength $H=dB$:
\begin{equation} \label{Action with H}
	S = \int d^4x \sqrt{g} \left[-\frac{1}{2}R + \frac{1}{2}\mathcal{F} H_{\mu \nu \rho} H^{\mu \nu \rho} \right] \ ,
\end{equation}
where $\mathcal{F}= 1/(3! K)$. The field strength $H$ is related to $d \theta$ via 
\begin{equation}
	H = K \star d \theta \ .
\end{equation} 
The dualisation from \eqref{Action with axion} to \eqref{Action with H} must be done under the path integral using Lagrange multipliers. We will explain this in the following subsection. 

\begin{figure}
	\centering
	\includegraphics[width=0.70\linewidth]{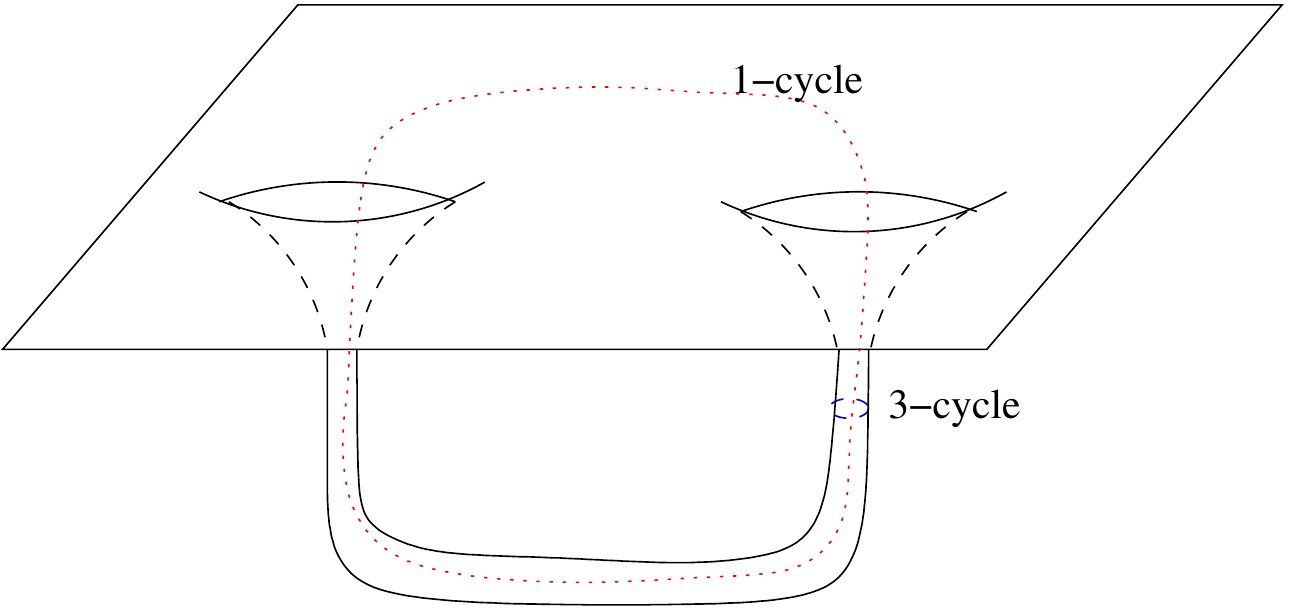}
	\caption{This picture illustrates a Euclidean wormhole, whose two ends are connected to the \textit{same} asymptotically flat space. Then there is a non-trivial 1-cycle (dotted line) passing through the wormhole. The cycle orthogonal to this 1-cycle is a $S^3$ (symbolised by the dashed line around the right-hand throat).}
	\label{Figure: Wormhole and 1-Cycle}
\end{figure}

In Euclidean space the theory of the 3-form $H$ coupled to gravity \eqref{Action with H} then has non-trivial solutions. In particular, \textit{gravitational instantons} are rotationally symmetric solutions with metric
\begin{equation} \label{Metric}
	ds^2 = \left( 1+ \frac{C}{r^4} \right)^{-1} dr^2 + r^2 d \Omega_3^2,
\end{equation}
where the parameter $C$ arises as a boundary condition or integration constant (see \autoref{Appendix: metric structure}).
For $C<0$ this is known as a \textit{Giddings-Strominger or Euclidean wormhole} \cite{Giddings:1987cg}: for large $r$ it approaches flat space, while for decreasing $r$ the geometry exhibits a throat with cross-section $S^3$. At $r=|C|^{1/4}$ one encounters a coordinate singularity, where another solution of this type can be attached (see e.g.~\autoref{Figure: Wormhole and 1-Cycle} and \ref{Figure: Types of Grav. Inst.}(a) for two possibilities). Gravitational instanton solutions for $C=0$ and $C>0$ can also be found if a dilaton-type field is included \cite{0406038,150906374}.

Before we extend our system to dilaton-type couplings, we review and discuss several subtleties involved in the aforementioned dualisation between $\theta$ and $B$ in Euclidean space.

\subsection{Dualisation} \label{Subsection: Dualisation}

For the sake of clarity, in this subsection we index the field variables by their rank, i.e.~we write $\theta_0$ and $B_2$. Those fields are sourced by an instanton and a microscopic string, respectively. We start from the two Euclidean actions in 4d:\footnote{The appearance of the $i$-factor in front of the coupling terms can be understood by writing these terms as $\int_M f_p \wedge j_{4-p}$ with $p$-form field $f_p$ and source current $j_{4-p}$.  One of the relevant tensor components of either $f_p$ or $j_{4-p}$ then always carries a zero-index and hence acquires an $i$-factor by Wick rotation.} 
\begin{align} \label{Action Theta}
	S[\theta_0] &= \int_M \frac{1}{2g_{\theta}^2} F_1 \wedge \star F_1 + iQ_{\theta} \int_I \theta_0, ~~~~~~~~F_1 = d \theta_0, \\ \label{Action B}
	S[B_2] &= \int_M \frac{1}{2g_{B}^2} H_3 \wedge \star H_3 + iQ_B \int_{\sigma} B_2, ~~~~H_3 = dB_2,
\end{align}
where $M$ denotes our 4-manifold, $I$ the set of points where the instantons are located, and $\sigma$ is the surface swept out by the string. One can identify the kinetic terms of \eqref{Action Theta} and \eqref{Action B} by imposing 
\begin{equation} \label{Dual relation for B and theta}
	H_3 =  g_B^2 \star F_1
\end{equation}
and $g_B^2 = 1/g_{\theta}^2$.  This now becomes a single theory with both strings and instantons allowed and either $\theta_0$ or $B_2$ to be used locally as the appropriate field variable. 

Note that the $H_3$-flux is quantised by 
\begin{equation} \label{Charge quantisation condition}
	\int_{S^3} H = n \in \mathbbm{Z},
\end{equation}
as we review in \autoref{Appendix: Charge Quantisation} in the context of the existence of fundamental strings and instantons. 

We now couple the 1-form/3-form theory to gravity. It is well-known that choosing either $\theta_0$ or $B_2$ as the fundamental field leads to Einstein equations differing by an overall sign \cite{Giddings:1987cg}. Indeed, the action of \eqref{Action Theta} gives the energy-momentum tensor 
\begin{equation} \label{Energy-Momentum tensor theta}
	T_{\mu\nu}^{(\theta)} = \frac{1}{g_{\theta}^2} \left(- \frac{1}{2} g_{\mu\nu}(\partial \theta_0)^2 + \partial_{\mu} \theta_0 \partial_{\nu}\theta_0 \right)\,,
\end{equation} 
while \eqref{Action B} leads to
\begin{align} \label{Energy-Momentum tensor B}
	T_{\mu\nu}^{(B)} &= \frac{1}{g_{B}^2} \left(- \frac{1}{2 \cdot 3!} g_{\mu\nu} H_3^2 + \frac{1}{2} H_{\mu \rho \sigma}H_{\nu}^{~\rho \sigma}\right) = \\ \nonumber
	&= -\frac{1}{g_{\theta}^2} \left(- \frac{1}{2} g_{\mu\nu}(\partial \theta_0)^2 + \partial_{\mu} \theta_0 \partial_{\nu}\theta_0 \right) = - T_{\mu\nu}^{(\theta)}\,.
\end{align}
In the second line we used \eqref{Dual relation for B and theta} together with $g_B^2 = 1/g_{\theta}^2$. 

The above sign difference implies that Euclidean wormholes exist in the $B_2$ but not in the $\theta_0$ formulation. Technically, this is due to the Hodge star being introduced before or after the variation w.r.t. the metric. Also at the intuitive level the difference is clear: The $H_3$-flux on the transverse $S^3$, which is fixed due to the Bianchi identity, supports the finite-radius throat. By contrast, the dual quantity $\theta^{\prime} \equiv \partial_r \theta$, i.e. the variation of $\theta$ along the throat, is \textit{not} fixed by the dual Bianchi identity and the solution is lost. 

We note that the \textit{Minkowski-space} Einstein equations remain the same on both sides of the duality. However, we are interested in the path integral in the \textit{Euclidean} theory with gravity, so this observation does not help.

Thus, one may wonder whether Giddings-Strominger wormholes do contribute to the action or whether the dual descriptions are really fully equivalent. This problem has been intensively investigated in the past, see e.g.~\cite{Giddings:1987cg, Grinstein:1988ja,Lee:1988ge,Abbott:1989jw,Brown:1989df,Burgess:1989da, Coleman:1989zu,9502069,9511080,9604038,9608065,9701093,0510048,Collinucci:2005opa, 07052768,10116301} and our present understanding mainly derives from \cite{0510048,Collinucci:2005opa,07052768}.

Indeed, it should be possible to resolve the problem by dualising under the Euclidean path integral and following the fate of the instanton solution. We review the dualisation following \cite{Burgess:1989da,0510048,Collinucci:2005opa}. To be specific, let $M$ be a cylinder, $M=S^3\times I$, with an interval $I\subset \mathbb{R}$. This is the simplest relevant topology since the $S^3$ can carry $H_3$-flux, supporting a narrow throat somewhere within $I$. 

Starting on the $B_2$-side, the partition function reads
\begin{equation}
	Z \sim \int_{\text{b.c.}} d[B_2] \exp \left(- \int_M \frac{1}{2g_{B}^2}  dB_2 \wedge \star dB_2\right)\,,
\end{equation}  
where ``b.c.'' denotes the boundary conditions $B_2(S^3_I) \equiv B_2^{(I)}$ and $B_2(S^3_F) \equiv B_2^{(F)}$ at the initial and final boundaries $S^3_I$ and $S^3_F$. The possibility of a non-trivial flux, $\int_{S^3} H_3\neq 0$, can as usual be implemented by defining $B_2$ in patches over the transverse $S^3$ and choosing appropriate transition functions. 

One can also express $Z$ as a path integral over $H_3$, imposing $dH_3=0$ with the help of a Lagrange-multiplier $\theta_0$:  
\begin{equation} \label{Partition function with Lagrange multiplier}
	Z \sim \int_{\text{b.c.}} d[H_3]d[\theta_0] \exp \left\{- \int_M \frac{1}{2g_{B}^2}  \left(H_3 \wedge \star H_3 + 2 ig_B^2 \theta_0 dH_3 \right) \right\}\,.
\end{equation}
The previous $B_2$-boundary conditions now translate into boundary conditions on the pullback\footnote{
	This is \textit{not} the same as $H_3$ at the position of the boundaries, which contains time-derivatives of $B_2$ and should not be constrained.
} 
of $H_3$ to the initial and final boundary, i.e.~$H_3(S^3_I) \equiv H_3^{(I)}$ and $H_3(S^3_F) \equiv H_3^{(F)}$. In this language, the information about a possible $H_3$-flux is simply part of the $H_3$ boundary conditions. The $\theta_0$-integral is unconstrained. The $i$ in front of the Lagrange-multiplier is needed to get a delta-functional $\delta(dH_3)$ in the path integral after integrating out $\theta_0$. Hence, we have $dH_3=0$ and Stokes theorem yields $H_3^{(I)} = H_3^{(F)}$. In other words $Z \sim \delta(H_3^{(I)} - H_3^{(F)})$.

Equation \eqref{Partition function with Lagrange multiplier} can be rewritten 
by integrating the second term by parts and completing the square: 
\begin{align} \label{Gaussian Path Integral H_3}
	Z \sim \int_{\text{b.c.}} & d[H_3]d[\theta_0]  \exp \left\{ - i \int_{\partial M} \theta_0H_3 \right\}  \\ \nonumber
	&\exp \left\{- \int_M \frac{1}{2g_{B}^2}  \left[ \left(H_3 - ig_B^2 \star d \theta_0\right) \wedge \star \left(H_3 - ig_B^2 \star d \theta_0\right) +g_B^4 d\theta_0 \wedge \star d \theta_0 \right]\right\}. 
\end{align}
According to \cite{Burgess:1989da,0510048,Collinucci:2005opa} one can now shift the variable $H_3 \to \tilde{H}_3\equiv H_3 - ig_B^2 \star d\theta_0$ and trivially perform the Gaussian integral. One may however also be concerned about this step since, for any fixed $\theta_0$, the boundary conditions, e.g. $\tilde{H}_3(S^3_I)=H_3(S^3_I)- ig_B^2 \star d\theta_0$, clash with the saddle point value $\tilde{H}_3=0$ of the Gaussian integral in the interior of $M$. 

To make this issue more explicit, let us write $H_3=\langle H_3\rangle + \delta H_3$, where $\langle H_3\rangle$ is constant along the $S^3$ but time dependent. Its boundary values are determined by the $H_3$-flux. Furthermore, decompose $\delta H_3$ into spherical harmonics on $S^3$. If the cylinder $M$ were flat and gravity non-dynamical, we would now simply have a quantum mechanical system of infinitely many, independent oscillators. The dualisation process sketched above would correspond, as is well known from $T$-duality for a scalar field on the cylinder $S^1\times\mathbbm{R}$, to a canonical transformation ($p\leftrightarrow q$) for each oscillator. In our case, the dual variables are coefficients of the spherical harmonic decomposition of $\theta_0$. 

Let us focus on the most interesting subsystem (see also the discussion in \cite{07052768}) with the variable $\langle H_3\rangle\sim p$ and the dual variable $\langle \theta_0\rangle\sim q$. Thus, we first restrict our attention to the question whether it is correct to naively integrate out $q$ in
\begin{equation}
	Z\sim \int_{\mathrm{b.c.}}d[p]\,\int d[q]\,\exp\left\{-\frac{1}{2}\int_{t_i}^{t_f} dt\,\left[(p-i\dot{q})^2+\dot{q}^2\right]\right\}\,.
\end{equation}
Based on an explicit, discretised calculation in \autoref{Appendix: Dualisation}, we claim this is indeed the case. One can now argue that, also for the full system \eqref{Gaussian Path Integral H_3} including all oscillators and gravity, this formal manipulation with path integrals is correct. It will then also remain correct if, as argued in \autoref{Appendix: Charge Quantisation}, $\langle H_3\rangle$ is initially quantised, i.e. $\int_{S^3}\langle  H_3 \rangle = n \in \mathbb{Z}$. Indeed, this quantisation is `neutralised' once the Lagrange multiplier is introduced and the now continuous variable $\langle H_3\rangle $ is integrated out as above.

As a result of all this the partition function can eventually be given as
\begin{equation} \label{Partition function theta}
	Z \sim \int d[\theta_0] \exp \left(-\int_M \frac{1}{2g_{\theta}^2} d\theta_0 \wedge \star d \theta_0 - i \int_{\partial M} \theta_0H_3\right),
\end{equation}
where $g_{\theta}^2=1/g_B^2$. We emphasise that the sign of the kinetic term is the one required for a well-defined Euclidean path integral. This sign will become important below. We also note that this procedure can be straightforwardly generalised to any $p$-form in arbitrary dimensions $d >p$.  Moreover, we observe that despite the shift $H_3 \to H_3 - ig_B^2 \star d\theta_0$, the field $\theta_0$ can be kept real (see also \cite{Collinucci:2005opa}).\footnote{In other references, e.g.~\cite{9701093,10116301}, the axion field was taken to be imaginary. Then, however, we do not see how to ensure $dH_3=0$ using \eqref{Partition function with Lagrange multiplier}.} 

Varying the action in \eqref{Partition function theta},
\begin{equation}
	\delta S =  \int_M\frac{1}{g_{\theta}^2} \delta \theta_0 d \star d \theta_0 - \int_{\partial M}\frac{1}{g_{\theta}^2} \delta \theta_0 \star d \theta_0 - i \int_{\partial M} \delta \theta_0H_3 \stackrel{!}{=}0\,,
\end{equation}
we find the equation of motion $d \star d \theta_0=0$ in the bulk and 
\begin{equation} \label{Complex saddle}
	H_3(\partial M) = \frac{i}{g^2_{\theta}} \star d \theta_0 (\partial M) 
\end{equation} 
at the boundary. Thus, the $\theta_0$ path integral has only complex saddle points \cite{0510048,Collinucci:2005opa}.\footnote{
	For 
	a treatment of path integrals with complex phase space or complex saddles, see e.g.~\cite{10096032} and \cite{151000978}, respectively.
}
Indeed, the possibility of taking $\theta_0$ imaginary at stationary points was discussed before, see e.g.~\cite{Burgess:1989da,Coleman:1989zu}.

To summarise, dualisation leads to a Euclidean path integral in which $\theta_0$ is a priori real and the kinetic term has the standard sign. However, a semi-classical evaluation is only possible on the basis of complex saddles. Crucially, the relevant field-theory solutions then also solve Einstein equations because imaginary $\theta_0$ flips the sign of $T_{\mu\nu}^{(\theta)}$ (cf. \eqref{Energy-Momentum tensor B}). Thus, one can expect gravitational instantons to contribute consistently both in the $B_2$ and the $\theta_0$ formulation. Nevertheless, it is natural to use the $B_2$ path integral to keep the saddle points real \cite{0510048,Collinucci:2005opa}, and we will do so in what follows.

\subsection{Gravitational Instantons in the Presence of a Massless Scalar Field}
\label{GravInstScalar}

One goal of this paper is to study the effect gravitational instantons can have on geometric moduli of string compactifications. In the 4-dimensional theory these moduli appear as scalar fields. Consequently, we will study systems of an axion $\theta$ and a scalar $\varphi$ coupled to gravity.\footnote{A string compactification will typically give rise to many axionic fields and many geometric moduli. We focus here on one, potentially super-Planckian, light axion which may be identified with the inflaton. Similarly, the scalar can be identified with the lightest modulus. Note that the analysis in this subsection neglects any mass term for the modulus $\varphi$, which will be included only later in \autoref{Section:Moduli stabilisation}.} The relevant Euclidean action then takes the form
\begin{equation} \label{Action with complex scalar}
	S = \int d^4x \sqrt{g} \left[-\frac{1}{2}R + \frac{1}{2} K(\varphi) g^{\mu\nu} \partial_{\mu} \theta \partial_{\nu} \theta + \frac{1}{2}  g^{\mu\nu} \partial_{\mu} \varphi \partial_{\nu} \varphi \right].
\end{equation}
Here we already canonically normalised the field $\varphi$. At 2-derivative level, the axion $\theta$ can only enter the action through a term $\partial_{\mu} \theta \partial^{\mu} \theta$ due to its shift symmetry. There is no such symmetry for $\varphi$ and hence the kinetic term for $\theta$ can in general depend on $\varphi$. This situation is typically encountered in string compactifications, see \autoref{Subsection: Dilaton coupling from string theory} for examples. In this subsection we consider a massless scalar field $\varphi$ and apply the subsequent results to the case of a massive scalar  in \autoref{Section:Moduli stabilisation}.

As we are interested in gravitational instantons, we should consider the dual formulation of the above theory. The relevant Euclidean action is then 
\begin{equation} \label{Action without potential}
	S = \int d^4x \sqrt{g} \left[-\frac{1}{2}R + \frac{1}{2} \mathcal{F}(\varphi)H^2 +  \frac{1}{2} g^{\mu\nu} \partial_{\mu}\varphi \partial_{\nu} \varphi \right],
\end{equation}
where $\mathcal{F}=1/(3!K)=1/(3!f_{\text{ax}}^2)$. Here $f_{\text{ax}}$ is the $\varphi$-dependent analogue of the familiar axion decay constant.   

In the following we will review explicit solutions of this system corresponding to gravitational instantons. Following \cite{150906374} we will construct solutions to the equations of motion for the metric, the 3-form $H$ and the scalar $\varphi$.

\subsubsection*{General solution}
\begin{figure}
	\subfigure[Wormhole connected to another universe]{\includegraphics[width=0.32\textwidth]{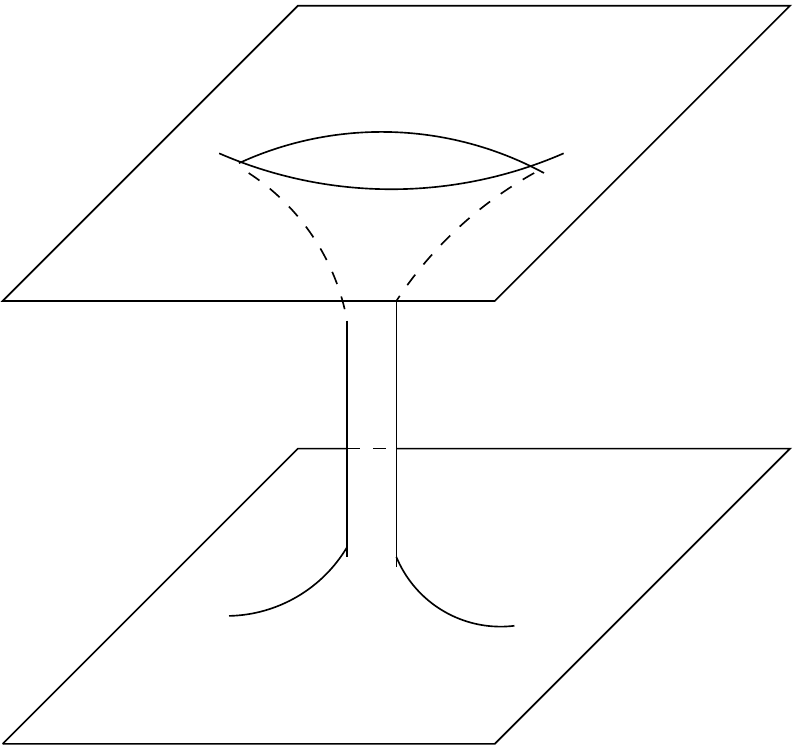}}
	\subfigure[Extremal gravitational instanton]{\includegraphics[width=0.32\textwidth]{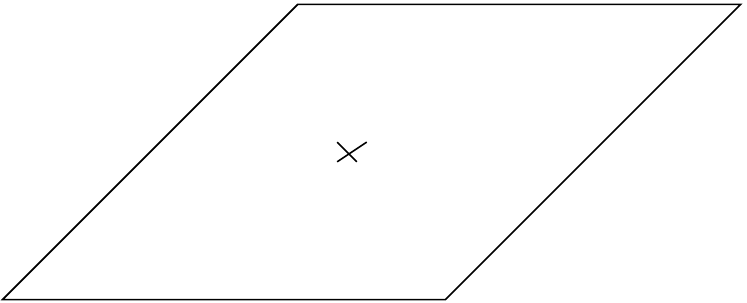}} 
	\subfigure[Cored gravitational instanton]{\includegraphics[width=0.32\textwidth]{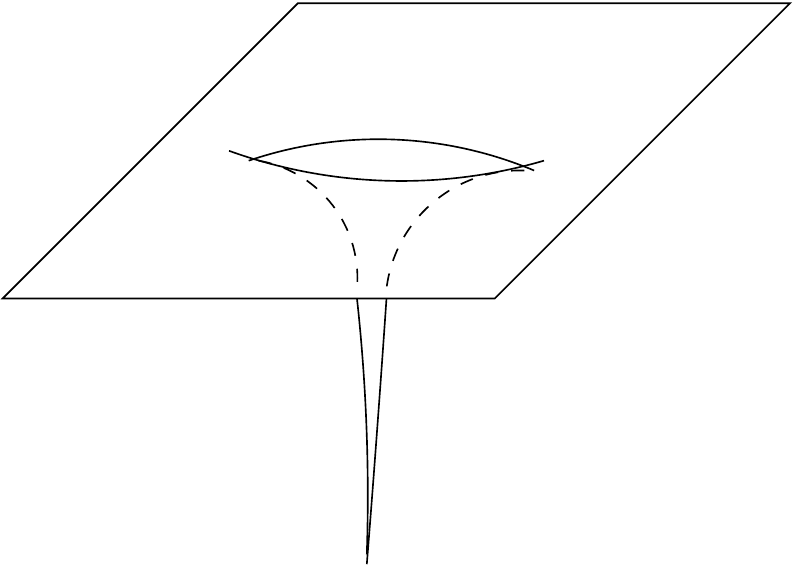}} 
	\caption{The three types of gravitational instantons are depicted. (a) Euclidean wormhole connecting two asymptotically flat spaces. It is also possible to connect both ends to the same space as shown in \autoref{Figure: Wormhole and 1-Cycle}. (b) Extremal gravitational instanton: in this case space is flat everywhere. The cross in the middle indicates the locus $r=0$. (c) Cored gravitational instanton: there is a curvature singularity at $r=0$.}
	\label{Figure: Types of Grav. Inst.}
\end{figure}

For completeness, let us recall the metric given in \eqref{Metric}:
\begin{equation} 
	\nonumber ds^2 = \left( 1+ \frac{C}{r^4} \right)^{-1} dr^2 + r^2 d \Omega_3^2 .
\end{equation}
The derivation of the functional form of $g_{rr}$ can be found in \autoref{Appendix: metric structure}.
There we show that the equation of motion for $g_{rr}$ decouples from the equations of motion of the massless fields $\varphi$ and $B$. In particular, the form of the metric \eqref{Metric} is independent of the functional form of the kinetic terms of these fields.  
The constant $C$ can a priori be negative, positive or zero. Depending on the sign of this parameter $C$, this solution has the following interpretations. Using the terminology of \cite{0406038,150906374} we can distinguish between three types of gravitational instantons (see \autoref{Figure: Types of Grav. Inst.} for an illustration).
\begin{itemize}
	\item \textit{Euclidean wormholes} ($C<0$):\newline
	The case $C<0$ leads to a geometry with a throat and we call this solution a Euclidean wormhole. The divergence of $g_{rr}$ at $r =r_0 \equiv |C|^{1/4}$ is only a coordinate singularity. The Ricci scalar $R$ is 
	\begin{equation}
		R = 6 \frac{C}{r^6}
	\end{equation}
	and thus it is finite for all $r \geq r_0$. The locus $r=r_0$ can then be interpreted as the end of one wormhole throat. We can then attach another solution of this type at $r=r_0$ which can either be attached to our universe (see \autoref{Figure: Wormhole and 1-Cycle}) or a different universe (see \autoref{Figure: Types of Grav. Inst.}). In this paper we will only consider wormholes which close again in our universe, i.e.~we are dealing with pairs of holes each connected by a ``handle''.
	\item \textit{Extremal instantons} ($C=0$):\newline
	The solution for $C=0$ is called an extremal gravitational instanton \cite{0406038,150906374}. Even though space is flat in that case, the fields $\varphi$ and $\theta$ still exhibit a nontrivial profile. This is possible due to a complete cancellation of terms in the energy-momentum tensor \cite{Rey:1989xj}. 
	\item \textit{Cored gravitational instantons} ($C>0$):\newline
	The case $C>0$ gives rise to a geometry with a curvature singularity at $r=0$. Such solutions are called \textit{cored gravitational instantons} \cite{150906374}.
\end{itemize}

Having reviewed the solution for the metric, we will now solve the equation of motion for $H$ without specifying $\mathcal{F}(\varphi)$.  From \eqref{Action without potential} we obtain the equation of motion:
\begin{equation} \label{EoM for B}
	d \star H = - \frac{\mathcal{F}^{\prime}(\varphi)}{\mathcal{F}(\varphi)} d \varphi \wedge \star H.
\end{equation}
We expect that solutions for $\varphi$ and $H$ should exist that respect the spherical symmetry of the background. We thus propose that $\varphi = \varphi(r)$. Similarly, following \cite{Giddings:1987cg}, we make the ansatz
\begin{equation} \label{Ansatz for H}
	H = h(r) \epsilon
\end{equation}
with $\epsilon$ the volume form on $S^3$ such that 
\begin{equation}
	\int_{S^3} \epsilon = 2 \pi^2 r^3 \ .
\end{equation}
From \eqref{Ansatz for H} it follows that $\star H \sim h(r) dr$ and the LHS of \eqref{EoM for B} vanishes.  As we have chosen $\varphi= \varphi(r)$ the RHS of \eqref{EoM for B} equally vanishes and the equation of motion for $H$ is satisfied. 

In addition, $H$ also has to satisfy the Bianchi identity $dH=0$. This enforces
\begin{equation} \label{Solution for h(r)}
	h(r) = \frac{n}{Ar^3},
\end{equation}
with $A \equiv A(S^3)=2\pi^2$ the area of the unit sphere. Charge quantisation \eqref{Charge quantisation condition} implies that $n \in \mathbbm{Z}$. 

In order to find the solution for $\varphi$ it is sufficient to consider the $rr$-component of the 
Einstein equations, $G_{rr} = T_{rr}$, which can be shown to be equivalent to the Klein-Gordon equation for $\varphi$. It reads 
\begin{equation} \label{Einstein equation without potential}
	\frac{1}{2} \left(1+ \frac{C}{r^4}\right) (\varphi^{\prime})^2 - \frac{3 \mathcal{F}(\varphi)n^2/A^2 + 3C}{r^6} = 0 \ ,
\end{equation}
where we already used the solution for $H$. We also defined $\varphi' \equiv \partial \varphi / \partial r$. The solution for $\varphi$ can then be found by integrating this differential equation.

\subsubsection*{Model-dependent solutions}
From \eqref{Einstein equation without potential} it is clear that explicit solutions for $\varphi$ will depend on the functional form of the term $\mathcal{F}(\varphi)$. In this subsection we will restrict our attention to functions of the form $\mathcal{F}(\varphi) \sim \exp(- \alpha \varphi)$, where we choose without loss of generality $\alpha >0$, as this functional form arises frequently in string compactifications. For example, this behaviour is observed when $\varphi$ is identified with the dilaton. Similarly, the same functional form appears if $\varphi$ corresponds to the volume modulus in setups with large compactification volume (e.g.~\cite{0502058}) or if $\varphi$ is a complex structure modulus at large complex structure. We will study such examples in \autoref{Subsection: Dilaton coupling from string theory}.

To be specific, we take
\begin{equation}
	\mathcal{F}(\varphi) =\frac{1}{3!f_{\text{ax}}^2 }  \exp(-\alpha \varphi) \ ,
\end{equation} 
where $f_{\text{ax}}$ is from now on a constant. The value of the parameter $\alpha$ will depend on the type of geometric modulus. We can assume $\lim_{r \to \infty} \varphi(r) =0$ without loss of generality. Then $f_{\text{ax}}$ will correspond to the asymptotic value of the axion decay constant.

In the following, we will summarise the explicit solutions for $\varphi$ for the Euclidean wormhole, the extremal instanton and for the cored instanton. Further details can be found in \autoref{Appendix: analytical solution to Einstein}.

\begin{figure}
\centering
\includegraphics[width=0.7\linewidth]{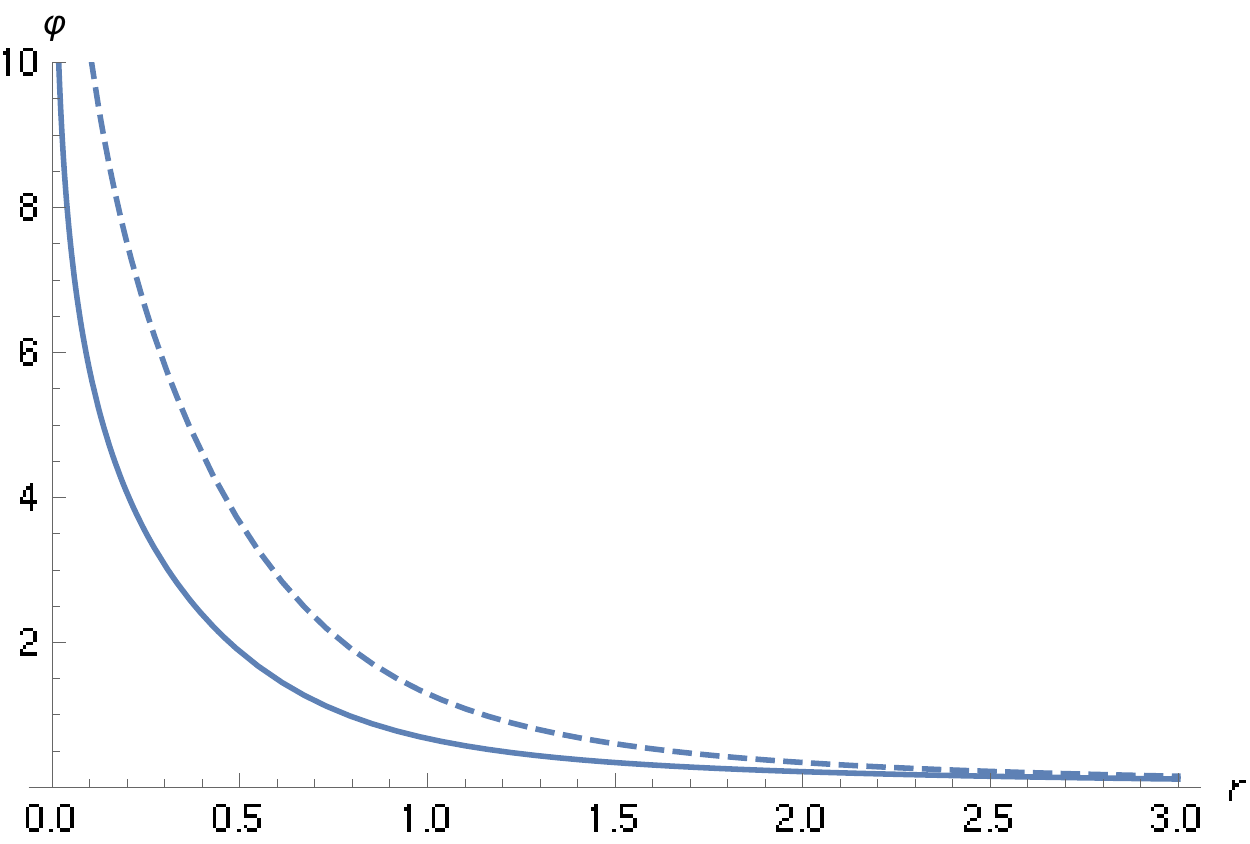}
\caption{This plot shows dilaton profiles for the cored gravitational instanton with $\alpha = 15$ (solid line) and $\alpha=0.1$ (dashed line). Again, $r$ and $\varphi$ are given in Planck units. For the purpose of illustration we have chosen $ K_+ =0.5$ and $n/f_{\text{ax}}$ such that $C/\sinh^2 K_+=1$.}
\label{Fig: Dilaton profile C>0}
\end{figure}

\begin{itemize}
	\item \textit{Euclidean Wormhole} ($C<0$):\newline 
	The analytical solution to \eqref{Einstein equation without potential} in this case is  \cite{Giddings:1987cg,0406038}
	\begin{equation} \label{Solution for C<0, with constant}
		e^{\alpha \varphi (r)} = \frac{1}{\cos^2(K_-)} \cos^2 \left(K_- + \frac{\alpha}{2} \sqrt{\frac{3}{2}} \arcsin \left(\frac{\sqrt{|C|}}{r^2}\right)\right) \ .
	\end{equation} 
	Here, we already implemented the boundary condition $\lim_{r \to \infty}\varphi(r) =0$, which also implies that
	\begin{equation} \label{C for wormholes}
		C = - \frac{n^2}{3!f_{\text{ax}}^2A^2} \cos^2(K_-).
	\end{equation}  
	The integration constant $K_-$ is not a free parameter. This can be seen as follows. When the field reaches the wormhole throat at $r=r_0 \equiv |C|^{1/4}$, the factor $(1+C/r^4)$ in \eqref{Einstein equation without potential} vanishes, hence 
	\begin{equation} \label{Condition finiteness of velocity}
		3 \mathcal{F}(\varphi(r_0))n^2/A^2 + 3C = 0 \ .
	\end{equation} 
	Using \eqref{C for wormholes}, this translates to 
	\begin{equation}
		\cos^2 \left(K_-+ \frac{\alpha \pi}{4}\sqrt{\frac{3}{2}}\right) =1,
	\end{equation}
	and thus
	\begin{equation}
		K_- = - \frac{\alpha \pi}{4}\sqrt{\frac{3}{2}}.
	\end{equation}
	Inserting this back into the solution yields 
	\begin{equation} \label{Solution for C<0 w/o constant K_-}
		e^{\alpha \varphi (r)} = \frac{1}{\cos^2(\sqrt{3/2} \alpha \pi/4 )} \cos^2 \left(\frac{\alpha}{2} \sqrt{\frac{3}{2}} \arccos \left(\frac{\sqrt{|C|}}{r^2}\right)\right).
	\end{equation}
	To see that one can take two wormhole solutions and glue them together, let us now change coordinates by writing $r= a(t)$ such that the metric becomes
	\begin{equation} \label{Metric with coordinate t}
		ds^2 = dt^2 + a^2(t) d \Omega_3^2 \ .
	\end{equation} 
	One can show that $a(t)$ and $\varphi(t)$ are symmetric under $t \to - t$. This implies the existence of a ``handle'' as shown in \autoref{Figure: Wormhole and 1-Cycle}, assuming also that the two throats are very distant in $\mathbb{R}^4$.
	
	Interestingly, not all values for $\alpha$ will lead to physically acceptable solutions. 
	Note that $\varphi(r)$ is regular everywhere on $r \in [|C|^{1/4}, +\infty)$ only for dilaton couplings in the range $0 \leq \alpha < 2 \sqrt{2/3}$.  For $\alpha > 2 \sqrt{2/3}$ there is always a value of $r> |C|^{1/4}$, where $e^{\alpha \varphi(r)}=0$, i.e.~$\varphi(r) \to - \infty$. This is consistent with \cite{Giddings:1987cg,9502069,0406038}.  In our case the field $\varphi$ corresponds to the string coupling or a geometric modulus of the string compactification. A runaway behaviour $\varphi(r) \to - \infty$ is then pathological as it would correspond to a limit of decompactification or vanishing string coupling. In all these cases new light states will appear resulting in a loss of control over the effective theory. This pathology is avoided for $\alpha = 2 \sqrt{2/3}$. However, in this case we obtain $C=0$ which will be discussed next. Overall, we find that only the range $0 \leq \alpha < 2 \sqrt{2/3}$ is physically allowed for Euclidean wormholes. 

Last, note that the limit $\alpha \to 0$ can be identified with the Giddings-Strominger wormhole \cite{Giddings:1987cg} which exhibits a constant dilaton profile. 
	
	\item \textit{Extremal Instanton} ($C=0$):\newline 
	For the case of an extremal instanton we find
	\begin{equation} \label{Solution for C=0}
		e^{\alpha \varphi(r)} = \left(1+ \frac{\alpha n}{4A f_{\text{ax}}} \frac{1}{r^2}\right)^2,
	\end{equation}
	which is valid for all $\alpha > 0$. (For $\alpha = 2 \sqrt{2/3}$ this solution agrees with \eqref{Solution for C<0 w/o constant K_-}. A plot of the dilaton profile in this case can be found in \autoref{Fig: Dilaton profile for C<0}.). The result can be obtained most easily by solving \eqref{Einstein equation without potential} for $C=0$. Notice that \eqref{Solution for C=0} with a minus sign in the bracket would in principle also be a solution (see \autoref{Appendix: analytical solution to Einstein}), but then there would again be a value of $r>0$ so that $e^{\alpha \varphi}=0$, leading to the same problems as described above. We hence exclude this possibility.
	\begin{figure}
		\subfigure[]{\includegraphics[width=0.50\textwidth]{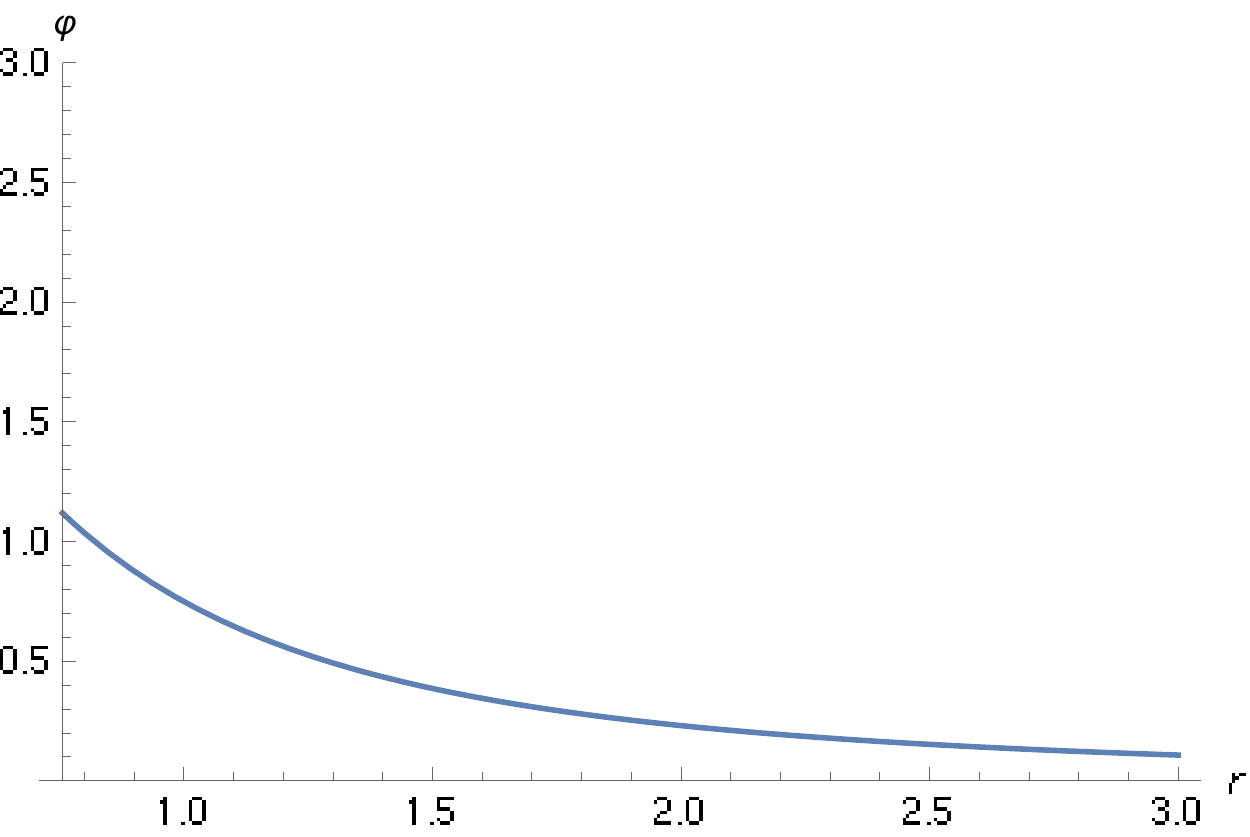}}
		\subfigure[]{\includegraphics[width=0.50\textwidth]{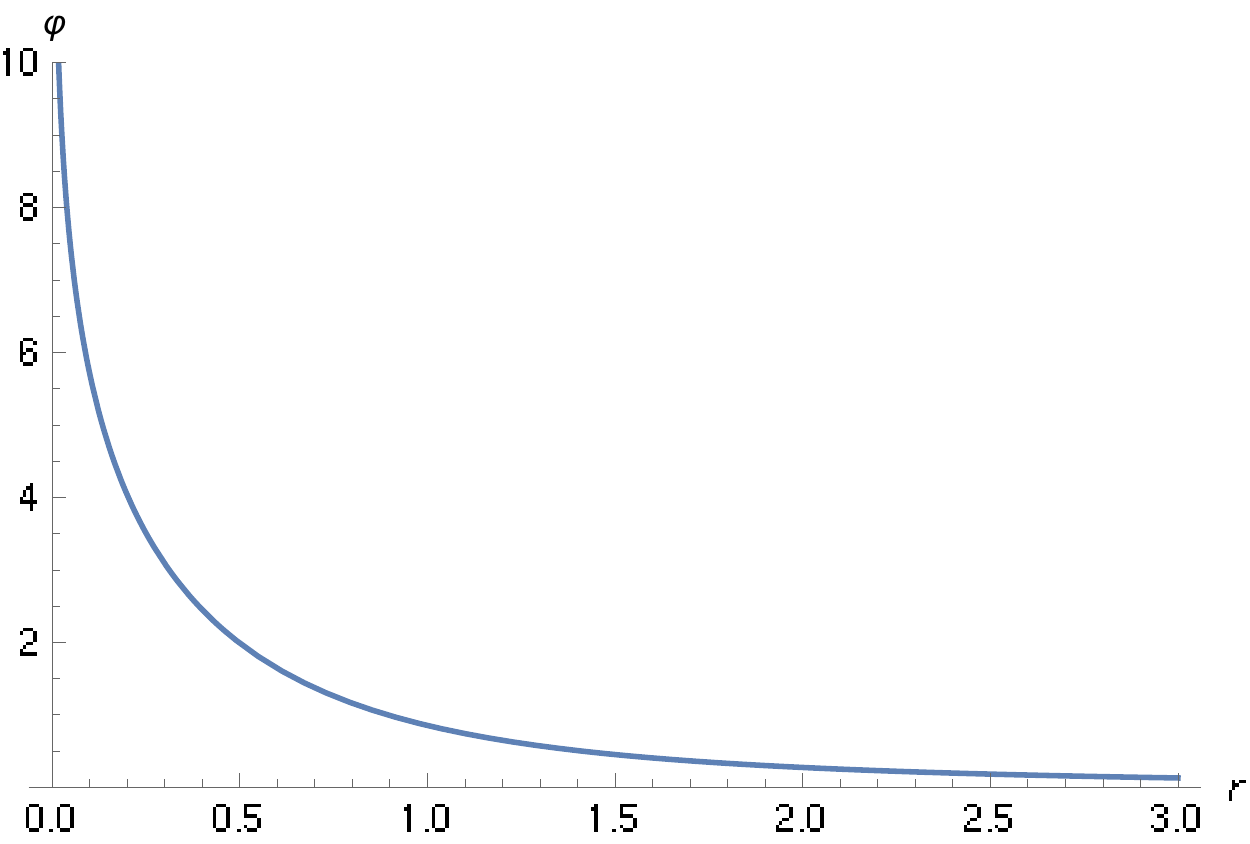}}
		\caption{Illustration of dilaton profiles. The values of $r$ and $\varphi$ are in Planck units. \newline (a) Euclidean wormhole ($C<0$): \newline Here we choose $n / f_{\text{ax}}$ such that $C= - \cos^2 \left( \alpha \pi \sqrt{3/2}/4\right)$ and plot for $\alpha=1$. \newline (b)  Extremal instanton ($C=0$) with $\alpha= 2 \sqrt{2/3}$.}
		\label{Fig: Dilaton profile for C<0}
	\end{figure} 
	
	\item \textit{Cored gravitational instantons} ($C>0$):\newline
	Finally, for the case of cored gravitational instantons $C>0$ one finds \cite{0406038,150906374} 
	\begin{equation} \label{Solution for C>0}
		e^{\alpha \varphi(r)} = \frac{1}{\sinh^2(K_+)} \sinh^2 \left(K_+ +  \frac{\alpha}{2} \sqrt{\frac{3}{2}} \text{arcsinh} \left(\frac{\sqrt{C}}{r^2}\right)\right),
	\end{equation}
	where we again ensured $\lim_{r \to \infty} \varphi(r) =0$ by demanding 
	\begin{equation} \label{C for C>0}
		C =  \frac{n^2}{3!f_{\text{ax}}^2A^2} \sinh^2(K_+) \ .
	\end{equation}
	In \autoref{Fig: Dilaton profile C>0} two plots of the dilaton profile are presented. The integration constant $K_+$ should be positive in order to again avoid a divergence of $\varphi$ for some $r>0$, but is otherwise unconstrained. This is different compared to wormholes or extremal instantons, which do not exhibit a free parameter. From this 4d effective theory one is lead to believe that there exists a whole family of cored instanton solutions parametrised by $K_+$. However, by considering the microscopic origin of gravitational instanton solutions, one finds evidence that only certain values of $K_+$ are allowed, as  we will now discuss.
\end{itemize}

\subsection{Interpretation of the Integration Constant $K_+$} \label{Subsection: Integration constant}

The integration constant $K_{+}$, or equivalently $C$, seems to be a free and continuous parameter giving rise to a family of solutions. We want to argue that this is not the case. Note that the cored gravitational instanton solutions are UV-sensitive and therefore a naive 4d field theory treatment is not sufficient. Instead, it is crucial to understand those solutions in a UV-complete theory, such as string theory. In this context the role of the integration constant $K_+$ becomes clear. Specifically, it was pointed out in \cite{0406038} that the parameter $C$ is determined by the mass $M$ and charge $Q$ of a dilatonic black brane wrapping internal cycles in a higher-dimensional theory, whose dimensionally reduced action coincides with \eqref{Action without potential}.  This holds true at least for some values of $\alpha$. Consequently, we conjecture that $C$ and $K_+$ generically take discrete and well-defined values determined by the underlying microscopic theory. Further, if the Weak Gravity Conjecture holds in 5d, cored gravitational instantons may not be stable. 

We support this conjecture by providing a specific toy-example borrowed from \cite{0406038}. 
Following their results, we can consider a five-dimensional model with Euclidean action in 5d Planck units
\begin{equation} \label{Action in 5d}
S = \int d^5x \sqrt{\hat{g}} \left[-\frac{1}{2}\hat{R} + \frac{1}{2} (\partial \hat{\phi})^2 + \frac{1}{4}e^{a \hat{\phi}} \hat{F}^2\right],
\end{equation} 
where $\hat{F}=d \hat{A}$ is a 2-form field strength tensor. For the dimensional reduction to a 4d theory we choose\footnote{For the purpose of compactification we switch to Euclidean time $\tau$ by Wick-rotation. For simplicity we choose the periodicity $\tau \sim \tau +1$. Later in this subsection we allow the circumference of the $S^1$ to have length $\ell >0$. This will then have to be taken into account in order to determine the axion-decay constant.} 
\begin{equation} \label{Metric ansatz in 5D}
d\hat{s}_{(5)}^2 = e^{2 \beta_1 \psi} d\tau^2 + e^{2\beta_2 \psi} ds_{(4)}^2 
\end{equation} 
together with $\hat{A}= \theta d\tau$ and $\hat{\phi}=\phi$, i.e.~the fields $\theta$ and $\hat{\phi}$ do not depend on the extra-dimensional coordinate $\tau$. In Einstein-frame with canonically normalised kinetic terms dimensional reduction fixes the constants $\beta_1$ and $\beta_2$ to 
\begin{equation}
\beta_1 = - 2 \beta_2, ~~~~~~\beta_2 = \frac{1}{\sqrt{6}}.
\end{equation}
After redefining the fields $\phi$ and $\psi$ via a rotation in the $(\phi, \psi)$-plane we get 
\begin{equation} \label{Action compactified to 4d}
S = \int d^4x \sqrt{g} \left[-\frac{1}{2}R + \frac{1}{2} (\partial \tilde{\phi})^2 + \frac{1}{2} (\partial \tilde{\psi})^2 + \frac{1}{2}e^{\alpha \tilde{\phi}} (\partial \theta)^2\right],
\end{equation}
where $g$ denotes the metric corresponding to the 4d-line-element. Setting $\tilde{\psi}=0$ in this action one obtains the model considered in \eqref{Action with complex scalar} with dilatonic dependence in the kinetic term of $\theta$. The 4d dilaton coupling $\alpha$ is related to the 5d dilaton coupling $a$ via\footnote{Notice that our normalisation of $\phi$ is such that the prefactors of the Ricci scalar $R$ and the kinetic term $(\partial \phi)^2$ are equal, while in \cite{0406038} the prefactor of the dilaton has a factor $1/2$ relative to $R$. This is why our dilaton-coupling $\alpha$ differs by a factor of $\sqrt{2}$.} 
\begin{equation}
\alpha ^2 = a^2 + \frac{8}{3}.
\end{equation} 
Therefore, the interpretation of the 4d theory in terms of a 5d theory is only possible if $\alpha \geq 2\sqrt{2/3}$. 

Let us now explicitly relate the integration constant $C$ to microscopic properties of a higher-dimensional theory. 
In \cite{0406038} it was shown that for $\alpha = 2\sqrt{2/3}$ the solutions of the 4d model \eqref{Action compactified to 4d} can be uplifted to  a five-dimensional Reissner-Nordstr\"om (RN) black hole solution
\begin{equation} \label{RN metric}
ds^2_{(5)} =  g_+(\rho)g_-(\rho)d\tau^2 + \frac{d\rho^2}{g_+(\rho)g_-(\rho)}  + \rho^2 d\Omega_3^2, ~~~~ \hat{F}_{\tau \rho} = \sqrt{6} \frac{Q}{\rho^3}
\end{equation}
where
\begin{equation}
g_{\pm}(\rho)= 1 - \frac{\rho_{\pm}^2}{\rho^2}
\end{equation}
with
\begin{equation}
\rho_{\pm}^2 = M \pm \sqrt{M^2-Q^2}.
\end{equation}
We take this as a simple toy-model to argue that $C$ is generically fixed by properties of a black brane wrapping internal cycles. The RN black hole can be interpreted as $N$ particles or 0-branes (or just one 0-brane wrapping the cycle $N$ times) of total mass $M$ and total charge $Q$. Note however that the ADM-mass $M_{\text{ADM}}$ is related to the mass parameter $M$ by $M_{\text{ADM}}=6 \pi^2 M$. Nevertheless, we henceforth call $M$ the mass of the RN-black hole. The charge $Q$ is defined such that $M=Q$ sets the extremality bound. That is, $Q=N\hat{q} \sqrt{6}/(6 \pi^2)$, where the charge $\hat{q}$ is defined by $N\hat{q}= 1/2 \int_{S^3} \star_5 \hat{F}$.\footnote{For the normalisation we found it useful to translate the conventions in \cite{Myers:1986un,Ortin:2015hya} to our situation.}  

 Upon toroidal dimensional reduction along the coordinate $\tau$ with the identification $\tau \sim \tau + \ell$ and the circumference $\ell >0$ of the compactified dimension, the 5d solution \eqref{RN metric} turns into an instanton solution \eqref{Metric}. Note that the coordinate singularity at $\rho=\rho_+$ of the 5d solution becomes a curvature singularity (at $r=0$) in the 4d solution \eqref{Metric}.  
In the subsequent computation we show that our integration constant $C$ is simply given by $C= \ell^2 (M^2-Q^2)$ in 4d Planck units.

Denote by $g_{MN}^{(5)}$ the RN-metric \eqref{RN metric}, where $M,N$ run over the coordinates of the 4d space and the extra-dimensional coordinate $\tau$. Now, rescale the metric as follows: $\tilde{g}_{MN}^{(5)}=g_{MN}^{(5)}/(g_+g_-)$. From the canonical Einstein-Hilbert term we then get:
\begin{align} \nonumber
\int d^5x \sqrt{g^{(5)}}R[g_{MN}^{(5)}] &= \int d^5x (g_+g_-)^{3/2} \sqrt{\tilde{g}^{(5)}}R[\tilde{g}_{MN}^{(5)}] + ... = \\ 
&= \int d^4x\ell  (g_+g_-)^{3/2} \sqrt{\tilde{g}^{(4)}}R[\tilde{g}_{\mu\nu}^{(4)}] + ...
\end{align}
The last term occurs in the compactified 4d theory using the identification $\tau \sim \tau+\ell$.
  We want to point out that for generic $\ell >0$ there is a conical singularity at the outer horizon $\rho=\rho_+$. In principle, one could avoid such a conical singularity by choosing the periodicity of $\tau$ appropriately (it would be the inverse of the Hawking-temperature \cite{Hawking:1982dh}), but this would mean to fix the compactification radius. Instead, we accept the conical singularity as a necessary feature of Euclidean branes wrapped on cycles of the compact space.\footnote{Note that in the so-called dual frame metric discussed in \cite{0406038} non-extremal instantons with $\alpha=2\sqrt{2/3}$ can be interpreted as sections of constant time of the RN black hole metric. In this frame one recovers a wormhole geometry connecting two asymptotically flat regions smoothly. One pays the price of rescaling by a divergent factor.  The above is technically different from our approach of obtaining gravitational instantons by compactification of a 5d black hole solution on an $S^1$. In our case the RN black hole solution \eqref{RN metric} in general yields conical singularities.} Since it is known that Euclidean branes wrapped on non-trivial cycles give rise to instantonic terms (see e.g. \cite{Wen:1985jz,Dine:1986zy}), we assume that the corresponding conical spacetimes are saddle-points of the Euclidean path integral.
  
We go to the Einstein frame (with 4d Planck mass $M_p=1$) by rewriting the Einstein-Hilbert term using the rescaled metric $g_{\mu\nu}^{(4)} = \ell  (g_+g_-)^{3/2}\tilde{g}_{\mu\nu}^{(4)}$. The compactified 4d line-element then reads
\begin{equation}
ds_{(4)}^2 = \ell  \frac{d\rho^2}{\sqrt{g_+g_-}} + \ell  \sqrt{g_+g_-} \rho^2 d\Omega_3^2. 
\end{equation}
For the comparison with the metric \eqref{Metric}, the obvious coordinate transformation to be made is simply $r^2 = \rho^2 \ell  \sqrt{g_+g_-}$. Using the definitions of $g_{\pm}$ it follows 
\begin{equation}
rdr = \ell^2  \frac{\rho(\rho^2-M)}{r^2}d \rho. 
\end{equation}
Together with $(\rho^2-M)^2 = r^4/\ell ^2 + (M^2-Q^2)$, this implies: 
\begin{equation}
\ell  \frac{d\rho^2}{\sqrt{g_+g_-}} = \frac{dr^2}{1+ \ell^2 (M^2-Q^2)/r^4}.
\end{equation}
Hence, we find the simple relationship 
\begin{equation}
C=\ell^2(M^2-Q^2)
\end{equation}
in 4d Planck units. Upon dimensional reduction of \eqref{Action in 5d} and using the periodicity of the Wilson line $\hat{A}_{\tau} \cong \hat{A}_{\tau} + \pi/(\hat{q}\ell)$ one can easily check that the axion decay constant reads $f_{\text{ax}} =1/(2\hat{q}\ell)$ for an axion $\theta$ with $2\pi$-periodicity. It follows that
\begin{equation}
C = \frac{N^2}{24 \pi^4 f_{\text{ax}}^2}\left[\left(\frac{M}{Q}\right)^2-1\right].
\end{equation} 
We can compare this result to our previous expression \eqref{C for C>0}. First, we can identify the wrapping number/number of 0-branes $N$ with the flux number $n$. We then find that the integration constant $K_+$ in \eqref{C for C>0} is completely determined by the parameters $M$ and $Q$ describing black holes/branes in the 5d theory. An immediate result is that $K_+$ and hence $C$ are not free parameters. The possible range of values is determined by the spectrum of black branes in the higher-dimensional theory. Furthermore, as $M$ and $Q$ are discrete quantities it follows that $C$ can also only take discrete values (for a given value of $f_{\text{ax}}$). This property is only important as long as $M$ and $Q$ are small. In the macroscopic regime of large $M$ and $Q$ the value of $C$ can be dialed to any positive value and it becomes effectively continuous. We come back to this in \autoref{Section:Cored Grav Inst}. 

Notice that the case of $M=Q$, which gives $C=0$, corresponds to an extremal Reissner-Nordstr\"om black hole. In this sense, the name \textit{extremal instanton} for flat $4$d solutions \eqref{Metric} is justified. In \autoref{Section:Cored Grav Inst} we comment on how to express the extremal instanton action in terms of $\ell$ and $M_{\text{ADM}}$, consistent with, for instance, \cite{09052844,10116301,150906374}.  

This example illustrates nicely how $4$d cored or extremal instanton solutions can be obtained from black holes/branes with mass $M$ and charge $Q$. Of course, one could also go beyond such simple toy-models we just discussed, allowing also for dilaton couplings $\alpha \neq 2\sqrt{2/3}$. We expect the relation $C=\ell^2(M^2-Q^2)$ to be modified by the corresponding parameter $a \neq 0$ in this more general case. Furthermore, one would expect that after SUSY-breaking extremal objects in string theory would appear as non-extremal instantons in the 4d effective theory.  

Last, let us remark on possible implications for cored gravitational instantons arising from the Weak Gravity Conjecture. In particular, if the Weak Gravity Conjecture holds in the 5d model we expect that objects with $M > Q$ can in principle decay. As cored gravitational instantons arise from such unstable objects upon dimensional reduction, one may wonder whether this instability is then inherited by the instantons. Here `unstable instanton' means that two instantons exist which cause the same flux change but have smaller total action. In this sense, the contribution of cored instantons to the Euclidean path integral is subdominant if cored instantons are `unstable'. This point will me made more precise in \autoref{Section: WGC}.

\section{Instanton Potentials from Euclidean Wormholes} \label{Section:Instanton Potentials from Wormholes}

The goal of this section is to show that the one-instanton action, describing a Giddings-Strominger wormhole, gives rise to an instanton potential of the structure $\cos \theta e^{-S}$.  

We begin with a brief review of Coleman's derivation \cite{ColemanBook,Vainshtein:1981wh} of the energy eigenvalues for a simple one-dimensional quantum mechanical system with periodic potential $V$, e.g.~$V(x) \sim \sin^2(2\pi x)$. These considerations can be applied to quantum field theory and in particular to our system as well. 

The Hamiltonian is  $H=p^2/2+V(x)$.  An instanton or an anti-instanton correspond to tunnelling events from $x$ to $x+1$ or $x-1$, respectively. Using the dilute-gas approximation we can distribute instantons and anti-instantons freely in time. Let us introduce a basis of states $\ket{j}$ in which the particle is localised at $x \simeq j$. Then for some time interval $T>0$, transition amplitudes are  \cite{ColemanBook} 
\begin{equation} \label{Coleman formula}
\braket{j_+|e^{-HT}|j_-} = \left(\frac{\omega}{\pi}\right)^{1/2} e^{-\omega T/2} \sum_{N=0}^{\infty} \sum_{\bar{N}=0}^{\infty} \frac{1}{N! \bar{N}!} (Ke^{-S_0}T)^{N+\bar{N}} \delta_{(N-\bar{N})-(j_+-j_-)},
\end{equation}
where $j_-$ and $j_+$ are the positions of the initial and final state, respectively.  $N$ and $\bar{N}$ count the number of instantons and anti-instantons.  Moreover, $\omega$ is defined by $\omega = V^{\prime \prime}(0)$. $K$ is the familiar determinant factor, which depends on details of the potential $V$. $S_0$ denotes the instanton action. The Kronecker delta can be rewritten as 
\begin{equation}
\delta_{ab} = \int_0^{2\pi} \frac{d\theta}{2\pi} e^{i(a-b)\theta},  
\end{equation} 
and thus, after performing the summation, 
\begin{equation}
\braket{j_+|e^{-HT}|j_-} = \left(\frac{\omega}{\pi}\right)^{1/2} e^{-\omega T/2} \int_0^{2\pi} \frac{d\theta}{2\pi}  e^{i(j_--j_+)\theta} \exp \left(2KT \cos \theta e^{-S_0}\right).
\end{equation}
From this we can read off that the system has an energy eigenbasis 
\begin{equation}
\ket{\theta} = \sum_j e^{ij \theta} \ket{j}
\end{equation}
with eigenvalues
\begin{equation}
E(\theta) = \frac{1}{2} \omega - 2K \cos \theta e^{-S_0}.
\end{equation}
This derivation reveals the logic behind the famous contribution $\sim \cos \theta e^{-S}$ to the axion potential in quantum field theory, where the centres of the instantons are not distributed on a time interval but instead in a region of spacetime with volume $\mathcal{V}$. One then simply has to replace the variable $T$ by the volume $\mathcal{V}$. 
\bigskip

In the following we explain how this computation can be used to derive an instanton potential induced by Euclidean wormholes. In the previous \autoref{Section:Gravitational Instanton Solutions } we reviewed that Euclidean wormholes exist in the presence of a non-vanishing $3$-form flux $H$ with quantised charge $n \in \mathbbm{Z}$. An instanton would then correspond to a transition from $n$ to $n+1$. By the logic of Coleman's computation above, this should induce a shift-symmetry breaking potential. 
 
In \cite{150906374} this was questioned, because Euclidean wormholes appear as conduits and charges would not disappear. In other words, one always has an instanton and an anti-instanton, thus preserving $n$.  

\begin{figure}[t]
	\centering
	\includegraphics[width=1.00\textwidth]{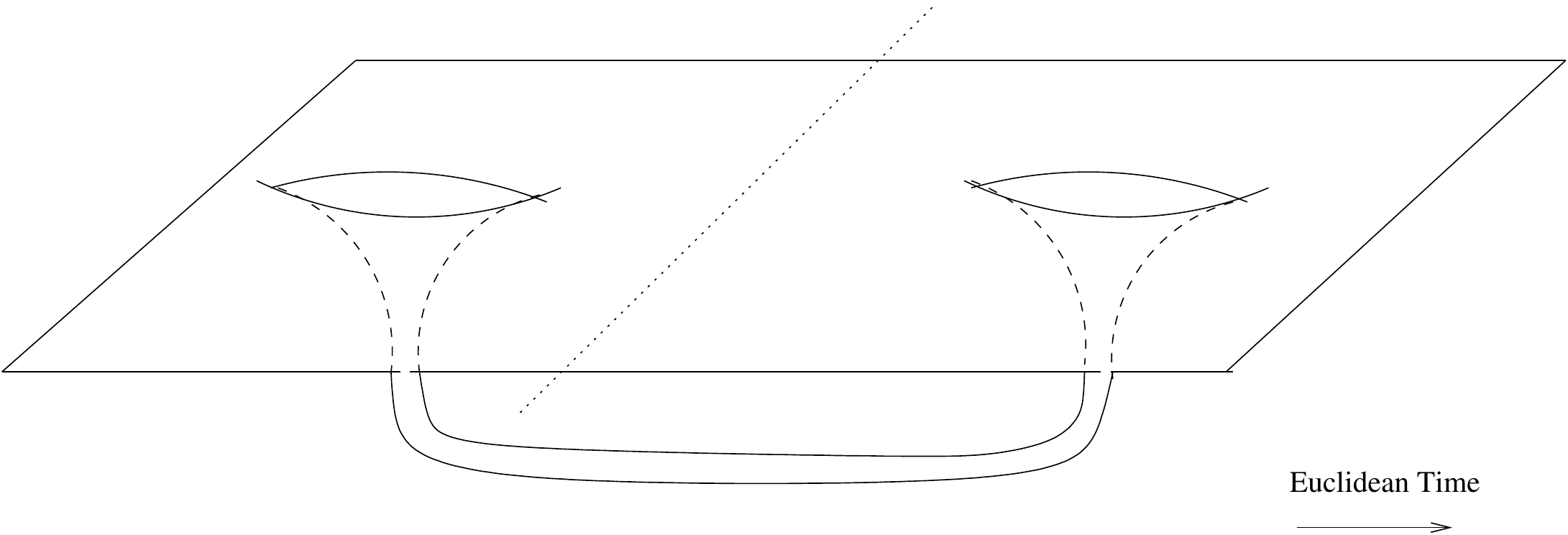}
	\caption{This picture presents a wormhole which opens at some initial time $t_i$ and closes at $t_f>t_i$. The dotted line indicates the separation of the two events. }
	\label{Figure: Wormholes}
\end{figure}

We argue that this issue is more subtle: the two ends of a Euclidean wormhole do not necessarily have to end at the same hypersurface of constant Euclidean time, but can also close on distant hypersurfaces. Similarly, the two ends can have very large spatial separation such that, from a local perspective, a potential \`a la Coleman should be induced. Then, a Minkowskian observer would only see either the instanton or anti-instanton part of the wormhole and thus find a change in the charge $n$, see \autoref{Figure: Wormholes}.  This invalidates the reasoning in \cite{150906374}, and hence we do not see any argument against the breaking of the shift-symmetry due to Euclidean wormholes.   

\begin{figure}[t]
	\centering
	\begin{overpic}[width=0.85\linewidth]{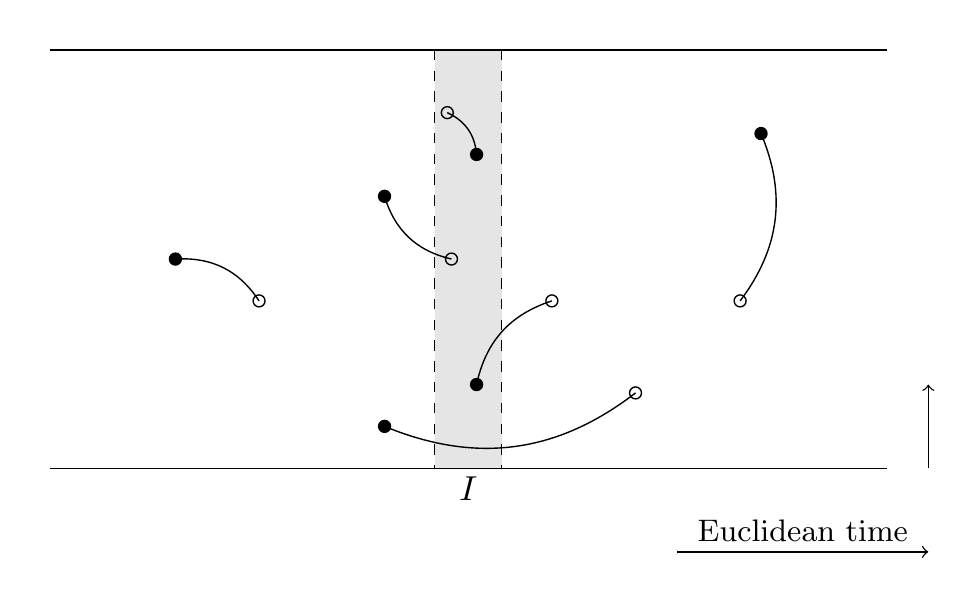}
		\put (97,20) {$\mathbbm{R}^3$}
	\end{overpic}
	\caption{This illustration shows pairs of connected black and white dots, each representing an end of a wormhole (black if the end corresponds to an instanton and white for an anti-instanton). Only few wormholes lie completely inside the shaded region $I \times \mathbbm{R}^3$.}
	\label{fig:wormholesummation}
\end{figure}

\begin{figure}[t]
\centering
\includegraphics[width=1.0\linewidth]{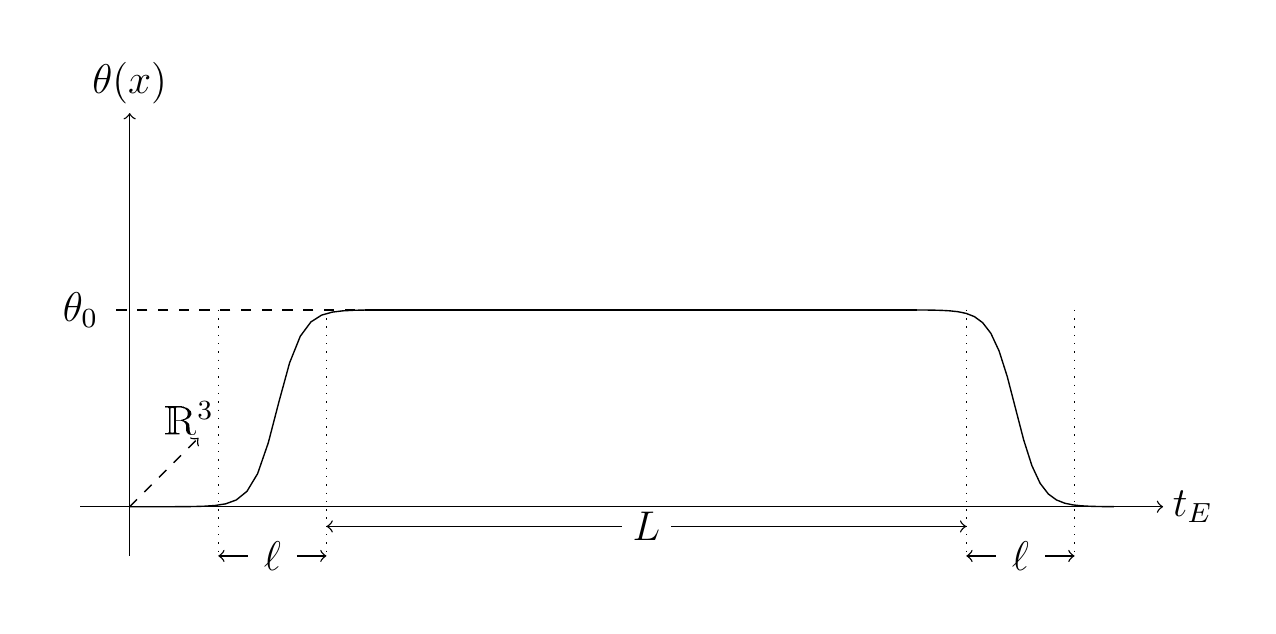}
\caption{We illustrate the smooth axion profile corresponding to the approximation in \eqref{Testfunction}. In the Euclidean time interval of length $L$ the axion field is constant (value $\theta_0$) and outside this interval the field decays smoothly. The characteristic length of this transition from $\theta_0$ to $0$ is denoted by $\ell$.}
\label{fig:testfunction}
\end{figure}

We want to make this mathematically more precise. This requires to compute the path integral contribution of all possible wormhole configurations. 
This allows us to infer the effective potential $V_{\text{eff}}(\theta)$ for the axion field $\theta$. The logic behind the computation of $V_{\text{eff}}(\theta)$ is the following. The expectation value of any observable $\mathcal{O}(\theta)$ is given by 
\begin{equation}
\braket{\mathcal{O}(\theta)} \sim \int d[\theta] \mathcal{O}(\theta) \exp \left(- \frac{1}{2}\int d^4x f_{\text{ax}}^2 (\partial \theta)^2\right) Z_{\text{wh}}[\theta] \ , 
\end{equation}	
where the path integral contribution of wormholes is schematically (i.e.~no combinatioral factors included yet) given by 	
\begin{equation} \label{Wormholesum}
	Z_{\text{wh}}[\theta] \sim \sum_{w} \prod_{n=1}^{w} \prod_{m=1}^{w} \int d^4x_n \int d^4x_m e^{-S}e^{i \theta(x_n)}e^{-S}e^{-i \theta(x_m)} \ . 
\end{equation} 
Here $w$ denotes the number of wormholes of a configuration.  Note that the phase difference occurs because one factor is for the instantons, the other for anti-instantons. Those factors arise from the second term of \eqref{Action Theta}, where $Q_{\theta} = \pm 1$ ($+$ for instantons, $-$ for anti-instantons). We write out these factors explicitly, because they finally give rise to the $\cos$-potential for $\theta$. The contribution $Z_{\text{wh}}[\theta]$ induces a change $\delta S_{\text{ind}}(\theta)$ of the action for the axion and we expect 
\begin{equation}
\braket{\mathcal{O}(\theta)} \sim \int d[\theta] \mathcal{O}(\theta) \exp \left(- \frac{1}{2}\int d^4x f_{\text{ax}}^2 (\partial \theta)^2 - \delta S_{\text{ind}}(\theta)\right) \ , 
\end{equation}	
where $\delta S_{\text{ind}}(\theta)$ contains by definition the effective potential of $\theta$ plus higher derivative corrections: 
\begin{equation}
\delta S_{\text{ind}}(\theta) = \int d^4x \left(V_{\text{eff}}(\theta) + \text{higher derivative terms}\right) \ .
\end{equation}
Hence, the effective potential $V_{\text{eff}}(\theta)$ can be determined by computing $Z_{\text{wh}}$ using any field configuration $\theta$ for which $V_{\text{eff}}(\theta)$ dominates all derivative terms. We choose a smooth version of the profile
	\begin{equation} \label{Testfunction}
	\theta(x) = \begin{cases}
	\theta_0 & \text{for}~ x \in I \times \mathbbm{R}^3 \\
	0 & \text{else} \ ,
	\end{cases} 
	\end{equation}
 i.e.~a profile which is only non-zero in a small Euclidean time interval $I$ (see \autoref{fig:wormholesummation}) and goes to zero smoothly at the boundary of $I$. This is illustrated in \autoref{fig:testfunction}. 
The volumes of $I \times \mathbbm{R}^3$ and of the remaining part of Euclidean space are denoted by $\mathcal{W}$ and $\mathcal{V}$, respectively. We assume that $\mathcal{W} \ll \mathcal{V}$ with $\mathcal{W}$ being large enough to typically contain many wormhole ends. 
We first check that the derivative terms can indeed be made subdominant with respect to the effective potential. For simplicity, we work near the minimum and use the approximation $V_{\text{eff}} \sim m^2\theta^2$. It is crucial that our axion profile at the boundary of $I$ features a smooth transition of characteristic length $\ell$ from 0 to $\theta_0$ with $\ell \ll L$, where $L$ is the length of the Euclidean interval $I$. 
	We then have $\partial \theta \sim \theta_0/\ell$ close to the boundary (and zero elsewhere) and the comparison of $\int d^4x f_{\text{ax}}^2 (\partial \theta)^2$ with $\int d^4x V_{\text{eff}}$ should yield 
\begin{equation}
\ell V_3 f_{\text{ax}}^2 \frac{\theta_0^2}{\ell^2} \ll L V_3 m^2 \theta_0^2 \ ,
\end{equation}
where $V_3$ denotes the corresponding 3-volume of the Euclidean spacetime regions we consider. It follows that
\begin{equation}
L \gg \frac{f_{\text{ax}}^2}{m^2 \ell} 
\end{equation}
has to be imposed. Note that $\ell$ cannot be arbitrarily small as we have to ensure that also higher-derivative terms must be subdominant. Comparing $f_{\text{ax}}^2 (\partial \theta)^2$ with $(\partial \theta)^4$ yields 
\begin{equation}
\ell \gg \frac{\theta_0}{f_{\text{ax}}} \sim \frac{1}{f_{\text{ax}}} \ .
\end{equation}
It is not hard to see that these two conditions together with $L \gg \ell$ can be satisfied simultaneously. Thus, our field configuration \eqref{Testfunction} is suitable for the calculation of the effective potential given below.
For this computation we find it useful to group the sum in the above expression \eqref{Wormholesum} according to the position of the wormhole ends, see again \autoref{fig:wormholesummation}. 
Let us assume that a wormhole corresponds, from the perspective of the $\theta$-field theory, simply to an instanton-anti-instanton pair. Then, denoting by $\theta$ the field configuration of \autoref{fig:testfunction}, we can write: 
\begin{equation} \label{Partition function wormhole naiv}
Z_{\text{wh}}[\theta]  \sim \sum_{n} \frac{1}{n!n!}\prod_{j=1}^{n} e^{-2S}K^2 \left(\mathcal{V} + \mathcal{W}e^{i\theta_0}\right) \left(\mathcal{V} + \mathcal{W}e^{-i\theta_0}\right) \ . 
\end{equation}	
Note that this differs from the toy model \eqref{Coleman formula} in the sense that the sector with $N \neq \bar{N}$ is not contained in \eqref{Partition function wormhole naiv}. But this is precisely our point: we want to find out whether a potential can still be generated if we impose $N=\bar{N}$, i.e.~an equal number of instantons and anti-instantons.
The combinatorial factor $1/(n!)^2$ is due to the indistinguishability of instantons and anti-instantons, respectively. The cross-terms $\mathcal{V}\mathcal{W}e^{\pm i \theta_0}$ correspond to wormholes where only one end is within $I \times \mathbbm{R}^3$. 
The sum can be expressed as a Bessel function $I_0$ 
\begin{equation}
Z_{\text{wh}}  \sim I_0(x)
\end{equation}
with
\begin{align} \label{Bessel sum}
I_0(x) &= \sum_{m=0}^{\infty} \frac{1}{m!m!} \left(\frac{x}{2}\right)^{2m} \\ 
x & \simeq 2Ke^{-S} \mathcal{V}\left(1+(\mathcal{W}/\mathcal{V})\cos \theta_0 + \mathcal{O}\left((\mathcal{W}/\mathcal{V})^2\right) \right) \ . 
\end{align}
Furthermore, there is an integral expression for $I_0$: 
\begin{equation} \label{Bessel integral}
I_0(x) = \frac{1}{2\pi} \int_{0}^{2\pi} d \phi e^{-x \cos \phi} \ . 
\end{equation}
Hence, we arrive at 
\begin{equation}
Z_{\text{wh}}  \sim \frac{1}{2\pi} e^{x} \int_{0}^{2\pi} d\phi e^{-x(1+\cos \phi)} \simeq \frac{1}{2\pi} e^x  \int_{0}^{2\pi} d\phi e^{-x(\phi-\pi)^2/2} \simeq \frac{1}{\sqrt{2\pi}} \frac{e^{x}}{\sqrt{x}} \ , 
\end{equation}
where we relied on the fact that in our case $x$ is large (because $\mathcal{V}$ and $\mathcal{W}$ are large).  Since we are interested in the effective potential $-\mathcal{W}V_{\text{eff}}(\theta_0) \simeq \ln Z_{\text{wh}} $, we can focus on the exponential factor: 
\begin{align} \label{Paritition function wormhole contribution}
Z_{\text{wh}}  \sim \exp \left[2Ke^{-S}\left(\mathcal{V} + \mathcal{W} \cos \theta_0\right)\right] \ . 
\end{align}
From here it is clear that, to explain the $\mathcal{W}\cos \theta_0$ term, the effective action for $\theta$ must contain a potential (which in our case contributes only in the region $I \times \mathbbm{R}^3$):
\begin{equation}
V_{\text{eff}}(\theta_0) \sim 2Ke^{-S} \cos \theta_0 \ .
\end{equation}
The $\theta$-dependency is as in the case of an instanton-anti-instanton gas without any constraints imposed. Actually, the term generated by wormhole instantons, which looks like a potential term for the field configuration \eqref{Testfunction}, is in fact a non-local interaction term. This can be seen by modifying \eqref{Testfunction} such that $\theta(x) = \theta_{\mathcal{V}} \neq 0$ for $x \in \mathcal{V}$. Then, $\cos(\theta_0)$ in \eqref{Paritition function wormhole contribution} is replaced by $\cos(\theta_0-\theta_{\mathcal{V}})$. Indeed the exponential now contains non-local terms. Globally, this non-local term preserves shift-symmetry. Nevertheless, we observe a crucial effect induced by Giddings-Strominger wormholes: The change of the action due to a local fluctuation, $S[\theta+\delta\theta] - S[\theta]$, corresponds to that induced by a potential $V (\delta\theta) \sim 2 K \exp(-S)\cos(\delta\theta)$.

This can also be seen by applying Coleman's computation \cite{ColemanBook} to our problem. We are then interested in computing the partition function $Z=\sum_{n}\braket{n|e^{-HT}|n}$. It is therefore sufficient to focus on transition functions $\braket{n|e^{-HT}|n}$, although transitions from $\ket{m}$ to $\ket{n}$ will occur as well (the Hamilitonian $H$ is in general non-diagonal is this basis). Due to the trace we do not need to consider the latter in our calculation. The result of this computation has to be compared with the partition function $Z=\int_{0}^{2\pi} d\theta \braket{\theta|e^{-HT}|\theta}/(2\pi)$ in the $\theta$-language. 

In the free theory $n$ is the dual variable to our axion field $\theta$, which can be seen as follows: The free theory action for the axion is given by $S= \int d^4x f^2 \dot{\theta}^2/2$ or, after integrating out the spatial directions, $S= \int dt A \dot{\theta}^2/2$ with $A \equiv f^2 V_3$, where $V_3$ is the 3-volume ($\theta$ is then the zero mode). The canonical momentum $p$ is then given by $p=A \dot{\theta}$. As it is well known from quantum mechanics on $S^1$, $p$ is quantised as $p=n \in \mathbbm{Z}$. One can therefore relate $\ket{\theta}$-states to $\ket{n}$-states via 
\begin{equation}
\ket{\theta} = \sum_n e^{in\theta} \ket{n}
\end{equation}
in the free theory (see also \cite{07052768}), where we chose the normalisation $\braket{\theta|\theta^{\prime}}= 2\pi \delta(\theta-\theta^{\prime})$. 
The Hamiltonian of the free theory is then given by $H=n^2/(2A)$ and in the free theory we have the transition amplitude
\begin{equation} \label{Transition amplitude free theory}
\braket{n|e^{-HT}|n}_{\text{free}} = e^{-Tn^2/(2A)} \ . 
\end{equation}
Let us now return to interacting theory and take into account the effects of the wormhole gas induced by the coupling of $\theta$ to gravity. We assume that instantons and anti-instantons are randomly distributed and that wormhole ends can have arbitrarily long separation with no physical effect. By applying Coleman's formula \eqref{Coleman formula} to our situation and taking into account \eqref{Transition amplitude free theory}, we find:
\begin{equation} \label{Transition amplitude with instantons}
\braket{n|e^{-HT}|n} = e^{-Tn^2/(2A)} \sum_{N=0}^{\infty} \frac{1}{N! N!} (Ke^{-S_0}T)^{2N}.
\end{equation}
We emphasise once more that off-diagonal elements $\braket{m|e^{-HT}|n}$ are in general non-zero.
For instance, if $T$ corresponds to half of the time interval of \autoref{Figure: Wormholes}, the instanton number clearly changes by unity. However, such off-diagonal elements never appear explicitly in our calculation, which relies solely on the partition function.  	 

We can once again express the sum in \eqref{Transition amplitude with instantons} by $I_0$ via \eqref{Bessel sum} and then use the integral expression \eqref{Bessel integral} with integration variable $\theta$. We find:
\begin{equation}
\braket{n|e^{-HT}|n} = e^{-Tn^2/(2A)} \int_{0}^{2\pi} \frac{d\theta}{2\pi} \exp \left(-2KT\cos \theta e^{-S_0}\right) \ . 
\end{equation}
For the partition function $Z=\sum_{n}\braket{n|e^{-HT}|n}$ one then obtains 
\begin{equation} \label{Partition function for free theory + instantons}
Z(T) \simeq \sqrt{\frac{2\pi A}{T}} \int_{0}^{2\pi} \frac{d\theta}{2\pi} \exp \left(-2KT\cos \theta e^{-S_0}\right) \ . 
\end{equation}
As we already mentioned, this should be compared with 
\begin{equation} \label{Partition function general}
Z=\int_{0}^{2\pi} \frac{d\theta}{2\pi} \braket{\theta|e^{-HT}|\theta} \ . 
\end{equation}
For a first naive comparison of \eqref{Partition function for free theory + instantons} and \eqref{Partition function general} we ignore the non-exponential $T$-dependence in the prefactor of  \eqref{Partition function for free theory + instantons}.\footnote{Our parametrisation is then equivalent to $Z(T)=\int dE \rho(E)e^{-ET}$. Thus, we found the partition function. The latter characterises the system unambiguously.} Then $\ket{\theta}$ is an eigenbasis of the Hamiltonian with eigenvalues $V(\theta)$. We see by comparison with \eqref{Partition function for free theory + instantons} that $ V(\theta) =2 \cos \theta Ke^{-S_0}$.

We can, however, be more precise and understand also the prefactor. To do so we observe that \eqref{Partition function for free theory + instantons} was derived on the basis of \eqref{Transition amplitude with instantons}, and in this equation a non-trivial approximation was made: Indeed, we placed the factor $\exp(-Tn^2/(2A))$ outside the instanton sum. In general, that is not justified for the following reason. If we start at $t=0$ with flux number $n$, and the first instanton occurs e.g. at $t=T_1$, we get a factor $\exp(-T_1n^2/(2A))$ from the kinetic term. If then the next instanton occurs at $T_2$, we get a further factor $\exp(-T_2(n+1)^2/(2A))$ and so on. The times $T_i$ have to be integrated over and these prefactors can not be extracted from the instanton sum. However, we can find conditions under which it is safe to approximate the  $(n+N_i)^2$ in the exponents (where $N_i$ is the number of instantons present at some given time) simply by $n^2$. To do so we note that, on the one hand, the instanton sum is dominated by instanton numbers of the order of 
\begin{equation}\
	\braket{N} \sim Ke^{-S_0}T \ .
\end{equation}
On the other hand typical values of $n$ dominating the sum over $\exp(-Tn^2/(2A))$ are of the order of $n\sim \sqrt{A/T}$. Thus, disregarding $N_i$ relative to $n$ in the $(n+N_i)^2$-terms will be justified if
\begin{equation}
\braket{N} \ll \sqrt{A/T}\qquad\mbox{or}\qquad A\gg \langle N\rangle^2 T \ .
\end{equation}
Given that we anyway choose $T$ large enough to ensure $\langle N\rangle\gg 1$, this implies in particular $A\gg T$.

With this in mind, we return to the corresponding quantum mechanical model. We conjecture that the instanton dynamics is captured by an effective potential $V(\theta)=2\cos(\theta)Ke^{-S_0}$. To confirm this, we calculate the partition function 
	\begin{equation} \label{Partition function quantum particle}
	Z(T) \simeq \frac{1}{Z(0)}\int d\theta \int_{\tilde{\theta}(0)= \theta}^{\tilde{\theta}(T) = \theta} d[\tilde{\theta}] \exp \left[ -\int_0^T dt \left(A \frac{\dot{\tilde{\theta}}^2}{2} + V(\tilde{\theta})\right)\right] \ . 
	\end{equation}
	 It has to be compared to \eqref{Partition function for free theory + instantons} to establish the correctness of the chosen effective description and, in particular, the potential. But working out \eqref{Partition function quantum particle} in the regime $A\gg T$ is easy. Indeed, if we first disregard the potential, we are simply dealing with a 1-dimensional system on the interval $(0,2\pi)$ and a kinetic-term prefactor $A$. This prefactor sets the minimum time by which any wave packet unavoidably spreads to an $\mathcal{O}(1)$ width due to quantum dynamics. In addition, the potential has a maximal steepness $|V'|\sim Ke^{-S_0}$, leading to a displacement of $Ke^{-S_0}T^2/A\sim  \braket{N} T/A$ during a time interval $T$. Our previously derived conditions on $A$, which underly our derivation of \eqref{Partition function for free theory + instantons}, are sufficient to ensure that the particle moves only by a distance $\Delta\theta\ll 1$ during the time $T$. Hence, in evaluating \eqref{Partition function quantum particle} we can approximate $V(\tilde{\theta}(t))\simeq V(\theta)$. The path integral then becomes that of free particle, to be evaluated on times too short for the periodicity of the configuration space to be relevant. One obtains the well-known time-dependence $\sim \sqrt{A/T}$ of the amplitude, to be multiplied by the integral over $\exp(-TV(\theta))$. This is now in perfect agreement with \eqref{Partition function for free theory + instantons}.\footnote{While our analysis establishes the quantum mechanical model with effective potential $V(\theta)=2\cos(\theta)Ke^{-S_0}$ only for a certain range of $T$, we expect it to be valid also for $T \to \infty$.} 

Thus, we find that Giddings-Strominger wormholes give rise to an effective potential $V(\theta)\sim 2Ke^{-S} \cos \theta$ in two independent approaches. We wish to remark that in both approaches we can be agnostic about details of the interpretation of wormholes connecting to baby-universes. Crucially, the axionic shift-symmetry is broken locally even if the condition of having equally many instantons and anti-instantons is imposed on the global space-time.
\bigskip

Finally, we wish to remark that the correct choice of the combinatorial factors is a subtle issue. We interpreted a configuration of $N$ wormholes as an instanton-anti-instanton-gas with (anti-)instantons randomly distributed. It is then plausible to include the combinatorial factor $1/(N!)^2$. However, one might argue that each instanton has a corresponding anti-instanton and therefore we should multiply by $N!$ to account for the number of possible pairings. If we assume that the right combinatorial factor is just $1/N!$, we can still do the computation starting with \eqref{Wormholesum}. We then still get $\cos \theta_0$, but this time the energy density in $I \times \mathbbm{R}^3$ scales with 
	\begin{equation} \label{Rho with 1/N!}
	V(\theta_0) \sim K^2\mathcal{V}e^{-2S} \cos \theta_0 \ , 
	\end{equation}
which diverges as $\mathcal{V} \to \infty$. Possibilities to avoid this divergence were discussed in \cite{Giddings:1988cx,Coleman:1988cy,Preskill:1988na,Klebanov:1988eh}, mostly in the baby-universe interpretation of Giddings-Strominger wormholes. It is possible to express the partition function as an integral over a parameter $\alpha$, which is an eigenvalue of a baby-universe operator \cite{Coleman:1988cy}.

We rather follow Preskill \cite{Preskill:1988na} to sketch the idea of how to evade the divergence:
For a combinatorial factor $1/N!$ the partition function reads 
\begin{equation}
Z \sim \sum_{N=0}^{\infty} \frac{C^N}{N!} = e^C \ ,
\end{equation}
where 
\begin{align}
C \sim \bar{z}z ~, ~~~~~ z  \equiv K\mathcal{V}e^{-S}e^{i \theta} \ . 
\end{align}
Clearly, $C \sim \mathcal{V}^2$. But formally, we can write
\begin{equation}
Z \sim \int d \alpha d \bar{\alpha} e^{-\bar{\alpha} \alpha + \alpha \bar{z} + \bar{\alpha} z } \ .
\end{equation}
If $\alpha$ is integrated out we obtain the divergent result \eqref{Rho with 1/N!}. (To see $\cos \theta_0$ coming in one has to group terms carefully as in our first computation.) If, as suggested by Preskill \cite{Preskill:1988na}, one has to fix $\alpha$ to a certain value, the energy density is simply given by\footnote{Nevertheless, the divergence remains disconcerting. For instance, the expectation value of the number of wormholes in a certain space-time region scales as $\braket{N} \sim \mathcal{V}^2$ and it is questionable whether the wormhole gas can be dilute in the limit $\mathcal{V} \to \infty$. See also discussions in e.g.~\cite{Preskill:1988na,Fischler:1988ia,Coleman:1989ky,Polchinski:1989ae}.}  
\begin{equation}
\rho \sim \alpha e^{-S} \cos \theta \ . 
\end{equation}
In any case, no matter which combinatorial factor is correct and no matter how to interpret $\alpha$, we always find that a term $\cos \theta$ arises in the effective action.

To summarise, we conclude that Euclidean wormholes are expected to induce an instanton potential $\sim \cos \theta e^{-S}$. Shift-Symmetry appears to be broken locally.
It would be interesting to study whether this can be seen more directly by building an analogy between gravitational and gauge instantons, where the role of the term $\theta \text{Tr}(F \wedge F)$ is played by $\theta \text{Tr}( R \wedge R)$.   
\bigskip

In the following we apply the presented derivation of the instanton potential to cases of $S=nS_0$, giving rise to potentials of the form $\sum_n \cos(n\theta)e^{-nS_0}$. 

\section{The Limit of Validity of Gravitational Instanton Actions}
\label{Section:Cored Grav Inst}

In this section we summarise the instanton actions for all cases $C<0$, $C=0$ and $C>0$ and find limits for the validity of the computation. Qualitatively, we have
\begin{equation}
S \sim \frac{n}{f_{\text{ax}}}
\end{equation}
in all three cases. This is of course already known for Euclidean wormholes, see e.g.~\cite{Giddings:1987cg,Grinstein:1988ja,Lee:1988ge,Abbott:1989jw,Brown:1989df,Burgess:1989da,Coleman:1989zu,0406038,150303886} and also for $C \geq 0$, see e.g.~\cite{0406038,150906374}.   

Furthermore, we address one concern raised in \cite{0406038}: the cored gravitational instanton solutions have a singularity at $r=0$ and hence it is unclear whether these solutions can be trusted all the way to the limit $r \to 0$. In fact, we expect a breakdown of the solutions at some radius $r = r_c >0$, which will be estimated in \autoref{Section:Inflation}. We expect such a cutoff radius to be present in any extra-dimensional theory independently of whether a curvature singularity exists or not. Therefore, even the extremal instantons, which do not have singularities, should only be trusted down to $r = r_c$. The situation is different for the Euclidean wormhole solutions, where we can have full control over the solution as long as $r_0 \gtrsim r_c$, with $r_0 \equiv |C|^{1/4}$ being the radius of the wormhole throat at the centre.  

The limit of validity affects the computation of the instanton action. In the case of $C \geq 0$ one would usually integrate from $r=0$ to infinity, but instead we can only rely on the contribution from the interval $(r_c,+ \infty)$. Whenever a significant fraction of the action comes from $(0,r_c)$, we cannot trust the instanton actions computed in \cite{0406038,150906374} and we will discard these cases.   
  
Thus, the initial task of this section is the evaluation of the on-shell contribution of the integral in \eqref{Action without potential}.
We proceed by using the equations of motion successively. Details of the computations are presented in \autoref{Appendix:Instanton Action}. 

At first, by tracing Einstein's equations, we can express the Ricci scalar by the trace of the energy-momentum tensor: 
\begin{equation}
R = - T.
\end{equation} 
One can then rewrite \eqref{Action without potential} as 
\begin{equation}
S =  \int_M d^4x \sqrt{g} \mathcal{F}(\varphi)H^2.
\end{equation}
However, this is not yet the full contribution to the instanton action, because the Gibbons-Hawking-York boundary term has to be taken into account. It is
\begin{equation} \label{Gibbons-Hawking-York}
S_{\text{GHY}} = - \oint_{\partial M} d^3x \sqrt{h}(K-K_0),
\end{equation}
where $h$ is the determinant of the induced metric on $\partial M$. $K$ and $K_0$ are the traces of the extrinsic curvatures of $\partial M$ in $M$ and flat space, respectively. 

Then, the instanton action is computed as
\begin{equation}
S_{\text{inst}} = S + S_{\text{GHY}} =  \int_M d^4x \sqrt{g} \mathcal{F}(\varphi)H^2 - \oint_{\partial M} d^3x \sqrt{h}(K-K_0).
\end{equation}  
 
Henceforth, we restrict to the case $\mathcal{F}(\varphi)= \exp(-\alpha \varphi)/(3!f_{\text{ax}}^2)$. Using this together with the equation of motion \eqref{EoM for B} and \eqref{Solution for C<0, with constant}, \eqref{Solution for C=0} or \eqref{Solution for C>0} depending on the choice of $C$, one can rewrite the first term, $S$, in the instanton action as an integral over $\varphi$. The contribution from $S_{\text{GHY}}$ is computed by considering a surface of constant $r$, see \cite{150906374} or \autoref{Appendix:Instanton Action}. 

In the following we analyse the instanton action case by case:

\subsubsection*{Case $C=0$:}

Extremal instanton solutions go along with a flat metric ($C=0$). Thus, we have 
\begin{equation}
S_{\text{GHY}}=0.
\end{equation}
However, the fields $\varphi$ and $B$ have a non-trivial profile giving rise to non-vanishing contributions to the instanton action. 
The full contribution from $r=0$ to $r=\infty$ is given by 
\begin{equation} \label{Instanton action C=0}
S_{\text{inst}} = - \frac{n}{f_{\text{ax}}} \int_{\varphi(0)}^{\varphi(\infty)} d \varphi \exp(-\alpha \varphi/2) = \frac{2}{\alpha} \frac{n}{f_{\text{ax}}}.
\end{equation}
As we explained previously, we cannot trust this computation for radii $r<r_c$. Nevertheless, as long as the main contributions to the action come from the regime $r>r_c$ the result \eqref{Instanton action C=0} can still be used to estimate contributions to the instanton potential in \autoref{Section:Inflation}. One should therefore compare the contribution $\Delta S$ from the regime $r<r_c$ with \eqref{Instanton action C=0}. Unfortunately, the contribution $\Delta S$ is UV-sensitive. We assume, however, that the actual UV-contribution $\Delta S$ can be parametrically estimated by the naive formula
\begin{equation}
\label{Delta S for C=0}
\Delta S = - \frac{n}{f_{\text{ax}}} \int_{\varphi(0)}^{\varphi(r_c)} d \varphi \exp(-\alpha \varphi/2) =  \frac{2}{\alpha} \frac{n}{f_{\text{ax}}} \left(1 + \frac{\alpha n}{4A f_{\text{ax}}}  \frac{1}{r_c^2}\right)^{-1} \ ,
\end{equation}
where we used \eqref{Solution for C=0} in the second step. 
Demanding that $\Delta S \ll S_{\text{inst}}$ implies
\begin{equation} \label{Condition dS << Sinst}
\frac{\alpha n}{4A f_{\text{ax}}} \gg r_c^2 \ ,
\end{equation} 
which in turn can be rewritten as a lower bound on $S_{\text{inst}}$:
\begin{equation} 
S_{\text{inst}} \gg \frac{8 A}{\alpha^2} r_c^2 \ .
\end{equation} 
This bound depends on the cutoff $r_c$ and the dilaton coupling $\alpha$. Interestingly, the bound gets weaker for larger $\alpha$ such that contributions from gravitational instantons become increasingly important with increasing $\alpha$. However, as we will describe in \autoref{Subsection: Dilaton coupling from string theory}, a regime of large dilaton coupling $\alpha$ may not be attainable in string theory. We find that only rather small values of $\alpha \sim \mathcal{O}(1)$ arise from the simplest string compactifications. 

Before addressing the next case, we want to point out that \eqref{Instanton action C=0} can be rewritten as 
\begin{equation} \label{Instanton action C=0 and ADM Mass}
S_{\text{inst}} = \ell M_{\text{ADM}}
\end{equation}
in the case of $\alpha = 2 \sqrt{2/3}$, where $M_{\text{ADM}}$ is the ADM-mass of our \textit{extremal} 5d RN-black hole of \autoref{Subsection: Integration constant}, which is consistent with e.g.~\cite{09052844,10116301,150906374}.\footnote{For the derivation of \eqref{Instanton action C=0 and ADM Mass} we used that the black hole charge $Q$ is related to $n$ by $n=2\pi^2 \sqrt{6} Q$. This can be obtained by dimensional reduction of the term $1/(2 \cdot 3!) \int (\star_5 \hat{F})^2$ together with $\hat{F}_{\tau \rho}=\sqrt{6}Q/\rho^3$ and \eqref{Charge quantisation condition}.} 

\subsubsection*{Case $C>0$:}

The Gibbons-Hawking-York boundary term yields 
\begin{equation}
S_{\text{GHY}} = \left.-3Ar^2 \left(\sqrt{1+ \frac{C}{r^4}}-1\right)  \right|_{r_c}^{\infty} = 3Ar_c^2\left(\sqrt{1+ \frac{C}{r_c^4}}-1\right),
\end{equation}
where $A=2 \pi^2$.

The contribution from \eqref{Action without potential} is given by the integral 
\begin{align} \nonumber
S &= - \frac{n^2}{Af_{\text{ax}}^2} \int_{\varphi(r_c)}^{\varphi(\infty)} \frac{\exp(-\alpha \varphi)}{\sqrt{n^2 \exp(-\alpha \varphi)/(A^2f_{\text{ax}}^2)+6C}} = \\ 
&= \left.\frac{2n}{\alpha f_{\text{ax}}} \sqrt{\exp(-\alpha \varphi)+\sinh^2 K_+} \right|_{\varphi(r_c)}^{\varphi(\infty)},
\end{align}
where we used \eqref{C for C>0}. Combining those two results and taking $r_c \to 0$, we obtain the instanton action 
\begin{equation} \label{Instanton action for C>0}
S_{\text{inst}} = \frac{2}{\alpha} \frac{n}{f_{\text{ax}}} \left(e^{-K_+} + \frac{\alpha}{2} \sqrt{\frac{3}{2}} \sinh K_+\right).
\end{equation} 
As before, we need to ensure that the integral from $r=0$ to $r=r_c$ only gives a minor contribution to the full instanton action \eqref{Instanton action for C>0}. This contribution is 
\begin{align} 
\Delta S &\equiv \left.(S+ S_{\text{GHY}})\right|_{r=0}^{r=r_c} = \\ 
&= \frac{2}{\alpha}\frac{n}{f_{\text{ax}}} \left[\sqrt{\exp(-\alpha \varphi(r_c))+\sinh^2 K_+}- \left(1- \frac{\alpha}{2} \sqrt{\frac{3}{2}}\right) \sinh K_+  \right] \nonumber \\
&~~~~~- 3Ar_c^2 \left(\sqrt{1+ \frac{C}{r_c^4}}-1\right). \nonumber
\end{align}
In the limit $r_c^2 / \sqrt{C} \ll 1$ this can be simplified to 
\begin{align}
\nonumber \Delta S & = \frac{4 n}{\alpha f_{\text{ax}}} \sinh K_+ \left(\frac{r_c^2}{2 \sqrt{C}}\right)^{\frac{\alpha}{\sqrt{2/3}}} + 3Ar_c^2 + \ldots \\
& =  \frac{2 n}{\alpha f_{\text{ax}}} \sinh K_+ \left[ 2{\left( \frac{r_c^2}{\sqrt{C}} \right)}^{\frac{\alpha}{2 \sqrt{2/3}}} + \frac{\alpha}{2} \sqrt{\frac{3}{2}} \left( \frac{r_c^2}{\sqrt{C}} \right)  \right] + \ldots \ ,
\end{align}
where omitted terms decrease as $r_c^4/C$. The condition $\Delta S \ll S_{\text{inst}}$ turns out to be self-consistent with the imposed regime $r_c^2 / \sqrt{C} \ll 1$. More precisely, by choosing $\sqrt{C}$ sufficiently large one can always ensure that $\Delta S \ll S_{\text{inst}}$. According to \eqref{C for C>0} this is equivalent to choosing $(n \sinh K_+)/f_{\textrm{ax}}$ sufficiently large. This is very similar to the parametric situation encountered above for $C=0$. 

To determine the strongest constraints on inflation we are interested in identifying the instantons with the smallest action. For a given value of  $n/f_{\textrm{ax}}$ and at a fixed dilaton coupling $\alpha$ cored gravitational instantons correspond to a family of solutions parameterised by $K_+$ (see \autoref{GravInstScalar}). We wish to identify the instanton with the smallest action in this family. As pointed out in \autoref{Subsection: Integration constant}, while $K_+$ is expected to take discrete values, it can be effectively treated as a continuous parameter in the limit of macroscopic objects. Hence we can determine the solutions with the smallest action by formally extremising \eqref{Instanton action for C>0} with respect to $K_+$ as it was done in \cite{150906374}. For $\alpha \geq 2 \sqrt{2/3}$ the action of cored instantons is always bigger than that of extremal instantons. If $0<\alpha < 2 \sqrt{2/3}$, the smallest cored instanton action is as big as the extremal instanton action for $\alpha = 2 \sqrt{2/3}$. To summarise, we obtain
\begin{equation}
S_{\text{cored}}(\alpha) \geq \begin{cases}
S_{\text{extremal}}(\alpha) & \text{for}~ \alpha \geq 2 \sqrt{2/3} \\ 
S_{\text{extremal}}(\alpha = 2 \sqrt{2/3}) & \text{for}~ \alpha < 2 \sqrt{2/3} 
\end{cases} \ ,
\end{equation}
where the extremal instanton action was computed above in \eqref{Instanton action C=0}.
The upshot is that the contributions to the axion potential due to cored gravitational instantons will always be subleading compared to the effects due to a suitable extremal instanton. As we are interested in determining the strongest constraints on axion inflation, we will hence neglect cored instantons in the following analyses and focus on extremal instantons and Euclidean wormholes instead.

\subsubsection*{Case $C<0$:}

For Euclidean wormholes the coordinate $r$ is defined on $r \in [r_0,+ \infty)$, where $r_0 \equiv |C|^{1/4}$ is the size of the wormhole at the centre. As long as $r_0 \gtrsim r_c$ one can safely integrate from $r=r_0$ to $r= \infty$. As $r_0 \equiv |C|^{1/4} \propto n/ f_{ax}$ (see \eqref{C for wormholes}) the condition $r_0 \gtrsim r_c$ can be fulfilled by choosing $n/f_{ax}$ sufficiently large.   

As pointed out in \autoref{GravInstScalar} we will only consider wormholes with dilaton couplings $\alpha < 2 \sqrt{2/3}$ in order to have regular solutions for $\varphi$. We then proceed with calculating the action. The Giddings-Hawking-York boundary term vanishes \cite{Giddings:1987cg},
\begin{equation}
S_{\text{GHY}} = 0,
\end{equation}
since two asymptotically flat regions are connected by a handle and thus the integral gives zero. 
The on-shell contribution from \eqref{Action without potential} for only half of the wormhole\footnote{To get the instanton action, we have to divide the full wormhole action by two, as the wormhole represents a pair of instanton and anti-instanton. For more details, see \autoref{Appendix:Instanton Action}.} is given by
\begin{align} \nonumber
S_{\text{inst}} &= - \frac{n^2}{Af_{\text{ax}}^2} \int_{\varphi(r_0)}^{\varphi(\infty)} \frac{\exp(-\alpha \varphi)}{\sqrt{n^2 \exp(-\alpha \varphi)/(A^2f_{\text{ax}}^2)-6|C|}} = \\ \label{Instanton action C<0}
&= \frac{2}{\alpha} \frac{n}{f_{\text{ax}}} \sin \left(\frac{\alpha \pi}{4}\sqrt{\frac{3}{2}}\right),
\end{align} 
where we used the solutions for $C<0$ from \autoref{GravInstScalar}. 
 Notice that the limit $\alpha \to 0$ corresponds to the Giddings-Strominger wormhole \cite{Giddings:1987cg}, and we have 
\begin{equation} \label{Instanton action Giddings-Strominger}
S_{\text{inst}} = \frac{\pi \sqrt{6}}{4} \frac{n}{f_{\text{ax}}}.
\end{equation}
Furthermore, in the limit $\alpha \to 2 \sqrt{2/3}$ we find the instanton action of an extremal instanton with $\alpha =2 \sqrt{2/3}$. 

\subsubsection*{Summary}

We summarise our results for the instanton action. 
For one, the instanton action $S_{\text{inst}}$ scales as $S_{\text{inst}} \sim n/f_{\text{ax}}$ for all three types of gravitational instanton. Results were obtained in an effective theory with a cutoff at a length scale $r_c$. The existence of this cutoff implies that not all gravitational instanton solutions can be trusted in the framework of the effective theory. One can derive a criterion for deciding which gravitational instantons to include. While numerical factors may vary, this condition exhibits the same parametric behaviour for all three types of gravitational instantons: given a cutoff at a length scale $r_c$ one has to choose $n /f_{\textrm{ax}} \gg r_c^2$ for being able to trust the instanton action computed in the effective theory. 

In order to determine the importance of such gravitational instantons it is crucial to estimate the size of the cutoff scale $r_c$. The first step is to see whether moduli stabilisation places a lower bound on $r_c$. 

\section{Gravitational Instantons and Moduli Stabilisation} \label{Section:Moduli stabilisation}

We now want to make progress towards realistic string 
compactifications. The pure Einstein-axion system is relevant only  
below the moduli scale. Above that scale, moduli can play the role of an additional scalar $\varphi$ with dilatonic coupling to the axion or 2-form kinetic term. We will make use of our detailed discussion of this extended system in \autoref{Section:Gravitational Instanton Solutions } and \autoref{Section:Cored Grav Inst}.

\subsection{Gravitational Instantons in the presence of a potential}
\label{sec:addmass}
We only consider the lightest modulus, which we will call $\varphi$. For instance, it could be the saxion associated with the axion $\theta$.
We will assume stabilisation at $\varphi = 0$. In the throat region of the instanton, the modulus will be typically driven away from this value. We will assume that this displacement is small enough so that the potential of the modulus can be approximated by a mass term, i.e.~$V=m^2 \varphi^2/2$. 

The obvious extension of \eqref{Action with H} is then
\begin{equation} \label{Action with potential}
S = \int d^4x \sqrt{g} \left[-\frac{1}{2}R + \frac{1}{2}\mathcal{F}(\varphi)H^2 +  \frac{1}{2} g^{\mu\nu} \partial_{\mu}\varphi \partial_{\nu} {\varphi} + V(\varphi) \right].
\end{equation}
We take ${\cal F}$ to be exponential, which is the case discussed in detail earlier and which is typical for string-derived models (see \autoref{Subsection: Dilaton coupling from string theory}). Nevertheless, due to the presence of the potential, solutions are more complicated than before. We make the most general ansatz respecting spherical symmetry
\begin{equation} \label{Metric Ansatz General}
ds^2 = \lambda(r)dr^2 + r^2 d\Omega_3^2,
\end{equation}
as in \autoref{Appendix: metric structure}. From the derivation therein it becomes clear that $\lambda(r)$ is no longer given by $(1+C/r^4)^{-1}$. 
However, we will see that for $r\ll r_*\equiv 1/m$ the mass term is negligible and we can use the approximation $\lambda(r) \simeq (1+C/r^4)^{-1}$ (cf. the related discussion in \cite{Abbott:1989jw}). Thus, the three types of gravitational instantons analysed above remain relevant. 

The fact that the mass term is negligible close to the centre of the instanton is intuitively clear: The field strength contribution to the energy-momentum tensor increases as one approaches the centre and hence, for sufficiently small $r$, the contribution from the mass term becomes subdominant. This will become more explicit below.

Employing \eqref{Metric Ansatz General}, the Einstein equation $G_{rr}=T_{rr}$ and the Klein-Gordon equation read
\begin{align}
\frac{1}{2}(\varphi^{\prime})^2  - \lambda(r) V(\varphi)+ \frac{3}{r^2}(\lambda(r)-1)  - 3 \lambda (r) \mathcal{F}(\varphi) \frac{n^2}{A^2 r^6}&= 0 \label{Einstein eq} \\
\varphi^{\prime \prime} + \left( \frac{3}{r} - \frac{\lambda^{\prime}(r)}{2 \lambda(r)} \right) \varphi^{\prime}  - \lambda(r) V^{\prime}(\varphi) - 3 \lambda(r) \mathcal{F}^{\prime}(\varphi) \frac{n^2}{A^2 r^6} &= 0. \label{KG eq}
\end{align}
Here we also used \eqref{Ansatz for H} and \eqref{Solution for h(r)}, which specify the profile of $H$.

\subsubsection*{Approximation}

As already sketched above, the strategy is as follows: Let $\varphi_0(r)$ and $\lambda_0 (r) \equiv (1+C/r^4)^{-1}$ be the field and metric profiles for $V\equiv 0$. Then we work out the conditions under which 
\begin{equation}
T_{rr}(\varphi_0, \lambda_0) \gg V(\varphi_0)\,.
\end{equation}
This specifies the regime where we can expect the $\varphi_0(r)$ and $\lambda_0(r)$ to provide good approximations to the true solutions $\varphi(r)$ and $\lambda(r)$.

We now go into more detail: The full energy-momentum tensor of \eqref{Action with potential} reads 
\begin{equation}
T_{\mu\nu} = - g_{\mu\nu} \left[\frac{1}{2}\mathcal{F}(\varphi)H^2 + \frac{1}{2} \partial_{\rho}\varphi \partial^{\rho}\varphi + V(\varphi)\right] + 3 \mathcal{F}(\varphi)H_{\mu\rho\sigma}H_{\nu}^{~\rho\sigma} + \partial_{\mu}\varphi \partial_{\nu}\varphi.
\end{equation} 
Taking $\varphi = \varphi_0$ and $\lambda = \lambda_0$ we find
\begin{equation}
T_{rr}(\varphi_0, \lambda_0) = \frac{3C}{r^6(1+C/r^4)} - \frac{V(\varphi_0)}{1+C/r^4},
\end{equation}  
where \eqref{Einstein equation without potential} was used. We see that the potential is negligible compared to the curvature contributions if (for $C \neq 0$)
\begin{equation} \label{Criterion for neglecting the potential}
\left|\frac{3C}{r^6}\right| \gg \frac{1}{2}m^2 \varphi_0^2\,.
\end{equation} 
Appealing again to \eqref{Einstein equation without potential}, we first consider the regime $r\gg |C|^{1/4}$. Then $\varphi_0^{\prime}(r) \sim 1/r^3$ and hence 
\begin{equation} \label{Phi asymptotically}
\varphi_0 (r) \sim \frac{1}{r^2}\,.
\end{equation}
Here we treat $n/A$ and $C$ as `${\cal O}(1)$ factors' and disregard them.
We explain this below.

With this, \eqref{Criterion for neglecting the potential} translates to
\begin{equation}
\label{rcriterion}
r \ll r_*\equiv \frac{1}{m} \ .
\end{equation}
Now, our interest is in the case $m \ll1$, i.e.~in moduli much lighter than the Planck scale. This implies $r_* \gg 1$ so that $r_* \gg |C|^{1/4}$, giving us a large validity range for our approximation $\varphi_0 \sim 1/r^2$. Crucially, while $|C|$ also figured as a large parameter in other parts of this paper, here the much stronger hierarchy $1/m\gg 1$ dominates and our crude approximation concerning `${\cal O}(1)$ factors' is justified. 

Next, we need to consider the region $r\lesssim |C|^{1/4}$. While here the profile $\varphi_0(r)$ is more complicated, we are now deeply inside the regime of large field strength. It is easy to convince oneself that the potential $\sim m^2\varphi^2$ remains subdominant. What is less obvious is whether the $m^2 \varphi^2$ approximation remains justified, given that the field now moves significantly away from zero. This will be discussed later. 

Finally, the extremal instanton with $C=0$ requires an extra comment. In this case the energy-momentum tensor vanishes and the criterion \eqref{Criterion for neglecting the potential} is no longer applicable. Instead, we require that the mass term in \eqref{Einstein eq} should be subdominant compared to every other term in this equation, i.e. 
\begin{equation}
m^2 \varphi_0^2 \ll \frac{3 \mathcal{F}(\varphi_0)n^2}{A^2r^6},
\end{equation} 
which yields again the condition $r \ll 1/m$ (here we used that $\mathcal{F}$ is approximately constant for large $r$). Note that in this case the behaviour of $\varphi_0$ at large $r$ is specified by \eqref{Solution for C=0} and the role of the `largish' parameter $|C|^{1/4}$ is taken over by $n/f_{\mathrm{ax}}$. 

To summarise, we have now argued rather generally that the gravitational solutions found in the absence of a potential are good approximations for $r \ll 1/m$. We will not need the behaviour of $\varphi$ outside that region, at $r \to \infty$. Indeed, by redefining $\varphi$ we can, as argued before, always ensure that the $\varphi_0$ asymptotically approaches the minimum of the potential at $\varphi=0$. Thus, even while the actual profile of $\varphi(r)$ can significantly deviate from $\varphi_0(r)$ at $r\gg 1/m$, there is no doubt that the fundamental property of $\varphi$ approaching zero at large $r$ will be maintained. Crucially, since $1/m\gg |C|^{1/4}$ and the action integral is dominated by the region $r\lesssim |C|^{1/4}$, we can also trust the zero-potential solutions for evaluating the action, independently of the large-$r$ region.

\subsection{Dilaton Couplings from String Compactifications} \label{Subsection: Dilaton coupling from string theory}
The gravitational solutions in \autoref{GravInstScalar} were obtained for scalars with dilatonic couplings, i.e.~where the prefactor of the kinetic term for the axion is given by $\mathcal{F}(\varphi)=e^{-\alpha \varphi}/(3! f_{\textrm{ax}}^2)$. This form frequently occurs for effective theories obtained from string theory compactifications. The value of $\alpha$ will depend on the precise identification of the axion and scalar with the corresponding fields in the string compactifications. In the following, we will provide relevant examples.

\subsubsection*{The Axio-Dilaton}

Let us first consider the case where both the axion and the scalar descend from the axio-dilaton field $S= C_0 + i/g_s$ with string coupling $g_s$ and universal axion $C_0$. It appears in the K\"ahler potential as 
\begin{equation}
\mathcal{K} = - \ln \left(-i(S-\bar{S})\right).
\end{equation}
The kinetic term of the Lagrangian $\mathscr{L} \supset  \mathcal{K}_{S\bar{S}} \partial_{\mu} S \partial^{\mu} \bar{S}$ then becomes 
\begin{equation}
\mathscr{L} \supset \frac{g_s^2}{4} (\partial C_0)^2 + \frac{1}{4g_s^2} (\partial g_s)^2 \ . 
\end{equation}
Canonical normalisation of our saxion gives $g_s = g_s^0 \exp(\sqrt{2}\varphi)$. Thus, the field strength coupling reads
\begin{equation}
\mathcal{F}(\varphi) = \frac{1}{2 \cdot 3!}\frac{1}{\mathcal{K}_{S\bar{S}}} = \frac{1}{3(g_s^0)^2} \exp(-2\sqrt{2}\varphi),
\end{equation}
so in our notation the dilaton coupling $\alpha$ is $\alpha=2 \sqrt{2}$. Notice that $\varphi \to \infty$ corresponds to the strong coupling limit, while the weak coupling limit is given by $\varphi \to -\infty$. 

\subsubsection*{K\"ahler Moduli at Large Volume}

Let us now consider the K\"ahler moduli sector at large volume. In particular, consider the case where the volume is dominated by one K\"ahler modulus $T$. For example, this arises in the scheme of moduli stabilisation known as the Large Volume Scenario (LVS) \cite{0502058}. The relevant part of the K\"ahler potential is
\begin{equation}
\mathcal{K} = -2 \ln \mathcal{V} = - 3 (T+ \bar{T}) + \ldots \ .
\end{equation}
Here we wish to identify the saxion with $\textrm{Re}(T)$ and the axion with $\textrm{Im}(T)$. The leading contribution to the kinetic term for the saxion and axion is then given by
\begin{equation}
\mathcal{K}_{T \bar{T}} = \frac{3}{(T+\bar{T})^2} \ .
\end{equation}
Canonical normalisation gives 
\begin{equation}
\mathrm{Re}(T) = \exp \left(-\sqrt{\frac{2}{3}} \varphi\right),
\end{equation}
and hence 
\begin{equation}
\mathcal{F}(\varphi) \sim \exp \left(-2\sqrt{\frac{2}{3}} \varphi\right) \ .
\end{equation} 
The dilaton coupling is thus $\alpha = 2 \sqrt{2/3}$. 

\subsubsection*{Complex Structure Moduli in the Large Complex Structure Limit (LCS)}

It is well-known that complex structure moduli in the LCS limit give rise to a shift-symmetric structure in the K\"ahler potential. Let $u$ be a complex structure modulus in the LCS regime and $z$ denote the remaining complex structure moduli. Then we have 
\begin{align} \nonumber
\mathcal{K} &= -\ln \left(\kappa_{uuu}(u+\bar{u})^3 + \kappa_{uui}(u+\bar{u})^2(z_i+ \bar{z}_i) + \frac{\kappa_{uij}}{2!}(u+\bar{u})(z_i+ \bar{z}_i) (z_j+ \bar{z}_j) \right.+ \\ 
&\left.+ \frac{\kappa_{ijk}}{3!}(z_i+ \bar{z}_i) (z_j+ \bar{z}_j)(z_k+ \bar{z}_k) + f(z_i) \right),
\end{align}
where the $\kappa_{ijk}$ denote the intersection numbers of the mirror-dual Calabi-Yau three-fold and $f$ is a function of the remaining complex structure moduli $z_i$ and accounts for instantonic corrections to the K\"ahler potential. For the moment only $u$ shall be stabilised in the LCS limit, i.e.
\begin{equation}
\mathrm{Re}(u) > 1 \ .
\end{equation}
Thus, one obtains 
\begin{equation}
\mathcal{K}_{u\bar{u}} = \frac{3}{(u+\bar{u})^2} 
\end{equation}
at leading order as long as $\kappa_{uuu} \neq 0$. Omitted terms scale as $(u+\bar{u})^{-3}$. Therefore, canonical normalisation yields
\begin{equation}
\label{eq:cannormsaxu} \mathrm{Re}(u) = \exp \left(-\sqrt{\frac{2}{3}} \varphi\right),
\end{equation}
and hence $\alpha = 2\sqrt{2/3}$.

In the situations studied so far the saxion and axion arose from the same complex scalar field. However, one may also consider the case where the saxion and axion originate from different moduli. To give just one example, let us again consider the complex structure sector of a CY threefold, but now we will assume that two complex structure moduli $u,v$ are in the LCS regime. Further, we assume the following hierarchy
\begin{equation}
\mathrm{Re}(u) \gg \mathrm{Re}(v) \gg 1 \ .
\end{equation}
We will now consider the axionic field $\textrm{Im}(v)$ and study the coupling to the saxion $\textrm{Re}(u)$.
As before, the leading contribution to the kinetic term of the saxion is
\begin{equation}
\mathcal{K}_{u\bar{u}} = \frac{3}{(u+\bar{u})^2} \ ,
\end{equation}
and the canonically normalised saxion is given by \eqref{eq:cannormsaxu}. The leading contribution to the kinetic term for the axion is 
\begin{equation}
\mathcal{K}_{v\bar{v}} \sim \frac{1}{(u+\bar{u})^2} \sim \exp \left(-2\sqrt{\frac{2}{3}} \varphi\right) \ ,
\end{equation}
where omitted terms decrease as $(u + \bar{u})^{-3}$. While both $\mathcal{K}_{u\bar{u}}$ and $\mathcal{K}_{v\bar{v}}$ scale as $(u + \bar{u})^{-2}$, this behaviour has different origins in the two cases. The leading contribution to $\mathcal{K}_{u\bar{u}}$ comes from $\kappa_{uuu} (u + \bar{u})^3$, whereas it is the terms $\kappa_{uvv} (u + \bar{u})(v + \bar{v})^2$ and $\kappa_{uuv} (u + \bar{u})^2(v + \bar{v})$ which contribute to $\mathcal{K}_{v\bar{v}}$ at leading order. 

Despite these differences we again find
\begin{equation}
\mathcal{F}(\varphi) \sim \exp \left(-2\sqrt{\frac{2}{3}} \varphi\right) \ ,
\end{equation}
and $\alpha = 2\sqrt{2/3}$.\footnote{This would be different if $\kappa_{uuu}$ and $\kappa_{vvv}$ were the only non-vanishing intersection numbers. Then we would still have $\mathcal{K}_{u \bar{u}} \simeq 3/(u + \bar{u})^2$ but now $\mathcal{K}_{v\bar{v}} \sim (v+ \bar{v})/(u+\bar{u})^3$ for $\text{Re}(u) \gg \text{Re}(v)$. Assuming that $\text{Re}(v)$ is stabilised such that we can take it as constant, we would now find a dilaton coupling $\alpha = \sqrt{6}$.}

Note that in all the cases examined above the dilaton coupling is just outside the range allowing for Euclidean wormhole solutions $0 \leq \alpha < 2 \sqrt{2/3}$. This observation and a possible way out have been pointed out in \cite{0412183, 07052768}. The idea is as follows. Even if wormholes charged under individual axions do not exist, one can nevertheless find solutions which are charged under more than one axion (see also \cite{Tamvakis:1989aq}). 

We will conclude this section with an example that allows for the existence of Euclidean wormhole solutions and may be useful to illustrate and develop the above idea. Let us consider both the axio-dilaton sector and the complex structure moduli sector of a CY 3-fold at LCS:
\begin{equation}
\mathcal{K} = - \ln \left(-i(S-\bar{S})\right) - \ln \left(\kappa_{uuu}(u+\bar{u})^3\right) \ .
\end{equation}
In the spirit of \cite{Tamvakis:1989aq, 0412183, 07052768} we could now investigate Euclidean wormhole solutions charged under both the universal axion as well as the complex structure axion. Alternatively, we may assume that we can stabilise moduli such that $S=iu$. Then we effectively have the theory of one 4-fold complex structure modulus and we obtain
\begin{equation}
\mathcal{K}_{u\bar{u}} = \frac{4}{(u+\bar{u})^2} \ .
\end{equation}
Taking the saxion as $\textrm{Re}(u)$ and the axion as $\textrm{Im}(u)$ we now find $\alpha = \sqrt{2}$ which lies within the range allowing for wormholes. We leave it for future work to investigate whether this pattern of moduli stabilisation can be realised in a realistic compactification.

\subsection{Maximal Field Displacements of Dilatonic Fields} \label{Subsection: Maximal field displacement}
In the previous sections we have made progress towards studying gravitational instantons in the presence of moduli. The results of \autoref{Subsection: Dilaton coupling from string theory} imply that a restriction to moduli with dilatonic couplings is well-motivated from string compactifications. We also made progress towards understanding the role of the potential stabilising the modulus in \autoref{sec:addmass}. In particular, in the regime $r \ll 1/m$ the potential can be ignored and gravitational instanton solutions for a massless dilaton will be good approximations.

There is another effect which we need to take into consideration. When approaching the core of a gravitational instanton, the value of the dilaton increases. In the case of a Euclidean wormhole it reaches a maximum at the wormhole throat, while for extremal and cored instantons the dilaton diverges for $r \rightarrow 0$. However, we cannot afford arbitrarily large field displacements, as this will take us outside the range of validity of our effective theory. 

To be specific, consider the effective theory of the axio-dilaton at weak string coupling. When approaching the centre of a gravitational instanton solution the string coupling increases compared to its asymptotic value. If it becomes too large the supergravity regime breaks down and we cannot trust our solutions. A similar argument can be made for any effective dilaton-axion theory from string compactifications. 

This gives us an additional criterion to decide which gravitational instantons to trust and which ones to disregard. We will analyse this condition focussing on Euclidean wormholes and extremal gravitational instantons. In the following, we will denote by $\varphi_{\textrm{max}}$ the threshold value at which the effective theory breaks down.

\subsubsection*{Euclidean Wormholes}

For Euclidean wormholes the displacement becomes maximal at the throat of the wormhole. Using our solution for $\varphi(r)$ \eqref{Solution for C<0 w/o constant K_-} one finds: 
\begin{equation}
\varphi(r_0) = - \frac{1}{\alpha} \ln \cos^2 \left(\frac{\alpha \pi}{4} \sqrt{\frac{3}{2}}\right).
\end{equation}
To trust the wormhole solution we require $\varphi(r_0) < \varphi_{\textrm{max}}$. The maximal displacement only depends on the dilaton coupling $\alpha$, which is not a free parameter, but a property of the physical system studied. As a result the maximal displacement is model-dependent. 

Recall that Euclidean wormhole solutions only exist for dilaton couplings in the range $0 \leq \alpha < 2 \sqrt{\tfrac{2}{3}}$. The maximal displacement at the wormhole throat is smallest for $\alpha =0$, grows when $\alpha$ is increased and eventually diverges for $\alpha \rightarrow 2 \sqrt{\tfrac{2}{3}}$. To give just one example, the value $\alpha = \sqrt{2}$ yields a displacement $\varphi(r_0) - \varphi(\infty) = \varphi(r_0) \simeq 2.2$ in Planck units, which may already be critical.

Another important result from this section is that the maximal displacement $\varphi (r_0)$ is independent of the ratio $n/f_{\text{ax}}$, or, equivalently, the wormhole radius at the throat $r_0$. Hence we do not get any additional constraints on these quantities due to the displacement of the saxion.

\subsubsection*{Extremal Instantons}

For extremal gravitational instantons the $\varphi$-profile exhibits a divergent behaviour for $r \to 0$. As laid out in \autoref{Section:Cored Grav Inst}, we will nevertheless trust such solutions as long as the dominant part of the instanton action arises from the region $r > r_c$, where $r_c$ is the length cutoff of our effective theory. This cutoff will be discussed in more detail in \autoref{sec:selfdual}. Here we will show that the displacement of the saxion gives an independent condition for the reliability of our action.

Let us be more precise. Given a threshold value $\varphi_{\textrm{max}}$ beyond which our theory breaks down, we can identify a radius $r_{\textrm{min}}$ at which the dilaton crosses this value: $\varphi(r_{\textrm{min}}) = \varphi_{\textrm{max}}$. This can be made explicit using \eqref{Solution for C=0}: 
\begin{equation}
		e^{\alpha \varphi_{\mathrm{max}}} = \left(1+ \frac{\alpha n}{4A f_{\mathrm{ax}}} \frac{1}{r_{\mathrm{min}}^2}\right)^2 \ ,.
\end{equation}
To trust our solution we need to ensure that $\Delta S / S \ll 1$, where $\Delta S$ is the contribution to the instanton action from the region $r < r_{\mathrm{min}}$. Using \eqref{Instanton action C=0} and \eqref{Delta S for C=0} we find 
\begin{equation}
\frac{\Delta S}{S} = \left(1+ \frac{\alpha n}{4A f_{\text{ax}}} \frac{1}{r_{\mathrm{min}}^2}\right)^{-1} = \exp \left(- \tfrac{\alpha }{2} \varphi_{\textrm{max}} \right) \,.
\end{equation} 
Hence the relevant condition is
\begin{equation}
\exp \left(- \tfrac{\alpha }{2} \varphi_{\textrm{max}} \right)  \ll 1 \,,
\end{equation} 
which gives an additional (model-dependent) constraint. 

Last, let us return to one aspect encountered for the case of Euclidean wormholes. There we observed that for $\alpha \rightarrow 2 \sqrt{\tfrac{2}{3}}$ the saxion displacement at the wormhole throat grows without bound and would exceed any finite value $\varphi_{\text{max}}$. Note that this does not necessary constitute a pathology. Rather, the behaviour observed for a wormhole becomes similar to that of an extremal instanton. In fact, in the limit $\alpha \rightarrow 2 \sqrt{\tfrac{2}{3}}$ the Euclidean wormhole becomes a pair of extremal instantons. We can then deal with the divergence of $\varphi$ as in the case of extremal instantons and cut our solution off at some $r= r_{\textrm{min}}$.

\section{Consequences for Large Field Inflation}
\label{Section:Inflation}

In this section we will analyse to what extent gravitational instantons constrain axion inflation. The idea is as follows: we will check whether the contribution to the axion potential $\delta V$ due to gravitational instantons can be large enough to disrupt inflation. 

To be specific, gravitational instantons contribute as
\begin{equation} \label{Delta V of gravitational instantons}
\delta V =  \mathcal{A} \cos (n\theta) e^{-S} \ ,
\end{equation}  
where $S \sim n/f_{\text{ax}}$ (see \autoref{Section:Cored Grav Inst}). Whether such instanton corrections can have significant influence on the slow-roll dynamics clearly depends both on the size of the instanton action $S$ and the prefactor $\mathcal{A}$. The latter is quoted to be of order $M_p^4$ \cite{150300795,150303886}. However, in \autoref{Appendix: Prefactor} we give arguments why the prefactor $\mathcal{A}$ can be significantly below the Planck scale in more realistic string compactification models. Specifically, we expect $\mathcal{A}$ to scale as $\mathcal{A} \sim \mathcal{V}^{-5/3}$ with compactification volume $\mathcal{V}$. 

We then compare $\delta V$ with the size of the axion potential during inflation. For large field inflation the scale of inflation is of the order \cite{150202114} 
\begin{equation}
V_{\textrm{inflation}} \sim 10^{-8}.
\end{equation} 
Hence, whenever we find $\delta V \sim 10^{-8}$ we will conclude that the effects of gravitational instantons on the axion potential are in principle large enough to spoil inflation. 

In what follows we compute $\delta V$ only for the case of a single axion, but our results can be straightforwardly extended to models of $N$-flation, kinetic alignment and the Kim-Nilles-Peloso mechanism, see \cite{150303886} for more details.  

\subsection{Action of the most `dangerous' Gravitational Instantons}
\label{sec:mostdanger}

To check whether gravitational instantons are dangerous for inflation, we want to focus on the instantons with the smallest action. At the same time, we need to ensure that these are solutions which we can trust within the framework of our effective theory. In brief, we are interested in the most relevant instanton within the regime of validity of our theory.

The breakdown of our effective gravity theory is crucial in this context, because it will put a lower bound on the instanton action $S$. As explained in \autoref{Section:Cored Grav Inst}, in a theory with length cutoff $r_c$ we can only trust gravitational instanton solutions with $n/ f_{\textrm{ax}} \gg r_c^2$. This translates into a lower bound on the instanton action as $S \sim n/ f_{\textrm{ax}}$.

To calculate the contributions of gravitational instantons to the axion potential we hence need to determine the cutoff $r_c$. In \autoref{sec:selfdual} we will estimate the smallest possible value of $r_c$ at which the description in terms of a 4-dimensional theory may hold. Before doing this it will be instructive to check how large $r_c$ can be so that gravitational instantons still induce a sizeable contribution to the inflaton potential.  

Note that gravitational instanton solutions for the case of a massless dilaton will be sufficient for our analysis, despite the fact that we are interested in the case of massive dilaton fields. As described in \autoref{sec:addmass} the non-zero potential does not affect the action significantly.

\subsubsection*{Euclidean Wormholes} 

For any $n$ and $f_{\text{ax}}$ the Euclidean wormhole action is computed in \eqref{Instanton action C<0}. At the same time the wormhole radius $r_0$ scales as $r_0 \sim (n/f_{\text{ax}})^{1/2}$ according to \eqref{C for wormholes}. As we require $r_0 \gtrsim r_c$ we get
 \begin{equation}
 S_{\text{inst}}  \gtrsim (2\pi^2) \sqrt{6}  \frac{2}{\alpha} \tan \left(\frac{\alpha \pi}{4} \sqrt{\frac{3}{2}} \right) r_c^2 \ .
 \end{equation}
On the allowed interval $0 \leq \alpha < 2 \sqrt{2/3}$ the instanton action as a function of $\alpha$ increases monotonically. Therefore, the most dangerous wormhole corresponds to the Giddings-Strominger instanton with $\alpha=0$. Hence
 \begin{equation}
\label{DangerWH}
 S_{\text{inst}} \geq S_{\text{inst}} (\alpha=0) \gtrsim 3 \pi^3 r_c^2. 
 \end{equation}
 Demanding that $e^{-S} \gtrsim 10^{-8}$ implies $r_c \lesssim 0.4$ (in Planck units). 
 
 In \autoref{Subsection: Dilaton coupling from string theory} we found that $\alpha=\sqrt{2}$ can be obtained from string compactifications and still lies in the allowed range of dilaton-couplings appropriate to allow for Euclidean wormholes. This example requires $r_c \lesssim 0.2$ in order to get a contribution of at least $\delta V \sim 10^{-8}$.
 
 Note that the prefactor $\mathcal{A}$ (see \autoref{Appendix: Prefactor}) may potentially lower the size of the contribution to the inflaton potential.
 
 \subsubsection*{Extremal Gravitational Instantons}

The action for extremal gravitational instantons is obtained from \eqref{Instanton action C=0}. However, we have to take into account the computability condition \eqref{Condition dS << Sinst} for the action. It follows that
\begin{equation}
\label{DangerExt}
 S_{\text{inst}} > \frac{8 \cdot (2\pi^2)}{\alpha^2}r_c^2 \ .
\end{equation}

In string theory $\alpha$ cannot be chosen arbitrarily large. The largest $\alpha$ we could obtain from string compactifications was $\alpha = 2\sqrt{2}$. Extremal gravitational instantons then become relevant if $r_c \lesssim 1$. Hence, extremal gravitational instantons may turn out to be somewhat more dangerous for axion-inflation than Euclidean wormholes. 

We do not consider cored gravitational instantons, for which a similar analysis could be made. The reason is that their action is always larger than that of a suitable extremal instanton (see \autoref{Section:Cored Grav Inst}). 
\bigskip

The question we want to address now is how small $r_c$ can be in any controlled model of quantum gravity. Knowing that moduli displacements are not an issue, one would naively expect that $r_c \simeq 1$ can be problematic as we reach already Planck regime. Notice however that it is important to determine $r_c$ as precisely as possible, because due to $\delta V \sim e^{-S} \sim e^{-r_c^2}$ the instanton contributions are very sensitive to every $\mathcal{O}(1)$-factorchange in the cutoff radius.  
 
 \subsection{Estimating the Critical Radius $r_c$}
\label{sec:selfdual}
 
Let us take string theory as our model of quantum gravity. String compactifications then yield a hierarchy of scales in the effective theory as depicted in \autoref{Fig:scales}. We expect that going beyond the Kaluza-Klein scale will render our effective description insufficient. The reason is that the gravitational instanton solutions we consider are obtained in a 4-dimensional effective theory which arises from a more fundamental description upon compactification. For the 4-dimensional picture to remain valid, we require the length scale $r_c$ associated with our 4-dimensional solution not to be smaller than the length scale associated to the compactified extra-dimensions. 

But how small can this length scale be? In string theory it cannot be arbitrarily small. String compactifications exhibit a property termed T-duality which states that a compactification with a small volume describes the same physics as another compactification with large volume. This gives rise to the notion of a smallest length scale at the self-dual value of the compactification volume. 

Putting everything together, we arrive at the smallest possible value $r_c$ where we can trust our effective 4-dimensional analysis. We find that $r_c$ should be related to the length scale of the compact dimensions at self-dual volume $\mathcal{V}_{\text{sd}}$ of the compactification space. In this way, we push the KK-scale as close to the Planck scale as possible, allowing us to consider the lightest gravitational instantons we can obtain within the regime of validity of our description. 

What we mean by ``related'' is at this naive level ambiguous. There are at least two ``canonical'' possibilities to make the definition of $r_c$ more precise. They differ by factors of $\pi$, which are unfortunately crucial when comparing $e^{-S}$ with $V_{\text{inflation}}$. Given the volume $\mathcal{V}_{\text{sd}}$ of the six-dimensional compact space at the self-dual point we can define a length scale as $\mathcal{V}_{\text{sd}}^{1/6}$ and a 3-volume by $\mathcal{V}_{\text{sd}}^{1/2}$. Two possible definitions of $r_c$ are then:
 \begin{enumerate}
 	\item The volume of the $S^3$ of our wormhole solution should not be smaller than $\mathcal{V}_{\text{sd}}^{1/2}$, i.e. 
 	\begin{equation} \label{Self dual radius - Def1}
 	2\pi^2r_c^3 = \mathcal{V}_{\text{sd}}^{1/2}.
 	\end{equation} 
 	\item More generously, the great circle of $S^3$ should not be smaller than $\mathcal{V}_{\text{sd}}^{1/6}$, i.e.
 	\begin{equation} \label{Self dual radius - Def2}
 	2\pi r_c = \mathcal{V}_{\text{sd}}^{1/6}.
 	\end{equation} 
 \end{enumerate}
 As a toy-model to compute $\mathcal{V}_{\text{sd}}$ we take $T^6$ and apply T-duality six times for each $S^1$ to get $\mathcal{V}_{\text{sd}}(T^6)  = \ell_s^6= (2 \pi)^6 (\alpha')^3$. To convert this into Planck units, recall that (see e.g.~\cite{Conlon:2006gv})
 \begin{equation}
 M_p^2 = \frac{4\pi \mathcal{V}}{g_s^2 \ell_s^8} \ .
 \end{equation}
In the following we also go to the S-self-dual point $g_s=1$.

The first criterion \eqref{Self dual radius - Def1} then gives $r_c M_p = \sqrt{4\pi} \cdot (2\pi^2)^{-1/3} \simeq 1.3$. Using \eqref{DangerWH} and \eqref{DangerExt}, which are both in 4d Planck units, the contributions to the axion potential due to gravitational instantons are then:
\begin{align}
\nonumber \textrm{Giddings-Strominger wormhole:}& \qquad e^{-S} \simeq 10^{-68} \ , \\
\nonumber \textrm{Extremal instantons:}& \qquad e^{-S} \lesssim 10^{-15} \quad \textrm{for } \alpha=2\sqrt{2} \ .
\end{align}
Hence, in both cases the gravitational instantons appear to be irrelevant for inflation.

If we apply the second criterion \eqref{Self dual radius - Def2} we have $r_c M_p = 1/\sqrt{\pi} \simeq 0.56$. This yields 
\begin{align}
\nonumber \textrm{Giddings-Strominger wormhole:}& \qquad e^{-S} \simeq 10^{-13} \ , \\
\nonumber \textrm{Extremal instantons:}& \qquad e^{-S} \lesssim 10^{-3} \quad \textrm{for } \alpha=2\sqrt{2} \ .
\end{align}
Again, Euclidean wormholes contribute to the axion potential too weakly to interfere significantly with inflation. However, extremal instantons can in principle be important, but this will depend on the value of the dilaton coupling $\alpha$. Note that for $\alpha = 2\sqrt{2/3}$ we still get $e^{-S} \lesssim 6 \cdot 10^{-9}$ for extremal instantons, which is marginal as far as the significance for inflation is concerned. However, we want to emphasise that our numerical results should be taken with a grain of salt. In particular, given a value of a length cutoff $r_c$ we only have a lower bound \eqref{DangerExt} for the action of the most important trustworthy extremal instanton. However, $\delta V$ is exponentially sensitive to the instanton action. Thus, unless the instanton action is close to saturating the inequality \eqref{DangerExt} the contributions from extremal instantons can quickly become irrelevant for inflation.

Of course, the instanton contribution $\delta V = \mathcal{A}e^{-S} \cos(n\theta)$ also involves the prefactor $\mathcal{A}$, which we estimate in \autoref{Appendix: Prefactor}. We expect $\mathcal{A} \sim \mathcal{V}^{-5/3}$, which is $\mathcal{O}(1)$ at the self-dual point. Note that in more realistic scenarios away from the self-dual point (i.e.~compactifications with a hierarchy of scales) it would suppress the gravitational instanton contributions even further. 
\bigskip 

Our results can be summarised as follows: overall, we find that gravitational instantons do not give rise to strong \emph{model-independent} constraints on axion inflation, even if we push the KK-scale as close to the Planck-scale as possible.  Extremal gravitational instantons may be important for inflation, but this is model-dependent, as the size of their contribution depends on the value of the dilaton coupling $\alpha$.

\section{Gravitational Instantons and the Weak Gravity Conjecture}
\label{Section: WGC}
Finally, we want to make further remarks on the relation between gravitational instantons and the Weak Gravity Conjecture (WGC) \cite{0601001, 150303886,150906374}. The original form of the WGC requires that the particle spectrum of a consistent, UV-complete gravitational theory with a $U(1)$ gauge field contains at least one particle whose charge-to-mass ratio is larger or equal to that of an extremal black hole \cite{0601001}. There exists a straightforward generalisation to gravitational theories with an axion coupling to instantons. In the following we will argue in analogy with the WGC for particles with U(1) charges, i.e.~we will treat instantons like particles with axion charge.

We start with the theory of an axion with an instanton-induced potential:
\begin{equation}
	\mathscr{L} = \frac{1}{2} (\partial \theta)^2 - \sum_i \Lambda_i^4 e^{-S_i} \cos (\tfrac{n_i}{f_{\text{ax}}} \theta) \ .
\end{equation}
The WGC then requires the existence of an instanton with
\begin{equation}
	z \equiv \frac{n_i/f_{\text{ax}}}{S_i} > z_{0} \ ,
\end{equation}
for some $z_0$ to be specified shortly.\footnote{The WGC can be made more precise by adding further attributes to the condition $z>z_0$ \cite{0601001}. A more careful definition becomes important when several U(1) group factors (or axion species) are present. See \cite{14022287, 150303886, 150307853, 150400659, 150906374, 160608437} for more details.} The quantity $z$ is the equivalent of the charge-to-mass ratio for the instanton, where the charge is given by $n / f_{\text{ax}}$ and the mass corresponds to $S$. When working with black holes an object satisfying $z > z_0$ is referred to as superextremal, while a black hole with $z < z_0$ is termed subextremal. It will be useful to extend this nomenclature to the case of instantons. The WGC then requires the existence of superextremal instantons.

To define the WGC for instantons it is hence important to determine $z_0$. In the black hole case $z_0$ is the charge-to-mass ratio of an extremal RN black hole. By analogy, we will define $z_0$ as the charge-to-mass ratio of an extremal gravitational instanton as suggested in \cite{150906374}. There is further support for this assertion. In \autoref{Subsection: Integration constant} we saw that gravitational instantons in 4d are related to RN black holes in 5d. More specifically, the relation $C= \ell^2(M^2-Q^2)$ implies that extremal black holes ($M^2=Q^2$) are in one-to-one correspondence with extremal instantons ($C=0$). It is thus plausible that extremal instantons play the role of extremal black holes in the WGC. Using our expression \eqref{Instanton action C=0} for the action of an extremal gravitational instanton we find
\begin{equation}
	z_{0} = \frac{n/f_{\text{ax}}}{S_{\text{extremal}}} = \frac{\alpha}{2} .
\end{equation}

\begin{figure}
		\subfigure[]{\begin{overpic}[width=0.45\textwidth]{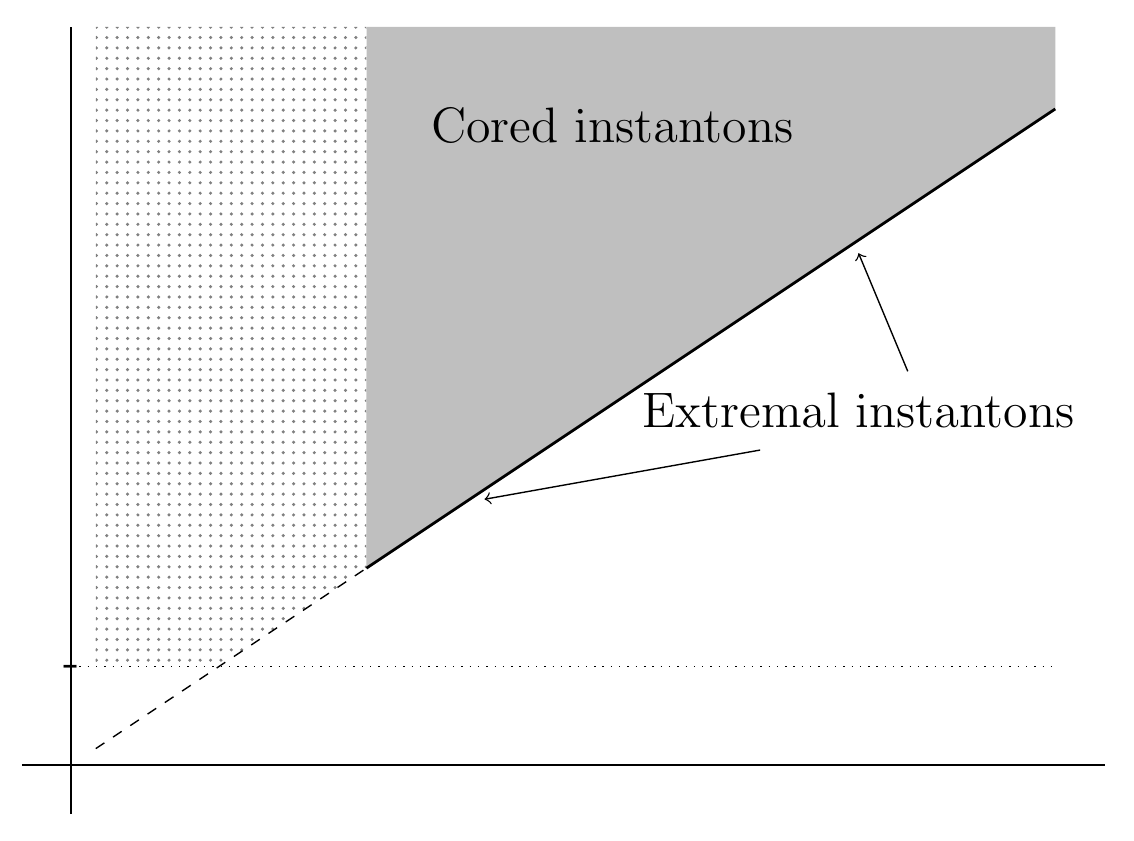} 
		\put (84,0) {$n/ f_{\textrm{ax}}$} \put (0,67) {$S$} \put (0,12) {$1$} \put (57,22) {$\alpha \geq 2 \sqrt{2/3}$} \end{overpic}}
		\subfigure[]{\begin{overpic}[width=0.50\textwidth]{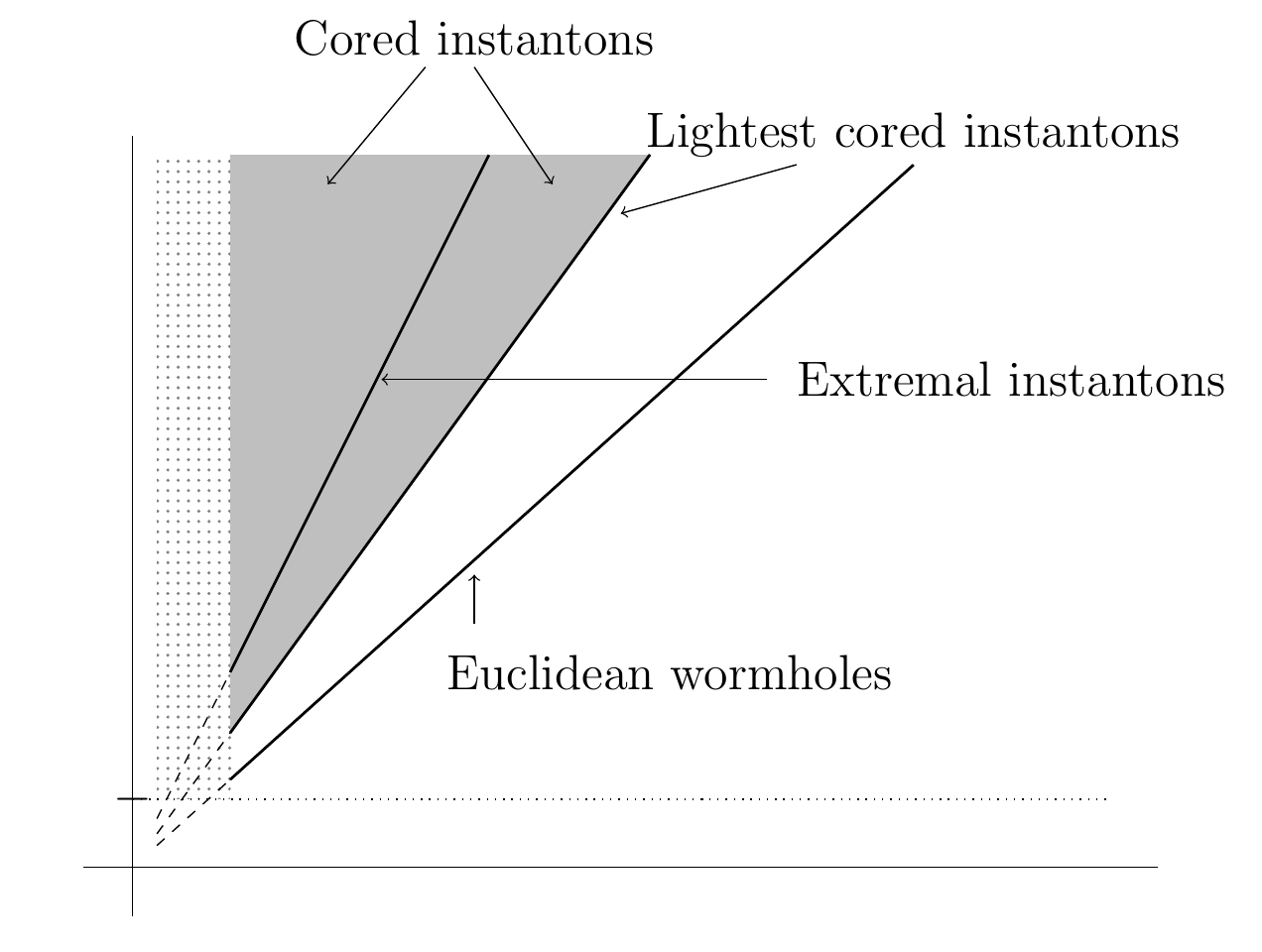}
		\put (80,0) {$n/ f_{\textrm{ax}}$} \put (3,60) {$S$} \put (3,11) {$1$} \put (57,32) {$\alpha < 2 \sqrt{2/3}$} \end{overpic}}
		\caption{Instanton action $S$ vs.~$n/ f_{\textrm{ax}}$ for \textbf{(a)} $\alpha \geq 2 \sqrt{2/3}$ and \textbf{(b)} $\alpha < 2 \sqrt{2/3}$.}
		\label{fig:ChargeToMass}
\end{figure}
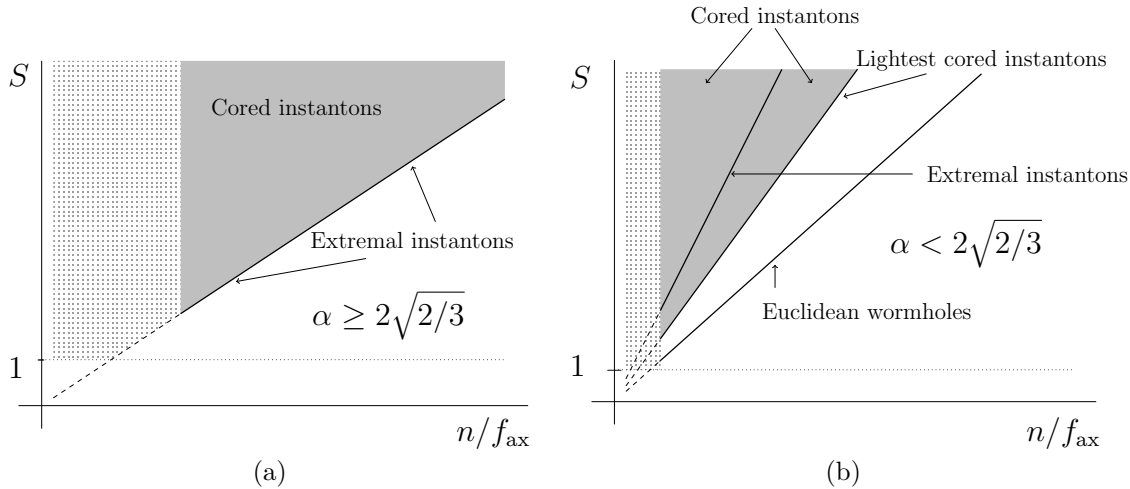

Let us now compute the charge-to-mass ratio $z$ for cored gravitational instantons and Euclidean wormholes to see how they fit into this picture. We begin with cored gravitational instantons. For fixed $n/f_{\text{ax}}$ we have
\begin{equation}
	S_{\text{cored}}(\alpha) \geq \begin{cases}
		S_{\text{extremal}}(\alpha) & \text{for}~ \alpha \geq 2 \sqrt{2/3} \\
		S_{\text{extremal}}(2 \sqrt{2/3}) & \text{for}~ \alpha < 2 \sqrt{2/3},
	\end{cases}
\end{equation}
and thus
\begin{equation}
	z_{\text{cored}} \leq \begin{cases} z_0 & \text{for}~ \alpha \geq 2 \sqrt{2/3} \\ \frac{2 \sqrt{2/3}}{\alpha} \, z_0 & \text{for}~ \alpha < 2 \sqrt{2/3} \end{cases}
\end{equation}
We can make the following observation. For $\alpha \geq 2 \sqrt{2/3}$ cored gravitational instantons are strictly subextremal and do not satisfy the WGC condition $z > z_0$. They hence play a role akin to subextremal black holes in the WGC for particles. This is consistent with the finding that for $\alpha \geq 2 \sqrt{2/3}$ cored gravitational instantons are related to subextremal black holes in higher dimensions (see \cite{0406038} and \autoref{Subsection: Integration constant}). The situation is different for $\alpha < 2 \sqrt{2/3}$. The lightest cored instantons are now superextremal. We illustrate our findings in \autoref{fig:ChargeToMass}.

Next, let us turn to Euclidean wormholes. From \eqref{Instanton action C<0} we find
\begin{equation}
	z_{\text{wh}} = \frac{n/f_{\text{ax}}}{S_{\text{wh}}} = \frac{\alpha}{2\sin \left(\frac{\alpha \pi}{4}\sqrt{\frac{3}{2}}\right)} > \frac{\alpha}{2} = z_{0}
\end{equation}
for $0 \leq \alpha <2 \sqrt{2/3}$, which is the allowed range for wormhole solutions. We find that Euclidean wormholes are strictly superextremal. In addition, one can also show that $z_{\text{wh}} > z_{\textrm{cored}}$. This is displayed in \autoref{fig:ChargeToMass} (b).

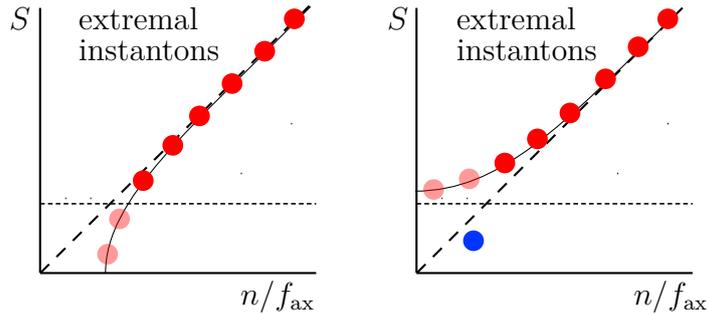
\begin{figure}
\centering
 \begin{overpic}[width=0.65\textwidth,tics=10]{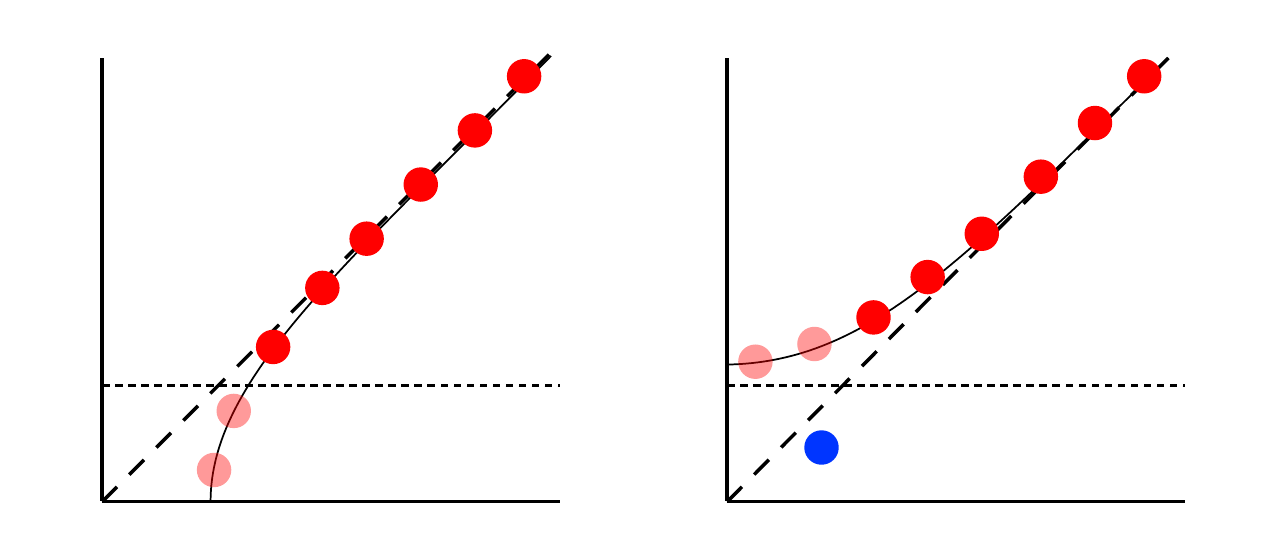}
 \put (34,-1) {$n / f_{\textrm{ax}}$} \put (13,35) {extremal} \put (13,31) {instantons} \put (62,35) {extremal} \put (62,31) {instantons} \put (4,35) {$S$} \put (85,-1) {$n / f_{\textrm{ax}}$} \put (53,35) {$S$} 
\end{overpic}
\caption{Possible realisations of the WGC for gravitational instantons \cite{ReeceTalk}. Red dots denote extremal gravitational instantons while blue dots correspond to additional superextremal instantons required by the WGC. }\label{fig:WGC}
\end{figure}

What can one learn from these results about the WGC? We will discuss this question for the two cases $\alpha \geq 2 \sqrt{2/3}$ and $\alpha < 2 \sqrt{2/3}$ separately. 

For  $\alpha \geq 2 \sqrt{2/3}$ the spectrum of gravitational instantons does not contain any superextremal objects that could satisfy the WGC. This is not surprising. Our analysis is restricted to macroscopic gravitational instantons, while it is expected that microscopic physics is responsible for satisfying the WGC. If the WGC is true, it could be realised in two different ways which are shown in \autoref{fig:WGC}. For one, extremal gravitational instantons (red dots in \autoref{fig:WGC}) could satisfy the WGC on their own. This occurs if quantum corrections decrease the instanton action for small $n$ such that they naively become superextremal (LHS of \autoref{fig:WGC}). If this is not the case (see RHS of \autoref{fig:WGC}) the WGC requires the existence of additional superextremal instantons (blue dot). At the moment it is not clear which implementation of the WGC, if any, is realised. Unfortunately, our analysis is not suitable for resolving this issue.

Let us move on to $\alpha < 2 \sqrt{2/3}$. Interestingly, the set of gravitational instanton solutions now contains superextremal objects in the form of wormholes and cored instantons. It thus seems that the WGC is satisfied in Einstein-axion-dilaton systems in virtue of cored instantons and Euclidean wormholes. Note that this is different to the situations shown in \autoref{fig:WGC}. Here the WGC would be satisfied by an infinite tower of superextremal macroscopic objects. 

Another interpretation of our findings is that the statement of the WGC has to be modified for $\alpha < 2 \sqrt{2/3}$. In this regime the `lightest' macroscopic object with given charge $n/f_{\textrm{ax}}$ is not the extremal instanton but the wormhole. Also the correspondence between extremal instantons and extremal black holes in higher dimensions is lost for $\alpha < 2 \sqrt{2/3}$. This may imply that the WGC condition is now set by the charge-to-mass ratio of the wormhole rather than that of the extremal instanton. To satisfy the WGC one would then require the existence of states with $z > z_{\textrm{wh}}$. We leave further investigations on this topic for future work.

Last, there may be further implications for gravitational instantons if the WGC for axions is true: gravitational instantons may not be `stable' in the following sense. To be specific, consider a cored instanton with action $S$ and axion charge $n$ in a theory with $\alpha > 2 \sqrt{2/3}$. This corresponds to a tunnelling process between two configurations differing by $n$ units of axion charge. Let us then assume that the WGC is true and implies the existence of instantons with charge-to-mass ratio $z > z_0$, where $z_0$ is the charge-to-mass of an extremal gravitational instanton as before. An immediate consequence is that a tunnelling process will then always be dominated by the instantons predicted by the WGC. For our example this works as follows. The instantons needed to satisfy the WGC have $z > z_0 \geq z_{\textrm{cored}}$. Let two such instantons have $n_1, S_1$ and $n_2, S_2$, such that $n_1+n_2=n$. Since $z > z_{\textrm{cored}}$ it follows that $S_1+S_2 < S$ and tunnelling via two such instantons will dominate over tunnelling via the cored instanton.\footnote{Note that this is equivalent to the statement that (sub-)extremal black holes can in principle decay if the WGC for particles holds.} 
Hence, we do not expect `unstable' gravitational instantons to be relevant in the path integral computation of the instanton potential as the major contributions should arise from the instantons satisfying the WGC. We leave a more rigorous analysis of this issue for future work.
 
\section{Conclusions and Outlook} \label{Section:Conclusions}

It is of great interest to understand whether quantum gravity forbids periodic scalars with large field range and flat potential. The obvious way in which this can happen is via instanton-induced corrections. In detail, there are two specific options: On the one hand, quantum gravity may demand the presence of instantons with certain actions and charges, via a generalized weak gravity conjecture. This is rather indirect: One tries to show that certain things `go wrong' unless the relevant particles (or instantons) exist. 

There is, however, also a more direct approach: gravity itself supplies, in a rather direct or `constructive' way the instantons which may lift the flat potential. In the present paper, we have tried to push this direct approach as far as possible, striving also for maximal model-independence. 

Our results are as follows. We observe that in a pure axion-gravity system a potential for the axion is generated by Giddings-Strominger wormholes and that this potential is parametrically unsuppressed if the cutoff is at the Planck scale. Trying to be more precise about this, we encountered a surprise: If, as a model of high-cutoff quantum gravity, we take string theory at self-dual coupling and self-dual compactification radius, we are still left with a purely numerical suppression factor of $\exp(-3\pi^2)\simeq 10^{-13}$. Such a result makes it hard to hope for a strong constraint on inflation, even after further refining the analysis. 

Furthermore, we continued to ask for generic 4d constraints, but assuming more concretely that the 4d theory arises from string theory with a potentially low moduli scale. First, we found that in this setting nothing too dramatic happens to gravitational instantons: One linear combination of the moduli acts as a 4d dilaton governing the axion coupling; the instantons become more diverse in that extremal and cored gravitational instantons exist in addition to wormholes; the calculation still breaks down only at the Kaluza-Klein scale, which can of course still be high.

Unfortunately, the predictions now become model dependent as the coupling strength of the 4d dilaton to the axion (an ${\cal O}(1)$ numerical factor) enters. Taking the highest value for this factor that we could obtain in the simplest models results in a less severe instanton suppression factor of $\exp(-2\pi)\simeq 10^{-3}$. This is of course highly relevant for inflation, but easily avoided by considering models with different dilaton coupling.

In both of the above approaches, the suppression factors start out small and further fall as $\exp(-r^2)$, with $r$ an appropriately normalised compactification radius in 4d Planck units. As a result, while we do believe that gravitational instantons are the most fundamental and model-independent way to constrain field ranges, the numbers appear to allow for enough room for realistic large-field inflation. 

Finally, we have attempted to connect our analysis of the various types of gravitational instantons, including their dependence on the axion-dilaton coupling, to the ongoing discussion of the weak gravity conjecture. In particular, we found a intriguing regime where wormholes are the objects with highest charge-to-mass ratio and may thus be sufficient to satisfy the instanton-axion weak gravity conjecture. 

\bigskip

There are many directions for further investigations. By limiting our analysis to gravitational instantons in 4-dimensional Einstein-axion-dilaton theories we were unable to arrive at strong constraints on inflation. While this approach allows us to remain ignorant about the detailed UV completion, we are forced to neglect potentially more important contributions. These would arise from gravitational instantons with low instanton numbers, which are incalculable in the 4-dimensional Einstein-axion-dilaton theory. However, a quantitative analysis may be possible if one assumes that UV physics is described by string theory. It is expected that gravitational instantons will correspond to non-perturbative effects such as D-brane instantons in string theory. To arrive at stronger constraints a better understanding of non-perturbative effects in string theory is desirable. In particular, it is expected that poorly understood non-BPS instantons may become important during inflation \cite{150303886}.

There is a related question that is worthy of further examination. While more important instanton contributions to the axion potential may exist, it is possible that the overall effect on the axion potential vanishes once all such contributions are included. To calculate contributions from `more important' instantons is equivalent to studying instantons in a theory at a higher energy scale. However, taking string theory as our UV completion, we would expect the theory to become supersymmetric and/or higher-dimensional at some scale. It is then possible that, once we work above the supersymmetry scale, there are cancellations between the various instanton contributions to the axion potential. This is somewhat analogous to the cancellation between boson and fermion loops in supersymmetric field theory. We regard it as important to determine whether such cancellations can occur.

While we were unable to arrive at strong model-independent constraints on inflation, gravitational instantons may be important for inflation in models where the dilaton coupling takes sufficiently large values. In the effective 4-dimensional theory the dilaton coupling is a free parameter. However, one would expect that its value is constrained by possible UV completions. Indeed, by considering simple axion-dilaton systems in string compactifications, we find that the dilaton coupling typically takes $\mathcal{O}(1)$ values, i.e.~it can neither be very small nor very large. It would be interesting to examine to what extent these results are generic.

The upshot of these points is clear: It is imperative to understand the ultraviolet end of the instanton spectrum.

\subsubsection*{Acknowledgements}

We thank T.~Bachlechner, A.~Collinucci, A.~Font, B.~Heidenreich, K.-M.~Lee, E.~Kiritsis, E.~Palti, M.~Reece, F.~Rompineve, T.~Rudelius, M.~Salmhofer, J.~Stout, I.~Valenzuela and T.~Weigand for helpful discussions. We are also grateful to the referee for valuable comments on the manuscript. This work was partly supported by the DFG Transregional Collaborative Research Centre TRR 33 ``The Dark Universe''. P.M.~acknowledges financial support by the Studienstiftung des deutschen Volkes. S.T.~would like to thank the Institute for Theoretical Physics in Heidelberg for hospitality during several visits and the Klaus-Tschira foundation for generous support. He also thanks A.~Kashani-Poor and R.~Minasian for hospitality at IHP in Paris.


\begin{appendices}
	\section{Derivation of the Metric Structure of Gravitational Instantons} \label{Appendix: metric structure}
		We present a derivation of the metric \eqref{Metric} following \cite{07052768}, which also shows that $C$ arises as an integration constant. The most general $4$d line element with rotational symmetry is 
		\begin{equation} \label{Metric ansatz general}
			ds^2 = \lambda(r)dr^2 + r^2 d\Omega_3^2,
		\end{equation} 
		where $d\Omega_3^2$ represents the metric on $S^3$. Let us be more generic than in \eqref{Action with axion} and consider a set of moduli $\phi^I$ on moduli space with metric $G_{IJ}$: 
		\begin{equation}
		S = \int d^4x \sqrt{g} \left[- \frac{1}{2} R + \frac{1}{2}G_{IJ}(\phi) g^{\mu\nu} \partial_{\mu} \phi^I \partial_{\nu}\phi^J\right].
		\end{equation}
		Due to the rotational symmetry of our system we take $\phi^I=\phi^I(r)$. Variation of $S$ with respect to $\phi^K$ yields the equation of motion 
		\begin{equation}
		\left(\sqrt{g}g^{rr}G_{KJ}\phi^{\prime J}\right)^{\prime} - \frac{1}{2}\sqrt{g}g^{rr} \partial_K G_{JL} \phi^{\prime J} \phi^{\prime L} = 0.  
		\end{equation}  
	Here, the prime denotes the derivative with respect to the coordinate $r$ and $\partial_K$ the derivative with respect to $\phi^K$. Let us introduce a new variable $\tau$ such that $dr/d \tau = \sqrt{g}g^{rr}$. The equation of motion above can then be rewritten as the geodesic equation on moduli space, i.e. 
	\begin{equation}
	\partial_{\tau}^2 \phi^I + \Gamma^{I}_{JL} \partial_{\tau}\phi^J\partial_{\tau}\phi^L = 0, 
	\end{equation} 
	with Christoffel-symbols $\Gamma^{I}_{JL}$ for the metric $G_{IJ}$. Along the geodesics we then have 
	\begin{equation}
	\partial_{\tau} \left(G_{IJ} \partial_{\tau}\phi^I \partial_{\tau}\phi^J\right)=0
	\end{equation}
	or, expressed in the coordinate $r$,
	\begin{equation} \label{Constant along geodesics}
	G_{IJ} \phi^{\prime I}\phi^{\prime J} = \frac{k}{(\sqrt{g}g^{rr})^2} = \frac{k \lambda(r)}{r^6},
	\end{equation}
	where we introduced a constant $k$ and used \eqref{Metric ansatz general}.
	
	Furthermore, the $rr$-component of the energy-momentum tensor is 
	\begin{equation}
	T_{rr} = \frac{1}{2} G_{IJ} \phi^{\prime I} \phi^{\prime J} \stackrel{\eqref{Constant along geodesics}}{=} \frac{k \lambda(r)}{2r^6}.
	\end{equation} 
	The algebraic form of $\lambda(r)$ can now be read off from the $rr$-component of Einstein's equations. The $rr$-component of the Einstein tensor is
	\begin{equation}
	G_{rr} = \frac{3}{r^2} (1-\lambda(r))
	\end{equation}
	and hence $G_{rr}=T_{rr}$ yields 
	\begin{equation} \label{Lambda}
		\lambda(r) = \left(1+ \frac{C}{r^4}\right)^{-1},
	\end{equation}
	where $C= k/6$ is indeed an integration constant. It is interesting to note that the metric component $g_{rr}$ is determined independently of the functional form of $G_{IJ}(\phi)$.
	
	Also note that the metric is asymptotically flat, because  $\lambda(r) \to 1$ as $r \to \infty$.
	
	Finally, we want to remark that for the creation of Euclidean wormholes ($C<0$) it is necessary to have $G_{IJ} \phi^{\prime I}\phi^{\prime J} <0$ (see \eqref{Constant along geodesics}). While one cannot simply put a wrong sign into the kinetic term of the scalar fields, one can instead consider a Lagrangian with a 2-form gauge field, whose dual field is an axion. According to our discussions of quantum mechanical dualisation in \autoref{Subsection: Dualisation}, this axion is imaginary at the saddle-point of the path integral and effectively obtains an opposite sign in the kinetic term. Moreover, for solutions with $C \geq 0$ one necessarily needs to include dynamical scalar fields so that $G_{IJ} \phi^{\prime I}\phi^{\prime J} >0$.

	\section{Charge Quantisation} \label{Appendix: Charge Quantisation}
	
	Let us first recall how flux and charge quantisation usually work in a $B_2$-/$\theta_0$-theory with strings and fundamental instantons. For any 3-cycle $S^3$ we have 
	\begin{equation} \label{Dirac quantisation H}
		Q_B \int_{S^3} H_3 = 2 \pi n 
	\end{equation}
	with integer $n$.\footnote{This follows from assuming gauge invariance of the coupling term in \eqref{Action B}, i.e.~one can define $B_2$ with either the south- or north pole of $S^3$ removed, getting the same result in both cases. This is another argument to see the necessity of the $i$-factor in front of the coupling terms.} 
	Analogously, for any 1-cycle $S^1$, we have
	\begin{equation}
		Q_{\theta} \int_{S^1} F_1 = 2 \pi m, \label{Dirac quantisation F}
	\end{equation}
	$m \in \mathbbm{Z}$. Obviously, $n$ and $m$ can only be non-zero if the relevant cycle is either non-trivial in $M$ or if it encloses the appropriate charged object.
	
	The above are just the familiar flux quantisation conditions. In order to derive charge quantisation, we temporarily go back to Minkowskian space and use the equations of motion of 
	\begin{equation}
		S= -\int_M \frac{1}{2g_B^2}  H_3 \wedge \star H_3 + Q_B \int_M B_2 \wedge j_2,
	\end{equation}
	where $j_2$ is the current modelling the distribution of strings. It can be defined explicitly by $\int_{\Sigma}j_2 = N$, where $N$ is the number of strings intersecting some surface $\Sigma$. Without loss of generality we choose $N=1$. From the equation of motion for $B_2$, 
	\begin{equation}
		d (1/g_B^2 \star H_3)  = -Q_Bj_2,
	\end{equation}
	we find, using Stokes theorem: 
	\begin{equation}
		Q_B= -\int_{\partial \Sigma} 1/g_B^2\star H_3 = - \int_{\partial \Sigma} F_1 = \frac{2\pi m}{Q_{\theta}}.
	\end{equation} 
	In the last step we used $F_1$-flux quantisation. Thus, we see that 
	\begin{equation} \label{Dirac quantisation - Full}
		Q_B Q_{\theta} = 2\pi m,
	\end{equation}
	which is the well-known Dirac quantisation condition. For the following, we take the freedom to choose $Q_{\theta}=1$, i.e.~the periodicity of the axion field is in this case $\theta_0 \to \theta_0 + 2\pi$. 
	Then, combining \eqref{Dirac quantisation - Full} with $m=1$ (here we assume that a string with smallest charge exists) and \eqref{Dirac quantisation H}, we find that the quantisation condition on $H_3$ can simply be expressed as 
	\begin{equation} \label{Charge quantisation condition - Appendix}
		\int_{S^3}H_3  = n.
	\end{equation}
	This flux quantisation condition \eqref{Charge quantisation condition - Appendix} is at the heart of gravitational instanton solutions.
	
	Now, we are actually interested in potentials introduced by gravitational instantons, i.e., in shift symmetry breaking by quantum gravity. Hence, assuming the existence of fundamental instantons defeats the purpose. So let us see how far we get with the logic above if we abandon the source term in (\ref{Action Theta}).
	
	First, if we allow for geometries with non-trivial 3-cycles, the $H_3$ flux quantisation condition (\ref{Dirac quantisation H}) can still be derived. All we need is the existence of strings coupled to $B_2$. This then also implies that $Q_B$ is quantised. By contrast, (\ref{Dirac quantisation F}) cannot be derived without assuming the existence of fundamental instantons. 
	However, if we allow for geometries which also have non-trivial 1-cycles (see \autoref{Figure: Wormhole and 1-Cycle}), and if we postulate that the dual potential $\theta_0$ is 
	a globally defined function taking values on $S^1$ (i.e.~$\theta_0 
	\equiv \theta_0+2\pi$)), then both (\ref{Dirac quantisation F}) and charge quantisation, (\ref{Dirac quantisation - Full}) and (\ref{Charge quantisation condition - Appendix}), follow.

	\section{Dualisation under the Path Integral} \label{Appendix: Dualisation}
	
	In \autoref{Subsection: Dualisation} we are interested in computing 
	\begin{equation}
	\bra{H_3^{(F)}}e^{-HT}\ket{H_3^{(I)}} \sim \int_{\text{b.c.}} d[H_3]d[\theta_0] \exp \left\{- \int_M \frac{1}{2g_{B}^2}  \left(H_3 \wedge \star H_3 + 2 ig_B^2 \theta_0 dH_3 \right) \right\},
	\end{equation}
	which is \eqref{Partition function with Lagrange multiplier}. Here, $T \equiv t_F-t_I$. At the end we want to obtain a path integral over the variable $\theta_0$, i.e. \eqref{Partition function theta}. This is nothing but dualising from a set of canonical momentum variables to their generalised coordinates. 
	
	Thus, we illustrate the subtleties of the computation leading to \eqref{Partition function theta} by considering the quantum-mechanical harmonic oscillator, i.e.~$H=q^2/2+p^2/2$. The momentum $p$ then corresponds to the background flux $\braket{H_3}$ or, more precisely, to the quantised charge $n$, while the position variable $q$ corresponds to $\theta_0$.  The transition amplitude from state $\ket{q_I}$ to $\ket{q_F}$ reads 
	\begin{equation} \label{Transition amplitude q}
	\bra{q_F} e^{-HT} \ket{q_I} = \int d[p] \int_{\text{b.c.}} d[q] \exp\left\{\int_{t_I}^{t_F} dt \left(ip \dot{q} - H(q,p)\right)\right\},
	\end{equation}
	with boundary conditions $q(t_I) = q_I$ and $q(t_F)=q_F$ imposed. 
	
	In fact we rather want to compute $\bra{p_F}e^{-HT}\ket{p_I}$, which is expressed similarly: 
	\begin{align} \label{Transition amplitued p}
	\bra{p_F}e^{-HT}\ket{p_I} &= \int d[q] \int_{\text{b.c.}} d[p] \exp\left\{\int_{t_I}^{t_F} dt \left(-iq \dot{p} - H(q,p)\right)\right\} \\ \nonumber
	&= \int d[q] \int_{\text{b.c.}} d[p] \exp\left\{\int_{t_I}^{t_F} dt \left(ip \dot{q} - H(q,p)\right)\right\} \exp \left\{-i(q_Fp_F - q_Ip_I)\right\}, 
	\end{align}
	where we impose again $p(t_I) = p_I$ and $p(t_F)=p_F$. In the second step we integrated the first term of the exponential by parts. In our case we have $H=q^2/2+p^2/2$ which allows us to complete the square. Integrating out $p$ without worrying about the boundary conditions to be imposed yields the desired result 
	\begin{equation} \label{Path integral over q}
	\bra{p_F}e^{-HT}\ket{p_I}  \sim \int d[q] \exp \left(-i(q_Fp_F - q_Ip_I)\right)e^{-S[q]},
	\end{equation}
	see also \cite{Feynman:1951gn} for comments on the integration over the momentum. 
	We wish to have a closer look at this decisive step.  
	
	To do so, we write the amplitude $\bra{p_F}e^{-HT}\ket{p_I}$ as: 
	\begin{equation} \label{Transition amplitute Fourier transform}
	\bra{p_F}e^{-HT}\ket{p_I} = \int dq_I dq_F \braket{p_F|q_F}\bra{q_F} e^{-HT} \ket{q_I} \braket{q_I|p_I}. 
	\end{equation}
	Now let us assume that the two dual relations \eqref{Transition amplitude q} and \eqref{Transition amplitued p} hold. Then, in particular, \eqref{Transition amplitude q} implies 
	\begin{equation}
	\bra{q_F} e^{-HT} \ket{q_I} = \int_{\text{b.c.}} d[q]e^{-S[q]},
	\end{equation}
	and the result \eqref{Path integral over q} follows immediately (use $\braket{p|q}=e^{-ipq}$). The operation of integrating out $p$ while disregarding its boundary conditions is thereby indirectly justified. 
	
	Finally, we can demonstrate this directly and explicitly by writing\footnote{We are grateful to K.-M.~Lee for pointing out this possibility and for further discussions on this issue. See also \cite{Coleman:1989zu}.}
	\begin{align} \nonumber
	& \bra{p_F}e^{-HT}\ket{p_I}   \\
\nonumber =& \int \prod_{m=0}^{N}dq_m \prod_{n=0}^{N-1} dp_n \braket{p_F|q_N}\braket{q_N|e^{-H \epsilon}|p_{N-1}}\braket{p_{N-1}|q_{N-1}}\braket{q_{N-1}|e^{-H \epsilon}|p_{N-2}} \ldots \\
	&\hphantom{\int \prod_{m=0}^{N}dq_m \prod_{n=0}^{N-1} dp_n} \ldots  \braket{p_1|q_1}\braket{q_1|e^{-H \epsilon}|p_0}\braket{p_0|q_0}\braket{q_0|p_I},
	\end{align}
	where $\epsilon \equiv T/(N+1)$ and $q_0=q_I$, $q_N=q_F$.
	This becomes the discretised version of \eqref{Transition amplitued p}:
	\begin{align} \nonumber
	\bra{p_F}e^{-HT}\ket{p_I} = \int \prod_{m=0}^{N}dq_m \prod_{n=0}^{N-1} dp_n \, & e^{-iq_N(p_F-p_{N-1})-H(q_N,p_{N-1})\epsilon} \ldots \\
	& \ldots e^{-iq_1(p_1-p_0)-H(q_1,p_0)\epsilon}e^{-iq_0(p_0-p_I)}.
	\end{align} 
	For the harmonic oscillator (and in fact for more general potentials $V(q)$) we can integrate out $p_0,...,p_{N-1}$ (after completing the square for each $p_m$). As a result we find
	\begin{align} \nonumber
	\bra{p_F}e^{-HT}\ket{p_I} \sim \int \prod_{m=0}^{N}dq_m \, &\exp \left\{-iq_Np_F\right\}  \exp \left\{-\frac{q_N^2}{2}\epsilon - \frac{(q_N-q_{N-1})^2}{2 \epsilon}\right\} \ldots \\ 
	&\ldots \exp \left\{-\frac{q_1^2}{2}\epsilon - \frac{(q_1-q_{0})^2}{2 \epsilon}\right\}  \exp \left\{iq_0p_I\right\}. 
	\end{align}  
	This is precisely the discretised version of \eqref{Path integral over q}. Hence, integrating out the momenta from \eqref{Transition amplitued p} to \eqref{Path integral over q} without considering the boundaries is indeed justified.

	\section{Analytical Solutions to Einstein's Equation} \label{Appendix: analytical solution to Einstein}
	
	Einstein's equation \eqref{Einstein equation without potential} which follows from the action \eqref{Action without potential} can be solved analytically. We explain how to arrive at solutions \eqref{Solution for C<0, with constant}, \eqref{Solution for C=0}  and \eqref{Solution for C>0} for $C<0$, $C=0$ and $C>0$, respectively. 
	\bigskip 
	
	First of all, \eqref{Einstein equation without potential} can be rewritten as: 
	\begin{equation}
	\pm \int d \varphi \frac{1}{\sqrt{ \mathcal{F}(\varphi)n^2/A^2 + C }} =  \sqrt{6}\int dr \frac{1}{r^3 \sqrt{1+C/r^4}}. 
	\end{equation} 
	Thus, integral representations can in principle be obtained for any $\mathcal{F}$. For $\mathcal{F}(\varphi)=1/(3!f_{\text{ax}}^2) \exp(-\alpha \varphi)$, explicit solutions exist.
	
	For the RHS, one finds
	\begin{equation}
	\sqrt{6}\int dr \frac{1}{r^3 \sqrt{1+C/r^4}} = \begin{cases}
	- \sqrt{\frac{3}{2|C|}} \arcsin \left(\frac{\sqrt{|C|}}{r^2}\right) + \text{const}  & \text{for} ~ C<0 \\
	- \sqrt{\frac{3}{2C}} \text{arcsinh} \left(\frac{\sqrt{C}}{r^2}\right) + \text{const} & \text{for} ~ C>0  \\
	- \sqrt{\frac{3}{2}} \frac{1}{r^2} + \text{const} & \text{for}~ C=0,
	\end{cases}
	\end{equation}
	where we use the substitution $y= \sqrt{|C|}/r^2$ for $C \neq 0$.
	The integral on the LHS can be rewritten as 
	\begin{equation}
	\pm \frac{1}{\sqrt{|C|}} \int d \varphi \frac{1}{\sqrt{k\exp(-\alpha \varphi) \pm  1 }},  
	\end{equation}
	with $k \equiv n^2/(3!|C| A^2 f_{\text{ax}}^2)$. If $C>0$ ($C<0$), the positive (negative) sign under the square root applies. In the case of $C>0$ we substitute 
	\begin{equation}
	\sinh y = \frac{1}{\sqrt{k}} \exp(\alpha \varphi /2), 
	\end{equation}
	and for $C<0$ we take 
	\begin{equation}
	\sin y = \frac{1}{\sqrt{k}} \exp \left(\alpha \varphi/2\right).
	\end{equation}
	Using appropriate identities for the hyperbolic or trigonometric functions, one arrives at 
	\begin{align}
	\pm \int d \varphi & \frac{1}{\sqrt{ \exp(-\alpha \varphi)n^2/(3!f_{\text{ax}}^2A^2) + C }} = \\ \nonumber
	&=   
	\begin{cases}
	\pm \frac{2}{\sqrt{|C|} \alpha} \left[\arcsin \left(\frac{1}{\sqrt{k}}\exp\left(\alpha \varphi/2\right)\right) - \text{const} \right] &  \text{for} ~ C<0  \\
	\pm \frac{2}{\sqrt{C} \alpha} \left[\text{arcsinh} \left(\frac{1}{\sqrt{k}}\exp(\alpha \varphi/2)\right) - \text{const}\right]  & \text{for} ~ C>0\\
	\pm \frac{2\sqrt{6}Af_{\text{ax}}}{n \alpha} \exp \left( \alpha \varphi /2\right) + \text{const} & \text{for} ~ C=0.
	\end{cases}
	\end{align}
	From here one can read off the solutions, which can be rewritten as \eqref{Solution for C<0, with constant}, \eqref{Solution for C=0} or \eqref{Solution for C>0}.
	
	\section{Computation of the Instanton Action} \label{Appendix:Instanton Action}
	
	We present further details of the computation of the instanton action in \autoref{Section:Cored Grav Inst}. The computation consists of determining the on-shell contribution from the action and the contribution coming from the Gibbons-Hawking-York boundary term. We begin by looking at the latter, where we follow \cite{150906374}.  
	
	\subsubsection*{Gibbons-Hawking-York boundary term:}
	
	The Gibbons-Hawking-York boundary term is  
	\begin{equation} 
	S_{\text{GHY}} = - \oint_{\partial M} d^3x \sqrt{h}(K-K_0),
	\end{equation}
	as described around \eqref{Gibbons-Hawking-York}. Starting from our metric ansatz \eqref{Metric} we choose hypersurfaces of constant $r$. The normal unit vector $n$ is then 
	\begin{equation}
	n = \sqrt{1+ \frac{C}{r^4}} \frac{\partial}{\partial r}.
	\end{equation}
	The trace of the extrinsic curvature is  
	\begin{equation}
	K = \nabla_{\mu}n^{\mu} = \partial_{\mu}n^{\mu} + \Gamma^{\mu}_{\mu\nu}n^{\nu}
	\end{equation}
	where $\nabla$ is the Levi-Civita connection on $M$. One finds 
	\begin{equation}
	\Gamma^{\mu}_{\mu r} = \frac{2C}{r^5} \left(1+ \frac{C}{r^4}\right)^{-1} + \frac{3}{r},
	\end{equation}
	and therefore
	\begin{equation}
	K=\nabla_{\mu}n^{\mu} = \frac{3}{r}\left(1+ \frac{C}{r^4}\right)^{1/2}.
	\end{equation}
	By taking $C=0$ we can also read off the  trace of the extrinsic curvature of $\partial M$ embedded in $\mathbb{R}^4$: 
	\begin{equation}
	K_0 = \frac{3}{r}.
	\end{equation} 
	It then follows 
	\begin{equation}
		S_{\text{GHY}} = - \oint_{\partial M} \epsilon_{S^3} (K-K_0) = - 3Ar^2 \left.\left[\left(1+ \frac{C}{r^4}\right)^{1/2}-1\right]\right|_{\text{boundary}},
	\end{equation}
	with surface area $A=2\pi^2$ of $S^3$. Recall that according to our conventions the volume form on $S^3$ contains a factor $r^3$. Clearly, for $C=0$ we have $S_{\text{GHY}}=0$. For $C>0$ the boundary is at  $r=0$ and at $r=\infty$, 
	\begin{equation}
	S_{\text{GHY}} = 3AC^{1/2}, ~~~~~~~~ C>0.
	\end{equation}
	In the case of $C<0$ the integral vanishes, because we always consider instanton-anti-instanton pairs, so $S_{\text{GHY}} =0$.
	
	These are the results used in \autoref{Section:Cored Grav Inst}.
	
	\subsubsection*{On-shell contribution:}
	
	We now evaluate the bulk action \eqref{Action without potential} on-shell, i.e.~we plug in the equations of motion successively. As described in \autoref{Section:Cored Grav Inst}, the first step is to express the Ricci scalar $R$ by the trace of the energy-momentum tensor using Einstein's equations: 
	\begin{equation}
	R=-T.
	\end{equation}
	The energy-momentum tensor $T_{\mu\nu}$ from the action \eqref{Action without potential} is 
	\begin{equation}
	T_{\mu\nu} = - g_{\mu\nu} \left[\frac{1}{2}\mathcal{F}(\varphi)H^2 + \frac{1}{2} \partial_{\rho}\varphi \partial^{\rho}\varphi \right] + 3 \mathcal{F}(\varphi)H_{\mu\rho\sigma}H_{\nu}^{~\rho\sigma} + \partial_{\mu}\varphi \partial_{\nu}\varphi.
	\end{equation}
	Consequently, 
	\begin{equation}
	T = g^{\mu\nu}T_{\mu\nu} =  \mathcal{F}(\varphi)H^2 - (\partial \varphi)^2, 
	\end{equation}
	and then \eqref{Action without potential} becomes simply
	\begin{equation}
	S = \int d^4x  \sqrt{g} \mathcal{F}(\varphi)H^2 = A \int dr \frac{r^3}{\sqrt{1+C/r^4}} \mathcal{F}(\varphi)H^2,
	\end{equation}
	where we used the rotational symmetry of our system. Next, we plug in the solution to \eqref{EoM for B}, 
	\begin{equation}
	H=\frac{n}{Ar^3} \epsilon,
	\end{equation} 
	and restrict ourselves to $\mathcal{F}(\varphi)=\exp(-\alpha \varphi)/(3!f_{\text{ax}}^2)$, for which we know the analytical solutions: 
	\begin{equation}
		S =  \frac{n^2}{Af_{\text{ax}}^2} \int dr \frac{1}{r^3\sqrt{1+C/r^4}} \exp(-\alpha \varphi).
	\end{equation}
	It is then convenient to rewrite the action as an integral over $d\varphi$ using Einstein's equation \eqref{Einstein equation without potential}. We consider only regular solutions. They are monotonically decreasing and therefore we have $\varphi^{\prime}(r)<0$ everywhere. Hence, 
	\begin{equation}
	S = - \frac{n^2}{Af_{\text{ax}}^2} \int d \varphi \frac{\exp(-\alpha \varphi)}{\sqrt{n^2 \exp(-\alpha \varphi)/(A^2f_{\text{ax}}^2)+6C}}.
	\end{equation}  
	The integral has to be evaluated case by case. 
	
	For extremal gravitational instantons with $C=0$ we have 
	\begin{equation}
	S = - \frac{n}{f_{\text{ax}}} \int_{\varphi(0)}^{\varphi(\infty)} d \varphi \exp(-\alpha \varphi/2) = \frac{2 n}{\alpha f_{\text{ax}}}.
	\end{equation} 
	In the case of $C>0$ we obtain 
	\begin{align} \nonumber
	S &= - \frac{n^2}{Af_{\text{ax}}^2} \int_{\varphi(0)}^{\varphi(\infty)} d \varphi \frac{\exp(-\alpha \varphi)}{\sqrt{n^2 \exp(-\alpha \varphi)/(A^2f_{\text{ax}}^2)+6C}} \\
	&= \frac{2n}{\alpha f_{\text{ax}}} \left.\sqrt{\exp(-\alpha \varphi)+ \sinh^2 K_+} \right|_{\varphi(0)}^{\varphi(\infty)} = \frac{2n}{\alpha f_{\text{ax}}} e^{-K_+},
	\end{align} 
	where we used \eqref{C for C>0} and took $K_+>0$. Combining this with the GHY boundary term yields the desired instanton action \eqref{Instanton action for C>0}.
	
	Finally, for Euclidean wormholes, i.e.~for $C<0$, we have 
	\begin{equation}
		S = \int d^4x  \sqrt{g} \mathcal{F}(\varphi)H^2 = 2 \times A \int_{r_0}^{\infty} dr \frac{r^3}{\sqrt{1-|C|/r^4}} \mathcal{F}(\varphi)H^2,
	\end{equation}
	where the factor of two occurs because the left integral is over the whole Euclidean space, and hence accounts for the whole wormhole and thus for the instanton and anti-instanton, while the integral on the RHS integrates from the centre of the wormhole to one end. The appearance of this factor may be seen more easily by evaluating the integral on the LHS using the $t$-coordinate \eqref{Metric with coordinate t} and then changing coordinates from $t$ to $r$.  As was noted in \cite{150307853}, this contribution has to be divided by two, because the instanton action $S_{\text{inst}}$ should only take into account half of the full wormhole action. Consequently, using the equations of motion as in the previous cases,
	\begin{align} \nonumber
	S_{\text{inst}} &= - \frac{n^2}{Af_{\text{ax}}^2} \int_{\varphi(r_0)}^{\varphi(\infty)} d \varphi \frac{\exp(-\alpha \varphi)}{\sqrt{n^2 \exp(-\alpha \varphi)/(A^2f_{\text{ax}}^2)-6|C|}}  \\
	&= \frac{2n}{\alpha f_{\text{ax}}} \left| \sin \left(\frac{\alpha \pi}{4}\sqrt{\frac{3}{2}}\right) \right|.
	\end{align}
	Hence,  \eqref{Instanton action C<0} follows, where we can drop the modulus due to the restriction to $0 \leq \alpha < 2 \sqrt{2/3}$.

		 \section{Estimating the Size of the Prefactor $\mathcal{A}$ in the Instanton Potential} \label{Appendix: Prefactor}
		 
		 The contribution of gravitational instantons to the axion potential is given by $\delta V = \mathcal{A}e^{-S}\cos (n\theta)$. While it has been proposed e.g.~in \cite{150303886,150300795} that $\mathcal{A} \sim 1$ (in Planck units), we attempt a somewhat more precise estimate. This is inspired by the analogies between gravitational instantons and instantons arising from Euclidean branes wrapping an internal cycle of the compactification manifold (see e.g.~\cite{150300795,150303886,150906374}). Let us start by recalling how the latter contributes to the supergravity $F$-term potential in a simple setup.
		 
		 We consider a Euclidean brane instanton modifying the perturbative superpotential $W_0$ as
		 \begin{equation}
		 	W = W_0 + A(z) e^{-a T},\label{ww0}
		 \end{equation}
		 where $z$ denotes the complex structure moduli and $T$ is a K\"ahler modulus. 
		 
		 Then the supergravity $F$-term potential 
		 \begin{equation}
		 	V_F = e^K \left(K^{i \bar{\jmath}} D_iW D_{\bar{\jmath}} \overline{W} - 3 |W|^2\right)
		 \end{equation}
		 is corrected at leading order by 
		 \begin{equation}
		 	\delta V \sim e^K W_0 A(z) e^{-a \tau} \ ,
		 \end{equation}
		 where $\tau$ is the real part of $T$. Recall that $K = - 2 \ln \mathcal{V} + ... $, which gives a suppression by $1/\mathcal{V}^2$. Furthermore, we rewrite the above expression in terms of the gravitino mass $m_{3/2} \sim W_0/ \mathcal{V}$ and the KK-scale $m_{\text{KK}} \sim 1/\mathcal{V}^{2/3}$: 
		 \begin{equation}
		 	\delta V \sim \frac{1}{\mathcal{V}^{5/3}} \frac{m_{3/2}}{m_{\text{KK}}} A(z) e^{-a \tau}.
		 \end{equation}
		 If we were allowed to compare this with \eqref{Delta V of gravitational instantons} then, using $m_{3/2} \lesssim m_{\text{KK}}$, we would conclude that 
		 \begin{equation}
		 	\mathcal{A} \lesssim \frac{A(z)}{\mathcal{V}^{5/3}}\label{calae}
		 \end{equation}
		 in Planck units. Here we identified $\exp(-a\tau)$ with $\exp(-S)$ motivated by the obvious analogy: Indeed, the Euclidean brane action is proportional to the brane tension and the volume of the cycle. Similarly, the action of a cored gravitational instanton is proportional to the ADM tension of a black brane wrapping a cycle in a higher-dimensional version of the gravitational instanton system, see e.g.~\cite{150906374} for an example.  
		 
		 Nevertheless, our proposal to estimate ${\cal A}$ by \eqref{calae} remains nontrivial. Indeed, we first need to consider a large wrapping number $n$ to relate to the calculable regime on the gravitational side. This is unproblematic in the present case since these higher instantons will contribute to $W$ analogously to \eqref{ww0}. Next, we are \textit{not} interested in Euclidean brane instantons (their effect is well-known) but in some possibly very different type of instanton arising in a string model and not having a simple microscopic description. The claim or proposal implicit in \eqref{calae} is then that this instanton may, conservatively, also be suppressed by a factor ${\cal A}$ which becomes small as the KK-scale and SUSY breaking scales go down. This appears to be reasonable since, beyond the simple Euclidean brane case discussed here, higher-dimensional and SUSY-based cancellations are expected to occur above those scales. 
		 
		 Accepting the above proposal, compactification volumes in the range $\mathcal{V} \sim 10^2$ to $10^3$ imply $\mathcal{A} \sim 10^{-4}$ and $10^{-5}$, respectively, assuming that $A(z) = \mathcal{O}(1)$. Note that in order to avoid destabilisation of the K\"ahler moduli the compactification volume is at most of order $\mathcal{O}(10^3)$, see e.g.~\cite{14043542,14043711,14112032}. Nevertheless, the suppression by $e^{-S}$ remains dominant in all regimes we considered.   
	 
\end{appendices}

\newpage 

\begingroup
\raggedright
\sloppy
\bibliographystyle{apsrev4-1}
 \nocite{apsrev41control}
\bibliography{apsrevcontrol,literature}
\endgroup

\end{document}